\documentclass[fleqn,10pt]{wlscirep}
\usepackage[utf8]{inputenc}
\usepackage[T1]{fontenc}
\usepackage{xspace}
\newcommand{\GeV}{\mathrm{GeV}}
\usepackage[page,title]{appendix}
\usepackage{url}
\usepackage{graphicx}
\usepackage{amssymb}
\usepackage{bm}
\usepackage{mathrsfs}
\usepackage{amsmath}
\usepackage{amsfonts}
\usepackage{color}
\usepackage{xspace}
\usepackage{tikz}
\usepackage{slashed}
\usepackage{cancel}
\usepackage{hyperref}
\usepackage[utf8]{inputenc}  
\usepackage{xcolor}
\usepackage{slashed}
\usepackage{fancyhdr}

\usepackage{graphicx}

\newcommand\UVIC{University of Victoria, Victoria, BC V8P 5C2, Canada}

\newcommand\FNAL{Fermi National Accelerator Laboratory, Batavia, IL 60510, USA}
\newcommand\MSU{Michigan State University, East Lansing, Michigan 48824, USA}
\newcommand\UM{University of Michigan, Ann Arbor, MI 48109, USA}

\begin{document}

\title{DarkQuest: A dark sector upgrade to SpinQuest at the 120~GeV Fermilab Main Injector
}

\author[1]{Aram Apyan}
\author[2]{Brian Batell}
\author[3]{Asher Berlin}
\author[4]{Nikita Blinov}
\author[5]{Caspian Chaharom}
\author[6]{Sergio Cuadra}
\author[5]{Zeynep Demiragli}
\author[7]{Adam Duran}
\author[3]{Yongbin Feng}
\author[8]{I.P. Fernando}
\author[9]{\mbox{Stefania Gori}}
\author[6]{Philip Harris}
\author[6]{Duc Hoang}
\author[8]{Dustin Keller}
\author[10]{Elizabeth Kowalczyk}
\author[2]{Monica Leys}
\author[11]{Kun Liu}
\author[11]{Ming Liu}
\author[12]{Wolfgang Lorenzon}
\author[13]{Petar Maksimovic}
\author[3]{Cristina Mantilla Suarez}
\author[14]{Hrachya Marukyan}
\author[13]{Amitav Mitra}
\author[15]{Yoshiyuki Miyachi}
\author[6]{Patrick McCormack}
\author[6]{Eric A. Moreno}
\author[11]{Yasser Corrales Morales}
\author[6]{Noah Paladino}
\author[2]{Mudit Rai}
\author[6]{Sebastian Rotella}
\author[5]{Luke Saunders}
\author[21]{Shinaya Sawada}
\author[17]{Carli Smith}
\author[5]{David Sperka}
\author[3]{Rick Tesarek}
\author[3]{Nhan Tran}
\author[18]{Yu-Dai Tsai}
\author[5]{Zijie Wan}
\author[12]{Margaret Wynne}

\affil[1]{Brandeis University, Waltham, MA 02453, USA}
\affil[2]{University of Pittsburgh, Pittsburgh, PA 15260, USA}
\affil[3]{\FNAL}
\affil[4]{\UVIC}
\affil[5]{Boston University, Boston, MA 02215, USA}
\affil[6]{Massachusetts Institute of Technology, Cambridge, MA 02139, USA}
\affil[7]{San Francisco State University, San Francisco, CA 94132, USA}
\affil[8]{University of Virginia, Charlottesville, VA 22904, USA}
\affil[9]{University of California Santa Cruz, Santa Cruz, CA 95064, USA}
\affil[10]{\MSU}
\affil[11]{Los Alamos National Laboratory, Los Alamos, NM 87545, USA}
\affil[12]{\UM}
\affil[13]{Johns Hopkins University, Baltimore, MD 21218, USA}
\affil[14]{Yamagata University, Yamagata, 990-8560, Japan}
\affil[15]{KEK Tsukuba, Tsukuba, Ibaraki 305-0801 Japan}
\affil[16]{Yerevan Physics Institute, Yerevan, 0036, Republic of Armenia}
\affil[17]{Penn State University, State College, PA 16801, USA}
\affil[18]{University of California Irvine, Irvine, CA 92697, USA}


\begin{abstract}
Expanding the mass range and techniques by which we search for dark matter is an important part of the worldwide particle physics program. 
Accelerator-based searches for dark matter and dark sector particles are a uniquely compelling part of this program as a way to both create and detect dark matter in the laboratory and explore the dark sector by searching for mediators and excited dark matter particles. This paper focuses on developing the DarkQuest experimental concept and gives an outlook on related enhancements collectively referred to as LongQuest. DarkQuest is a proton fixed-target experiment with leading sensitivity to an array of visible dark sector signatures in the MeV-GeV mass range. Because it builds off of existing accelerator and detector infrastructure, it offers a powerful but modest-cost experimental initiative that can be realized on a short timescale. 
\end{abstract}

\maketitle

\vspace{1cm}
\noindent\makebox[\linewidth]{\rule{0.5\paperwidth}{0.4pt}}
\begin{center}
\normalsize \textit{Submitted to the Proceedings of the US Community Study on the Future of Particle Physics (Snowmass 2021)} 
\end{center}
\noindent\makebox[\linewidth]{\rule{0.5\paperwidth}{0.4pt}}

\clearpage

\tableofcontents
\clearpage

\section{Introduction}
\label{sec:intro}
Although overwhelming astrophysical and cosmological evidence supports the existence of Dark Matter (DM), its identity, interactions, and origin remain elusive. The last few years have seen tremendous progress in experimental searches for weakly interacting dark matter with masses $\mathcal{O}$(10--1000 GeV). The lack of discovery in this traditional mass range has generated interest in searching for DM candidates with masses below a few GeV. DM candidates with masses between MeV and GeV are particularly well motivated since they can be in thermal equilibrium with the Standard Model (SM) and get the measured relic abundance through the freeze-out mechanism.

Because of the Lee-Weinberg bound, thermal DM in the MeV-GeV range generically requires the existence of a \emph{dark sector}, composed of one or more particles not charged under the SM gauge symmetries. Some (not too small) coupling between the dark mediator and the SM is required to guarantee dark sector-SM thermal contact and/or to obtain the measured DM relic abundance.  As a result, the experimentally viable models feature small couplings, leading to small cross-sections and often macroscopic decay lengths of new states. In addition, enhanced symmetries of DM models often lead to long-lived dark particles giving further motivation to search for long-lived visibly-decaying dark particles. 

Expanding the mass range and techniques by which we search for dark matter is an important part of the worldwide particle physics program and has been highlighted in the Basic Research Needs (BRN) Report for Dark Matter Small Projects New Initiatives~\cite{BRNreport}.  Accelerator-based searches for dark matter are a uniquely compelling part of this program as a way to both create and detect dark matter in the laboratory and explore the dark sector by searching for mediators and excited dark matter particles. Recently, there has been an extensive community effort to explore dark sector physics from a model-building, phenomenological, as well as experimental perspective~\cite{BRNreport,battaglieriUSCosmicVisions2017,alexanderDarkSectors20162016c}. This has spurred a worldwide low-mass dark matter search program at the LHC, RHIC, Fermilab, JLab, BaBar/SLAC, Belle/KEK, and other facilities. 

\subsection*{DarkQuest}
This paper details the DarkQuest experimental concept as a high sensitivity, near-term, modest-cost opportunity to explore extended regions of new parameter space in dark sector physics scenarios. DarkQuest is a proton fixed target beam dump spectrometer experiment on the neutrino-muon beamline of the Fermilab Accelerator Complex 
receiving a high-intensity beam of 120~GeV protons from the Main Injector. 

\begin{figure}[tbh]
    \centering
    \includegraphics[width=0.9\textwidth]{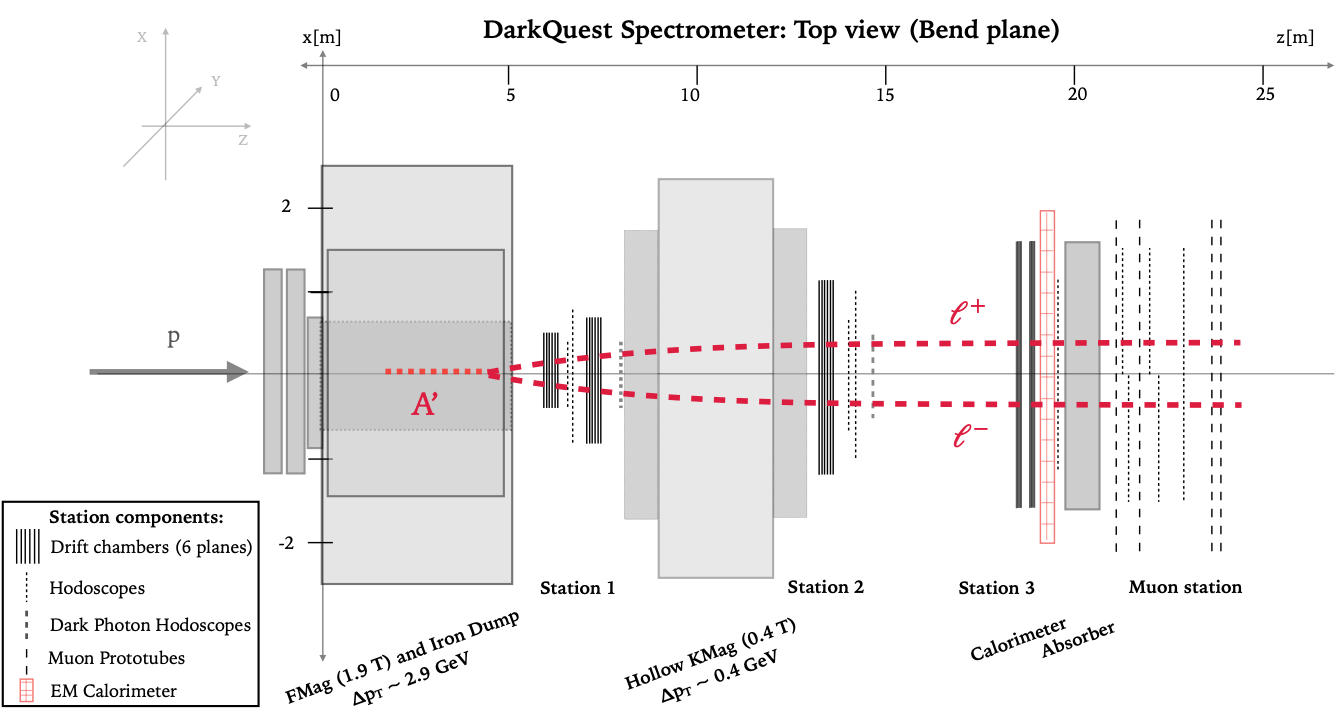}
    \caption{Top view of the proposed DarkQuest detector upgrade to the SpinQuest experiment adapted from Ref.~\cite{seaquestcollaborationSeaQuestSpectrometerFermilab2017a}. Different components of the spectrometer are drawn approximately to scale: a focusing magnet (FMag) and beam dump placed $\approx$1~m downstream the target, an open-aperture magnet (KMag) to sweep away soft SM radiation, a system of drift chambers and scintillator hodoscopes grouped into stations that serve for tracking and triggering, respectively and an absorber placed upstream the muon prototube station. The DarkQuest detector upgrade has an electromagnetic calorimeter (in red) to extend the detection capability to electrons, pions and photons. The calorimeter is drawn as currently placed in the simulation (before the absorber), but, when installed, it will require the drift chambers in Station 3 to be moved further upstream.}
    \label{fig:darkquest}
\end{figure}

DarkQuest distinguishes itself from other experimental initiatives in that it can cover these physics opportunistically on a short timescale with very modest resources. DarkQuest takes advantage of the long history of investment in the existing E906/E1039 SeaQuest/SpinQuest spectrometer experiments at Fermilab, which focus on proton parton distribution function measurements in Drell-Yan events. The experimental apparatus is illustrated in Fig.~\ref{fig:darkquest}.

SpinQuest is planning to run from 2022-2023 and is focused on a nuclear physics program. Leveraging the significant existing infrastructure and past investment in both the beamline and detectors, this paper describes a program to extend the SeaQuest/SpinQuest detector capability and launch a synergistic dark matter and dark sector search program on the timescale of 2024 and beyond. In particular, to gain full sensitivity to dark sector searches, we reuse sectors from the decommissioned electromagnetic calorimeter (EMCal) of the PHENIX experiment to extend the DarkQuest detection capability to electrons, pions, and photons. By recycling the EMCal, we significantly reduce the overall cost and allow for efficient prototyping and installation. 

This paper is structured as follows.  In Sec.~\ref{sec:physics}, we introduce the physics goals and impact of this proposal, including the proton fixed target beam dump technique, the connection to other experimental initiatives, and how the proposal fits with the Dark Matter Small Projects Initiative.  In Sec.~\ref{sec:detector}, we discuss the experimental apparatus and its key performance parameters.  We detail the existing detector elements of the SeaQuest/SpinQuest spectrometer and the expected performance of the EMCal for the DarkQuest physics program. Section~\ref{sec:performance} provides a detailed description of simulation studies performed to understand the performance of the DarkQuest detector for dark sector searches. Section~\ref{sec:sensitivity} discusses the sensitivity of the DarkQuest experiment to certain physics scenarios. In Section~\ref{sec:outlook}, we will lay out the path forward and discuss further longer term ideas on an expanded physics program using the SpinQuest/DarkQuest detector apparatus.

\newpage
\section{DarkQuest Science Goals and Physics Signatures}
\label{sec:physics}
The SNOWMASS RF6 group has identified three Big Ideas that can be tested at various high-intensity experiments, corresponding to probing different classes of dark sectors. These directions are organized by their main signature: (1) production of invisible BSM states that could be responsible for the DM of the universe, (2) production of a mediator particle to the dark sector and detection of its decay back into SM states and (3) production of mixed visibly-decaying and invisible states, signaling the existence of a rich dark sector.

In this Section we argue that DarkQuest can test theoretical scenarios that exemplify each of these main ideas. We begin by describing dark sector searches at fixed target experiments in general and outline the specific capabilities of DarkQuest. We then define several specific benchmark scenarios that embody the key RF6 directions and show that DarkQuest can reach unexplored parameter space in each case. The purely-SM signals are studied in the context of dark photon, Higgs portal scalars, sterile neutrinos and axion-like particle models (ALPs), while the DM and rich dark sectors are captured by models of inelastic dark matter (iDM) and  Strongly Interacting Massive Particles (SIMPs). The latter models offer the possibility of explaining the dark matter of the universe in a predictive framework with concrete experimental targets.

\subsection{Dark sector searches at accelerators}
In a spectrometer-based experiment such as DarkQuest, weakly coupled dark sector states with lifetimes of order 1i\,m, once produced, can be detected through their displaced decays to visible SM particles. Fixed target proton beam dump experiments, in particular, offer several key advantages in probing light dark sectors. First, enormous luminosities can be attained, allowing access to tiny couplings. Second, light particles are typically produced with a significant boost enhancing their decay length in the laboratory frame. This ensures that the decays of these particles can occur behind a shield that filters out SM backgrounds. Third, in many scenarios, including those with dark photon mediators, high-intensity {\it \bf proton} beams dumps provide the largest production rates of dark sector particles, relative to lepton beams. Secondary fluxes of hadrons, photons, and muons produced in the proton-target reactions can lead to production mechanisms for a variety of dark - visible sector couplings. In particular, high-energy protons beams, such as the 120 GeV Fermilab Main Injector, offer kinematic access to heavier dark particles in the mass range of $\sim$10 GeV and below. 

\begin{figure}[tbh!]
\centering
\includegraphics[width=0.44\textwidth]{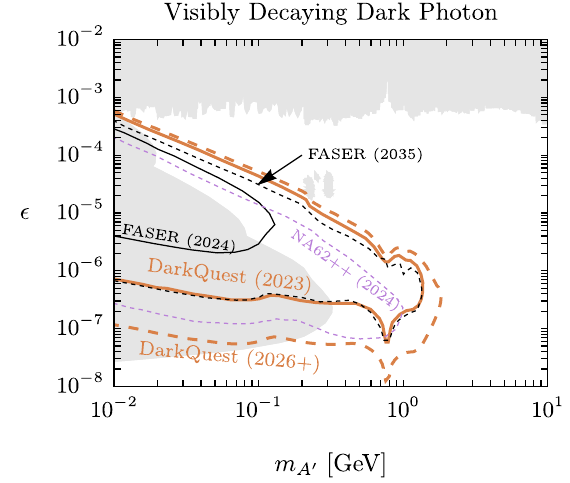} ~~~~~~~
\includegraphics[width=0.37\textwidth]{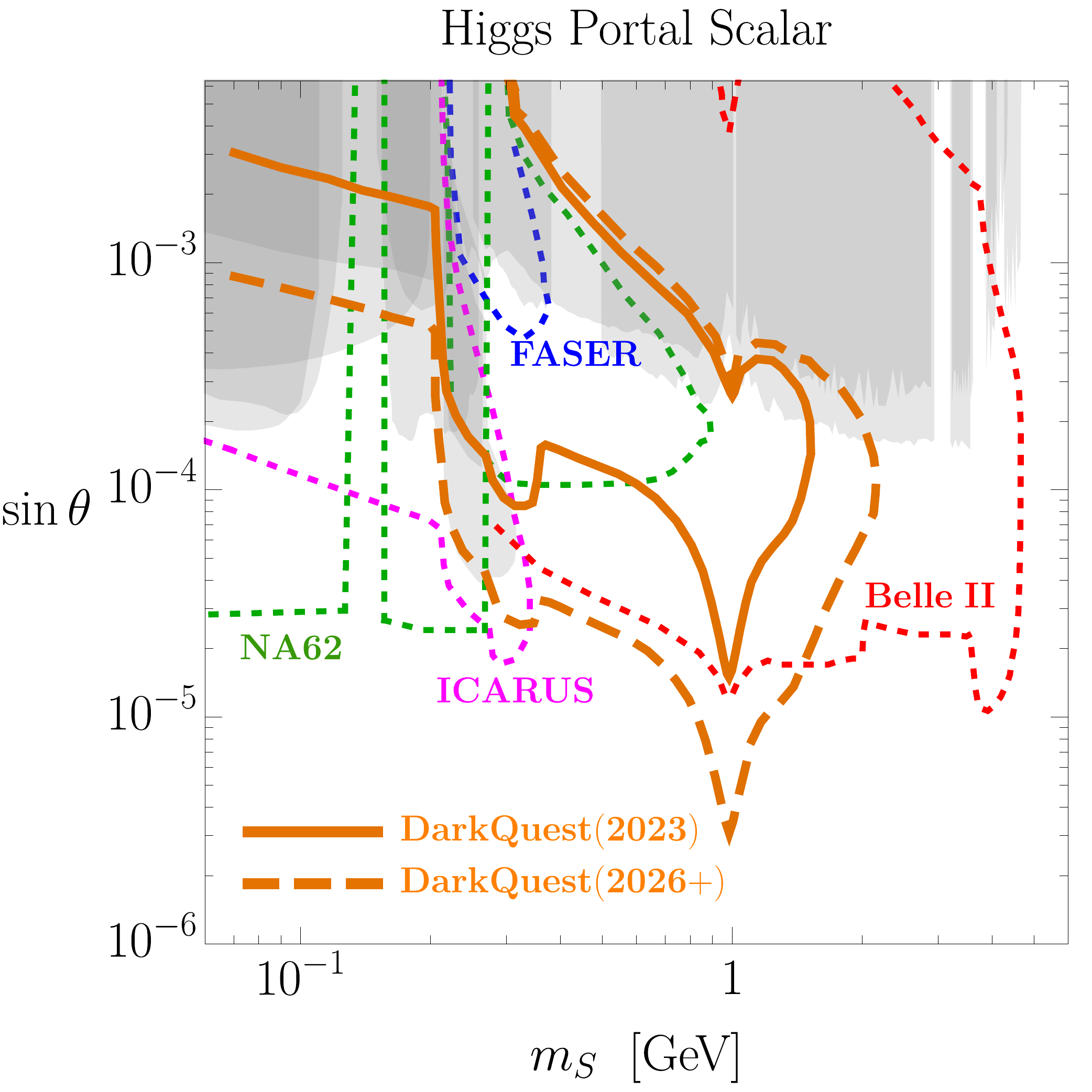}~~~\\
\vspace{0.1cm}
~~~~~~\includegraphics[width=0.37\textwidth]{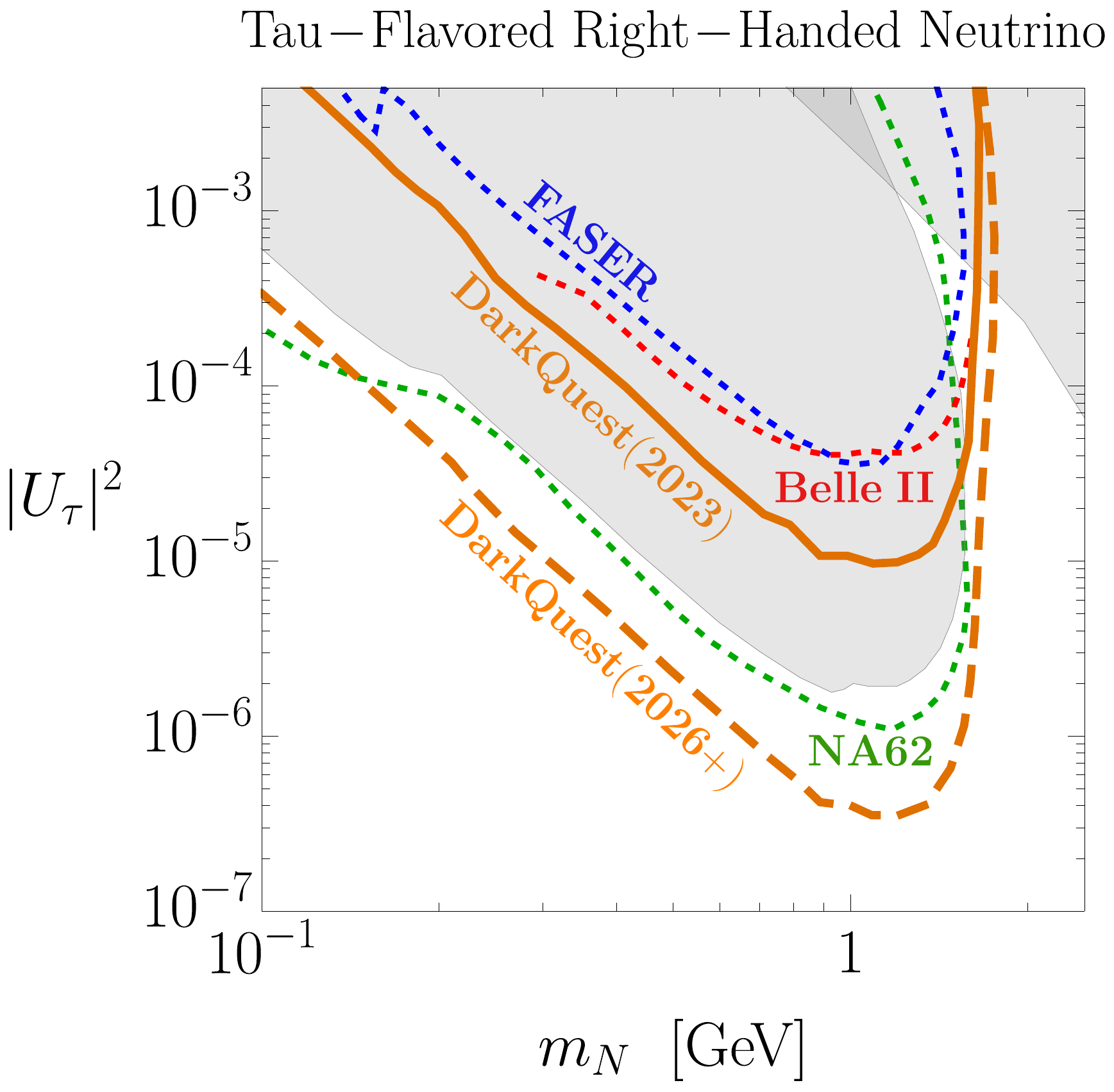}~~~~~~~~~~~
\includegraphics[width=0.41\textwidth]{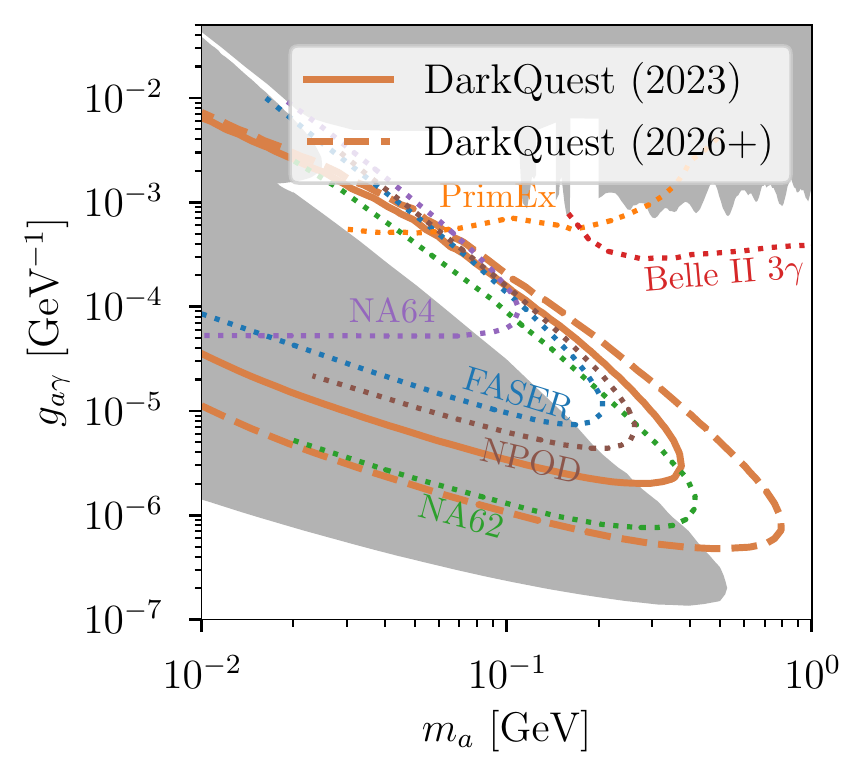}
\caption{Comparison of the DarkQuest reach and the reach of other proposed experiments 
with the corresponding timescale for models that encapsulate RF6 Big Idea 2: production and detection of an unstable dark sector mediator. Upper left panel: visible dark photon model~\cite{Berlin:2018pwi}. Upper right panel: Higgs portal scalar model~\cite{Batell:2020vqn}. Lower left panel: Right-handed neutrinos mixing with the SM tau-neutrino~\cite{Batell:2020vqn}. Lower right panel: axion-like particles coupled to photons~\cite{Blinov:2021say}. In gray we show the bounds from past experiments.
}
\label{fig:sesitivity_comparison}
\end{figure}

Fixed-target beam-dump experiments have already been used to test dark sector models. However, the sensitivity of past experiments has been limited by the thickness of their shield material and their luminosity. For example, the experiments with the most intense beams, such as E137 and CHARM had effective shield thickness of $\mathcal{O}$(100 m). As a result, there are vast parts of dark sector parameter space that remain open where the typical lifetimes are  $\mathcal{O}$(10 cm - 1 m). This gap in sensitivity is illustrated for the dark photon model (described below) in the top panel of Fig.~\ref{fig:sesitivity_comparison}, as the white area between the gray shaded collider constraints at the top and the gray shaded old beam-dump results in the bottom-left of the figure. 

DarkQuest is well-suited to probe this difficult-to-reach parameter space. The large proton beam energy and DarkQuest's relatively short baseline (or shield thickness) is well matched to study dark sector particles in the previously-untested region. The Fermilab Main Injector is ready to deliver intense proton beams to accumulate large numbers of protons on target over a relatively short time. The Injector is also one of the most energetic proton beams (and the most energetic at a DOE-funded National lab), enabling the reach for more massive (few-GeV) dark particles. DarkQuest also differs from previous beam dump experiments in that it can characterize potential signals, going beyond setting limit setting from a highly displaced energy deposition. The DarkQuest spectrometer with the EMCal upgrade will be able to perform vertexing, invariant mass and energy measurements. A schematic of the DarkQuest experiment is illustrated in Fig.~\ref{fig:darkquest} which includes the proposed EMCal and the spectrometer.  A more detailed discussion of the detector design and the performance is given in Sec.~\ref{sec:detector}.

\subsection{Benchmark dark sector scenarios}\label{Sec:2.2}
The DarkQuest experiment is an ideal laboratory for probing dark sectors through their production and decay into SM particles. Several DM and dark sector models can be tested at DarkQuest; all scenarios share the common feature of containing a ``portal'' interaction with the SM that connects the SM to the dark sector. The particle that mediates these interactions is called a mediator. DarkQuest can search for two broad classes of dark particles: (1) the mediators; (2) the dark particles participating in the process of thermalization and annihilation of DM in the early Universe (we will refer to these as ``DM excited states"). In the first scenario, the mediator is the lightest state in the hidden sector, and DarkQuest can search for the mediator decay into visible standard model particles; this discovery mode encapsulates RF6 Big Idea 2.  In the second scenario, the hidden sector contains additional particles (such as the DM itself and its excited states) at a similar mass scale as the mediator. In this case, DarkQuest can search for a rich set of signals from the DM excited state decay cascades; this discovery mode simultaneously encapsulates RF6 Big Ideas 1 and 3. 

We organize the discussion of the benchmark scenarios in terms of these Big Ideas. We outline the prospects of DarkQuest probes of Big Idea 2 in the context of dark photon mediators, Higgs portal scalars, sterile neutrinos and axion-like particles, while Big Ideas 1 and 3 are exemplified by inelastic and strongly-interacting DM models. In several reach plots, we highlight the luminosity impact on the reach. Particularly, we present two luminosity scenarios: $10^{18}$ POT and $10^{20}$ POT.

\begin{figure}[tbh!]
\centering
\includegraphics[width=0.48\textwidth]{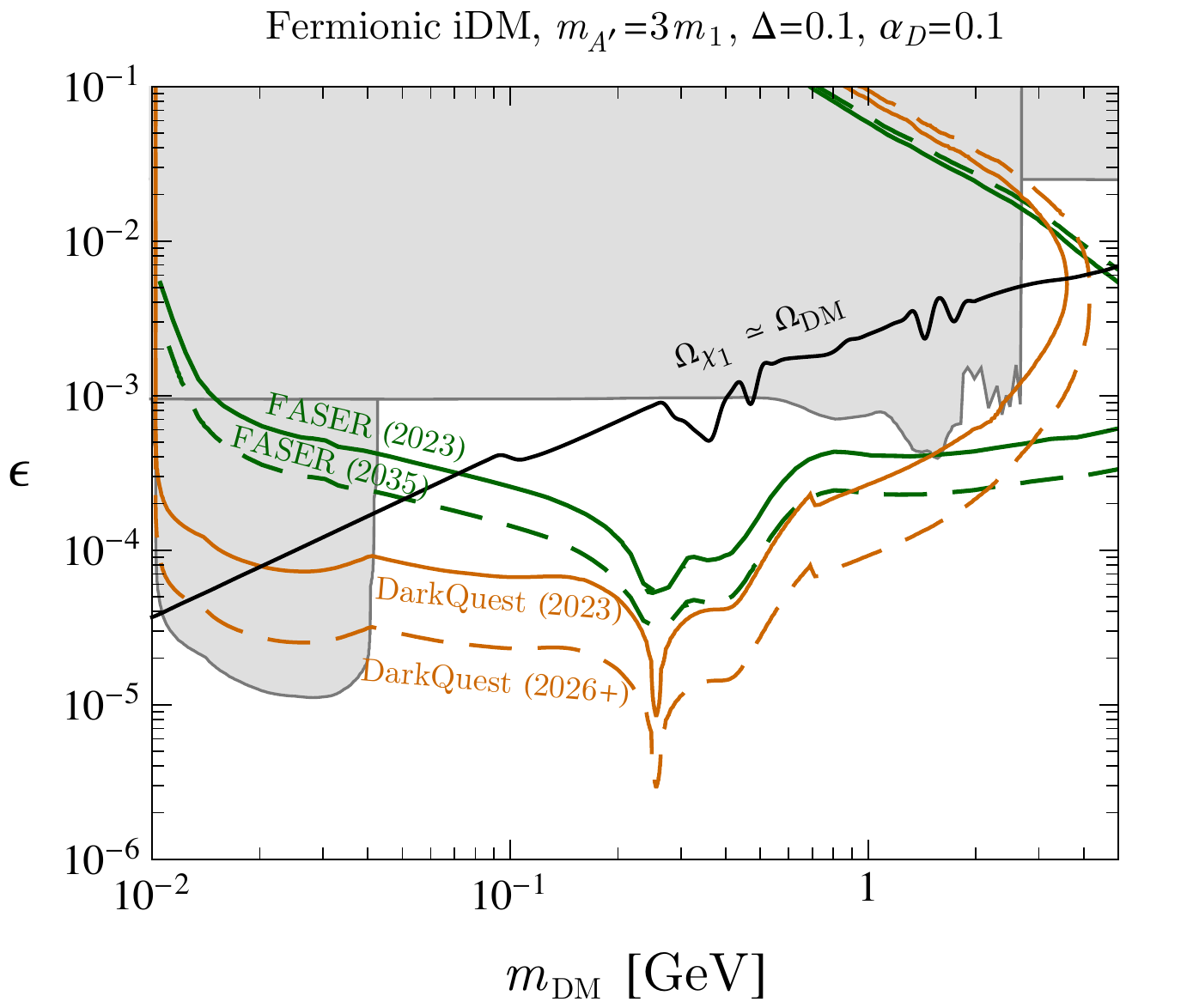}
\includegraphics[width=0.48\textwidth]{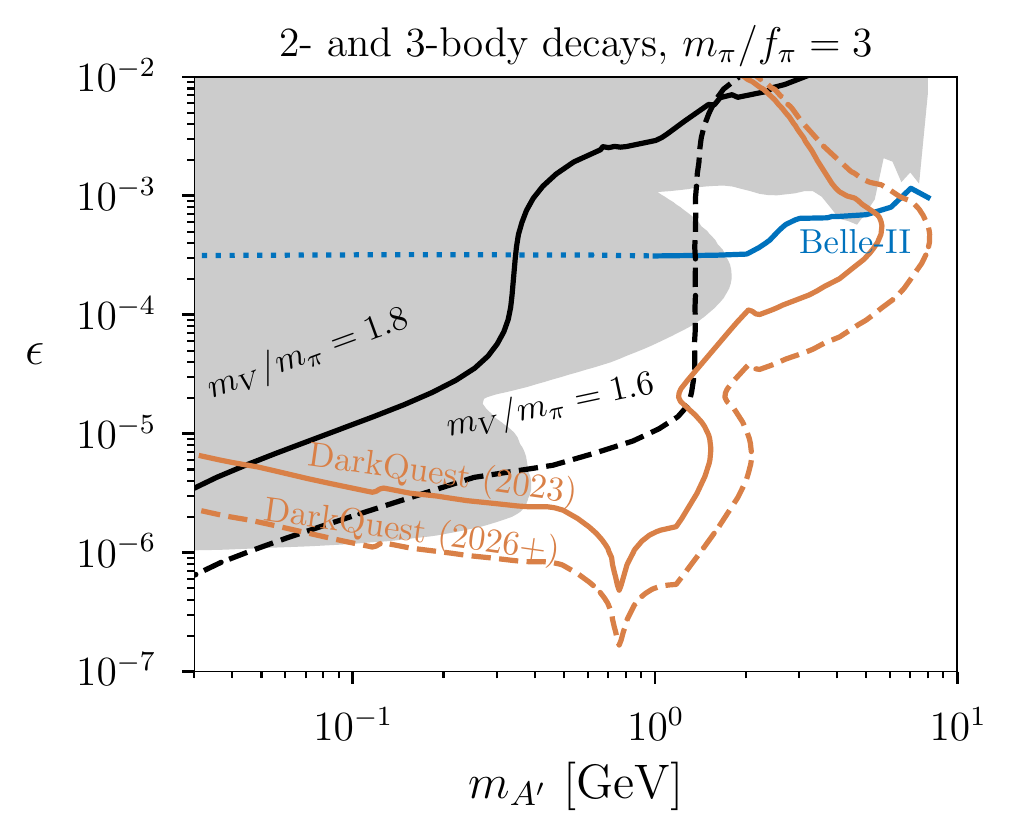}
\caption{Comparison of the DarkQuest reach and the reach of other proposed experiments for models representative of RF6 Big Ideas 1 and 3: production of DM and other particles associated with a rich dark sector. Left: Inelastic dark matter models \cite{Berlin:2018pwi}. Right: Strongly interacting massive particle (SIMP) models~\cite{Berlin:2018tvf}. In each case the black lines represent parameters for which the models saturate the observed relic density of DM. In gray we show the bounds from past experiments.
}
\label{fig:sesitivity_comparison_rich_ds}
\end{figure}

\subsubsection*{RF6 Big Idea 2: Searching for the DM-SM mediators}
\paragraph{Dark Photons}
The dark photon, $A'$, is a well-motivated mediator particle that couples to SM electromagnetic currents with a suppressed electromagnetic charge $\epsilon e$, where $\epsilon \ll 1$ and $e$ is the electric charge. Dark photons are produced at DarkQuest from the collisions of the high-energy protons with nuclei in the iron dump via the decay of SM mesons (pions, eta, ...), and the Bremsstrahlung and Drell-Yan processes~\cite{Gardner:2015wea,Berlin:2018pwi}. If the dark photon is the lightest hidden sector state, as it happens in secluded DM models \cite{Pospelov:2007mp}, then it can only decay back to SM particles. Its branching ratio to electrons and muons is sizable and of $\mathcal O(10\%)$ for all masses above threshold ($m_{A^\prime}>2 m_e\simeq 1$ MeV for electrons, and $m_{A^\prime}>2 m_\mu\simeq 200$ MeV for muons). The dark photon proper lifetime is macroscopic ($\gtrsim\text{cm}$) for $\epsilon \sim 10^{-6} \times (m_{A^\prime} / \text{GeV})^{-1/2}$. Therefore, a search for displaced electron and muon decays at DarkQuest will be an effective probe of this model. The reach of DarkQuest for dark photons is shown in the upper left panel of Fig.~\ref{fig:sesitivity_comparison}.

We emphasize that the EMCal enables the detection of electrons and will allow DarkQuest to cover previously-untested parameter space below the dimuon threshold ($m_{A^\prime}\lesssim 200$ MeV). Furthermore, in the case of a discovery, searching for both a $e^+e^-$ and a $\mu^+\mu^-$ resonance above 200 MeV offers the opportunity of characterizing the nature of the newly discovered particle. As an example, a dark photon has comparable branching ratios into electrons and into muons when both are kinematically open, whereas a dark scalar does not. The discovery of an electron and a muon signature with a similar signal strength would be strong evidence for the existence of a dark photon over another type of dark mediator. 

Finally, as shown in the upper left panel of Fig. \ref{fig:sesitivity_comparison}, DarkQuest will be able to cover a sizable region of dark photon parameter space not probed by past experiments (in gray in the figure). This exploration will be possible on a much shorter time scale than other proposed experiments at CERN (FASER) and on a comparable timescale as the proposed dump mode run of CERN NA62++ experiment. 
As shown, due to the shorter baseline of DarkQuest, the NA62++ and DarkQuest regions are highly complementary, with DarkQuest probing regions with larger $\epsilon$. The reach of DarkQuest with $10^{20}$ POT (denoted DarkQuest (2026+) in the plot) extends the reach of the proposed FASER experiment at CERN after the ultimate collection of the High-Luminosity LHC data (denoted FASER (2035) in the plot).



\paragraph{Higgs Portal Scalars} 
Perhaps the simplest extension of the SM involves a new singlet scalar field, $S$, that couples through the Higgs portal. Such a dark scalar may mediate interactions with a dark sector as well as play an important role in addressing a variety of outstanding questions, such as inflation, Higgs naturalness, etc.  Dark scalars inherit the couplings of the Higgs boson through their mixing, with interactions strengths scaled by $\sin\theta$, where the $\theta$ is the dark scalar-Higgs mixing angle. Thus, dark scalars can be produced at DarkQuest through rare kaon and $B$ meson decays, proton Bremsstrahlung, and gluon-gluon fusion~\cite{Berlin:2018pwi,Batell:2020vqn}. Given that it couples to SM particles in proportion to their masses, the dark scalar typically decays to the heaviest available final states and can be naturally long-lived. For low masses below the dimuon threshold $S$ dominantly decays to $e^+e^-$, while for scalar masses above the dimuon and hadronic  thresholds the $S$ branching ratio to $e^+ e^-$ is negligible and $S$ mainly decays to $\mu^+ \mu^-$, $\pi \pi$, or other hadronic states. For $m_S \gtrsim 2 m_\mu$, the scalar decay length is longer than ${\cal O}$(1~m) provided the mixing angle is smaller than about $10^{-3}$. This implies that searches for displaced dimuon and hadronic final states can provide interesting sensitivity to the dark scalar model in this mass range, as is highlighted in the upper right panel of Fig.~\ref{fig:sesitivity_comparison}. It is seen that DarkQuest will be able to explore substantial new regions of parameter space, which nicely complements the reach of several other experiments that will take data during the next decade.

\paragraph{Sterile Neutrinos} Sterile neutrinos are well-motivated BSM particles with no ordinary weak interactions except those induced by the mixing with the SM neutrinos, through the neutrino portal $H \bar L N$, where $H$ is the SM Higgs doublet and $L$ the SM lepton doublet.
Sterile neutrinos can be copiously produced at DarkQuest through meson and $\tau$ decays. They will subsequently decay to a range of different visible signatures including a pseudoscalar meson and a lepton, a vector meson and a lepton, a lepton and two or more pions, or three leptons. The sterile neutrino lifetime can be macroscopic depending on the value of the mixing with the SM neutrinos and on its mass. For example, for a 1 GeV sterile neutrino and mixing with the tau neutrino $|U_\tau|\sim 10^{-2}$ the lifetime is roughly a few meters. 

In the lower left panel of Fig. \ref{fig:sesitivity_comparison}, we show the reach of DarkQuest on the parameter space of $\tau$-mixed sterile neutrinos. The gray region in the figure has already been probed by past experiments. Both stages of DarkQuest ($10^{18}$ and $10^{20}$ POT) will be able to cover uncharted parameter space. For comparison, we also show the reach of several other proposed experiments expected to collect data in the near term. 

\paragraph{Axion-Like Particles}
Axion-like particles (ALPs) are a generalization of the hypothetical axion which was introduced to address the strong-CP problem. Compared to axions, ALPs have a mass and coupling to the SM that are independent of each other. ALPs can interact with SM particles via a wide variety of dimension-five couplings. Here we consider only the possibility of their coupling to photons via the operator $\mathcal{L}\supset g_{a\gamma} F_{\mu\nu} \widetilde{F}^{\mu\nu}/4$, where $F$ is the EM field strength tensor (and $\tilde{F}$ its dual) and $g_{a\gamma}$ is the dimensionful coupling coefficient. The sensitivity of DarkQuest to photon-coupled ALPs have been studied in Refs.~\cite{Berlin:2018pwi,Dobrich:2019dxc,Blinov:2021say}. A different variant of this scenario with gluon interactions was also considered in Ref.~\cite{Blinov:2021say}. The dominant photon-coupled ALP production mechanism at proton beam-dumps is via the collisions of secondary photons (arising from the decays of mesons) with target nuclei. The ALPs can traverse the beam dump and decay into a pair of photons, which are detected in the EMCal. While the proposed DarkQuest configuration is sensitive to such photons, one expects significant backgrounds from long-lived hadrons like $K_L$ and $\Lambda$, which can decay into final states with $\pi^0$'s. Additional $\sim 17$ nuclear interaction lengths of shielding are required to realize a background-free search for photon-coupled ALPs. With this caveat, DarkQuest offers leading sensitivity to ALPs, already with $10^{18}$ POT as shown in the lower right panel of Fig.~\ref{fig:sesitivity_comparison}.

\subsubsection*{RF6 Big Ideas 1 an 3: Production and Detection of DM in a Rich Dark Sector}
The SM structure is complex and there is no reason for a dark sector to be substantially simpler than the visible matter of the SM. When DM is part of a larger dark sector, its experimental signatures can differ substantially from minimal DM scenarios. Particularly, dark sector mediators can be still produced at the DarkQuest experiment through their portal interaction, but they can then decay to a combination of DM and SM visible particles. Below we discuss two example scenarios: inelastic dark matter (iDM) and strongly interactive massive particles (SIMP) dark matter.

\paragraph{iDM}
Models of inelastic DM \cite{TuckerSmith:2001hy} represent a very interesting framework for DM co-annihilation. These models contain two nearly mass-degenerate states $\chi_{1,2}$ (with masses $m_{1,2}$), which couple to a mediator, e.g. the dark photon, via a off-diagonal interaction $A' \bar \chi_1 \chi_2$. The suppression of the mass splitting originates from the approximate $U(1)$ gauge symmetry.
The lighter state, $\chi_1$, is cosmologically stable and, therefore, can make 
up all of the DM, while $\chi_2$ decays to $\chi_1$ through on-shell or off-shell emission of an $A'$. The DM abundance is regulated by the process $\chi_1 \chi_2\to A^{\prime *}\to \mathrm{SM}$ and is exponentially sensitive to the fractional mass splitting $\Delta = (m_2-m_1)/m_1$. In viable DM models, 
observational constraints force $\Delta \ll 1$ and the decay proceeds as $\chi_2\rightarrow \chi_1 A^{'*} \rightarrow \chi_1 \; \mathrm{SM}^+ \, \mathrm{SM}^-$, where $\mathrm{SM}^\pm = e^\pm, \; \mu^{\pm}, \dots$ are the kinematically-allowed SM final states. The requirement of obtaining the correct relic abundance of $\chi_1$ defines a target region (``thermal targets'') in the model parameter space that will be accessible with the DarkQuest experiment~\cite{Berlin:2018pwi}.

At DarkQuest, the dark photon is produced in beam-target collisions; it then promptly decays to $\chi_1 \chi_2$. $\chi_1$ leaves the detector as missing momentum, while $\chi_2$ undergoes a 3-body decay through an off-shell $A'$, $\chi_2 \to \chi_1 \ell^+ \ell^-$. Generically, in the thermal region of interest, $\chi_2$ travels macroscopic lengths before decaying, since its width is suppressed by the fifth power of the mass splitting, $\Delta$. 
A search for displaced non-resonant leptons at DarkQuest would have unprecedented reach for such models, as shown in the left panel of Fig.~\ref{fig:sesitivity_comparison_rich_ds}, for a generic benchmark scenario with 10$\%$ mass splitting. The dark line across the plot is the region where the DM candidate has the measured relic abundance.
DarkQuest will have the opportunity to completely test the model for DM masses up to $\sim 4$ GeV. In particular, sensitivity to electron 
final states is {\bf{key}} in these scenarios since, because the small $m_1-m_2$ mass splitting forces $\chi_2$ to decay to electrons for the mass range of interest at DarkQuest (only the region $m_1\gtrsim 2$ GeV has the $\mu^+\mu^-$ channel kinematically open).

\paragraph{SIMPs}
The dark sector interactions described above are \emph{weak}, in the sense that they do not lead to confinement, akin to the SM Weak and electromagnetic interactions. If the dark sector instead interacts \emph{strongly} with itself (like QCD in the SM), then the particle content and cosmological history can be qualitatively different. 

In the strongly-interacting massive particle (SIMP) paradigm, the dark sector consists of dark quarks charged under a confining gauge group. The DM is then composed of the pions, $\pi_D$, of this dark QCD. In addition to these states, SIMP models also give rise to hidden sector spin-1 vector mesons, $V_D$, the analogs of the $\rho$, $\phi$, and $\omega$ of the SM. These vector mesons contribute sizably to the cosmological history of the DM states, $\pi_D$.
A viable cosmology where $\pi_D$ constitutes all of the DM requires a coupling to the SM mediated, e.g. by the dark photon. 
In these models, the dark photon generically decays to a stable dark pion, $\pi_D$ , and an unstable vector, $V_D$, that subsequently decays to SM leptons after traveling a macroscopic distance $A^\prime\to \pi_DV_D,V_D\to e^+e^-$ or $V_D\to\pi_D e^+e^-$. 

The strong interactions in the dark sector give rise to qualitatively new early-universe DM production mechanisms~\cite{Hochberg:2014dra,Hochberg:2014kqa}, which favor different regions of the model parameter space compared to weakly-coupled scenarios~\cite{Berlin:2018tvf}. 
These target regions are shown as black lines in the right panel of Fig.~\ref{fig:sesitivity_comparison_rich_ds}. DarkQuest can test these models through cascade decays that terminate in SM particles. The dark photon can be produced as before, but here will decay promptly into $V_D \pi_D$. The hidden vector meson $V_D$ will then decay to displaced leptons. Additional mesons with different charges under the hidden sector gauge group decay with vastly different decay lengths, enabling 
DarkQuest to probe a vast range of couplings. This leads to the two-lobe 
structure in the DarkQuest sensitivity shown in the right panel of Fig.~\ref{fig:sesitivity_comparison_rich_ds}. As shown in the figure, depending on the value of $\epsilon$, the DarkQuest reach could extend the reach of past experiments (shaded in gray in the figure) by up to almost one order of magnitude in mass.

\paragraph{Additional signatures and other opportunities}
While we focused on models where the dark sector is coupled to the SM through the dark photon portal, DarkQuest can probe a wide range of other DM scenarios, such as all the other possible mediators between the dark sector and the SM: scalars, axion-like particles~\cite{Berlin:2018pwi}, as well as right-handed neutrinos \cite{Batell:2020vqn}. 

An (incomplete) list of search channels that will be available at DarkQuest is shown in Tab.~\ref{tab:channels}. As explained above, the EMCal upgrade will enable DarkQuest to search for dark sector particles over a significantly broader mass range by enabling electron final state measurements.
In some cases, the DarkQuest sensitivity lies \emph{entirely} below the dimuon threshold, meaning the EMCal is critical to probe these DM models (the light iDM case discussed above, or the leptophilic Higgs case of Ref.~\cite{Berlin:2018pwi} are both examples). A final set of signatures are diphoton or meson final states (such as from axion-like particles or right-handed neutrinos, respectively). To demonstrate the importance of these additional signatures, in the lower panel of Fig.~\ref{fig:sesitivity_comparison} we also show the reach for right-handed neutrinos, as obtained via searches for 
vertices containing two or more charged particles and assuming $K_L$ and other backgrounds can be reduced or rejected.

Thanks to the EMCal upgrade, DarkQuest will be a  multi-purpose experiment uniquely able to probe a broad range of GeV-scale and below DM-motivated frameworks. As a by-product, DarkQuest will also be able to tackle other open questions in particle physics beyond  dark matter including understanding the origin of neutrino masses and oscillations, finding a solution to the strong CP-problem, explaining the origin of the baryon-antibaryon asymmetry of the Universe, and how to address present anomalies in data like the one in $(g-2)_{\mu}$ or those from $B$-physics. This will be achieved thanks to the many searches for visible dark sectors that will be possible at DarkQuest after the EMCal upgrade.

\renewcommand{\arraystretch}{1.8}
\begin{table}
\centering
\begin{tabular}{|c||c|c|c|}
\hline
\textbf{Signature} & \textbf{Model}\\
\hline\hline
$e^+ e^-$ & \renewcommand{\arraystretch}{1} \begin{tabular}{@{}c@{}}dark photon\\ dark Higgs \\ leptophilic scalar$^*$ \end{tabular}  \\
\hline
$e^+ e^- e^+ e^-$ & Higgsed dark photon \\
\hline
$e^\pm \pi^\mp, e^\pm K^\mp, \cdots$ & sterile neutrino \\
\hline
$e^+ e^- + \text{MET}$ & \renewcommand{\arraystretch}{1} \begin{tabular}{@{}c@{}}inelastic dark matter \\ strongly interacting dark matter \\ hidden valleys \end{tabular} \\
\hline
$\pi^+ \pi^-, K^+ K^-, \cdots$ & dark Higgs$^*$ \\
\hline
$\gamma \gamma$ & axion-like particle$^*$ \\
\hline
\end{tabular}
\caption{Various experimental signatures and the relevant models that can be searched for at DarkQuest with the EMCal upgrade. In each scenario, the EMCal is 
necessary to attain physics reach over broad mass range, especially below $2m_\mu$. In some cases (marked with a *), the search would be impossible without an EMCal for \emph{any} mass.
We do not add the several muon signatures, as they can be searched for also with the present SeaQuest setup.}

\label{tab:channels}
\end{table}

\subsection{Comparison with other proposed experiments and techniques}

The EMCal upgrade will enable DarkQuest to attain world-leading sensitivity to dark sector physics. This sensitivity overlaps with and is complementary to other proposed visible dark sector experiments, such NA62++, FASER, Codex-b, MATHUSLA, and SHiP~\cite{Beacham:2019nyx} at CERN, as well as other techniques called out in the BRN report particularly electron/muon missing momentum and proton/electron beam dumps. DarkQuest's direct competition, the CERN fixed-target experiment NA62++, has a much longer baseline, allowing DarkQuest to probe shorter lifetimes. This is precisely the region of parameter space that has eluded previous collider and fixed target experiments. The DarkQuest reach also significantly surpasses that of the initial version of the FASER experiment, 
slated to operate during LHC Run 3 (see the solid lines denoted ``FASER (2024)'' in Fig. \ref{fig:sesitivity_comparison}). The ultimate FASER program requires the completion of the High-Luminosity LHC program in the 2030s (see the dashed lines denoted ``FASER (2035)'' in Fig. \ref{fig:sesitivity_comparison} for the latter reach). 


DarkQuest is also complementary to other accelerator-based fixed-target techniques, such as longer baseline beam-dump and missing momentum experiments. The longer baseline proton beam-dump experiments like the Short Baseline Near Detector (SBND) will use lower energy proton beam but will achieve a higher luminosity compared to DarkQuest. 
These facts ensure that in the visible channel, SBND will be sensitive to lower couplings and longer dark sector particle lifetimes. However, a large part of this parameter space has already been probed by old electron and proton beam-dump experiments. Similar statements hold for the electron Beam-Dump eXperiment (BDX). On the other hand, the proton and electron beam dump program will also have the sensitivity to a qualitatively different set of DM targets compared to DarkQuest, by producing DM in the target and observing its scattering in a downstream detector.

Missing momentum experiments such as the LDMX (electron beam) and $M^3$ (muon beam) seek to produce dark particles in fixed target collisions and detect them indirectly by measuring the momentum of the recoiling beam particle~\cite{Berlin:2018bsc,Kahn:2018cqs} and large missing momentum. Such experiments are complementary to visible dark sectors searches and are sensitive to other qualitatively different dark matter signatures and vice versa. 
For those common dark sector scenarios, the size of the detectors and the energy of the beam mean that experiments like LDMX will be sensitive to smaller masses, larger couplings and, correspondingly, shorter lifetimes compared to DarkQuest. Moreover, LDMX will probe dark sector couplings to leptons, while DarkQuest is testing hadronic couplings. Thus both experiments are needed to achieve complete coverage of new physics scenarios and to characterize the signal in the event of a discovery.

DarkQuest is highly complementary to direct detection efforts. In the SIMP and inelastic DM models discussed above, direct detection experiments will not be able to probe the thermal relic targets that are accessible at DarkQuest. In the SIMP case, this is because the parameter space that direct detection is sensitive to (the upper left region of the right panel of Fig.~\ref{fig:sesitivity_comparison_rich_ds}) is already excluded by existing searches. For models with inelastic DM, the scattering rate at direct detection experiments is loop-suppressed, and therefore orders of magnitude below the sensitivity of any proposed direct detection experiment.

\subsection{Dark sector search strategy}


In this subsection, we describe the production and decay of dark sector signatures in DarkQuest which result in the sensitivity presented in the scenarios above. We will describe the production mechanisms and show the anticipated kinematics of those dark sector particles.  This will inform and define the requirements of the detector performance of the DarkQuest experiment to optimize its sensitivity.

\begin{figure}[tbh!]
    \centering
    \includegraphics[width=0.48\textwidth]{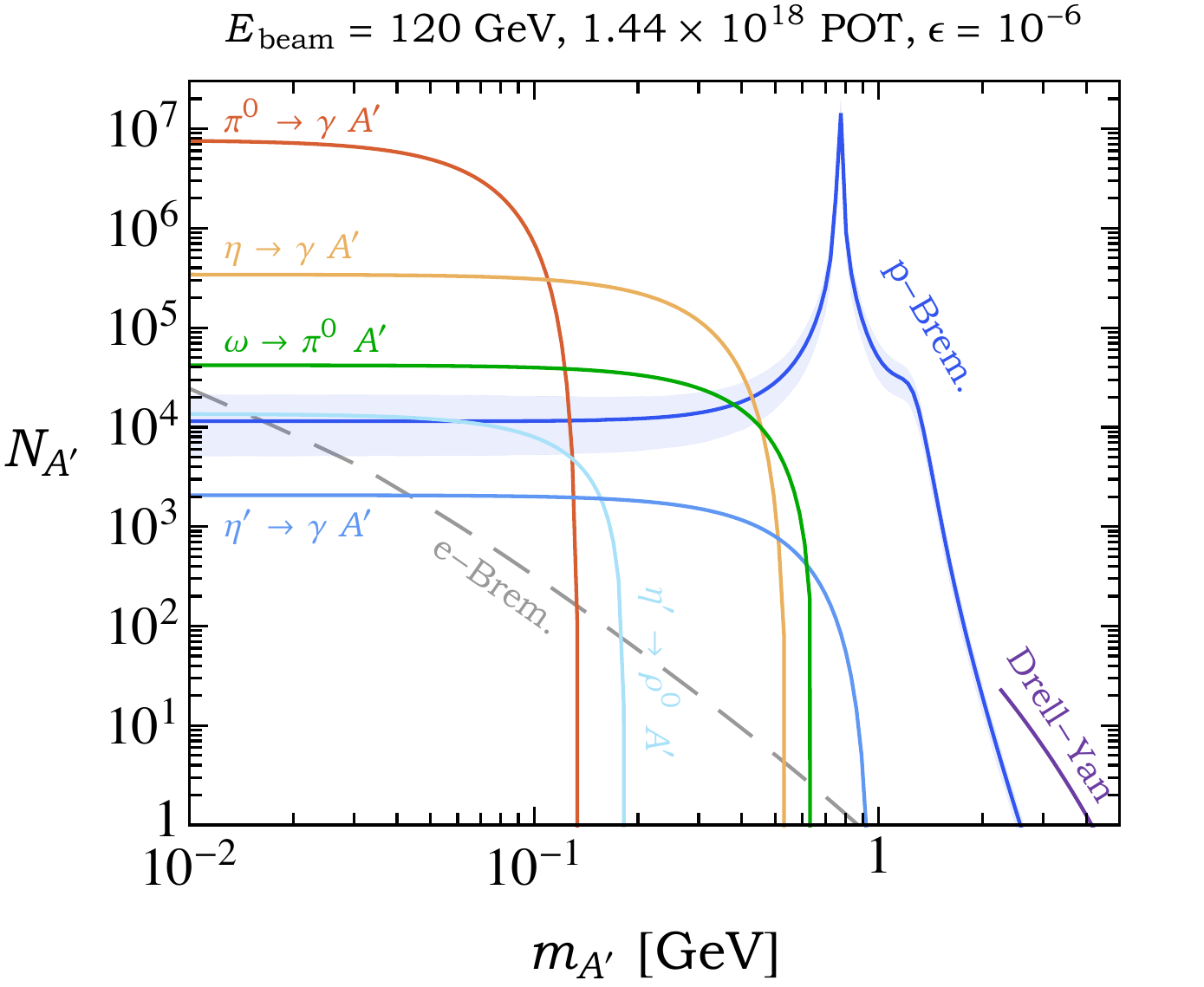}
    \caption{Number of dark photons (solid colors) produced at DarkQuest with $\sim 10^{18}$ POT in various production channels for $\epsilon=10^{-6}$. For comparison, we also show the analogous production rate for electron bremsstrahlung (dashed gray), assuming a 120 GeV electron beam, $\sim 10^{18} \text{ EOT}$, and production within the first radiation length of a tungsten target. Figure taken from Ref.~\cite{Berlin:2018pwi}.}
    \label{fig:ProductionAPrime}
\end{figure}

\begin{figure}[tbh!]
    \centering
    \includegraphics[width=0.43\textwidth]{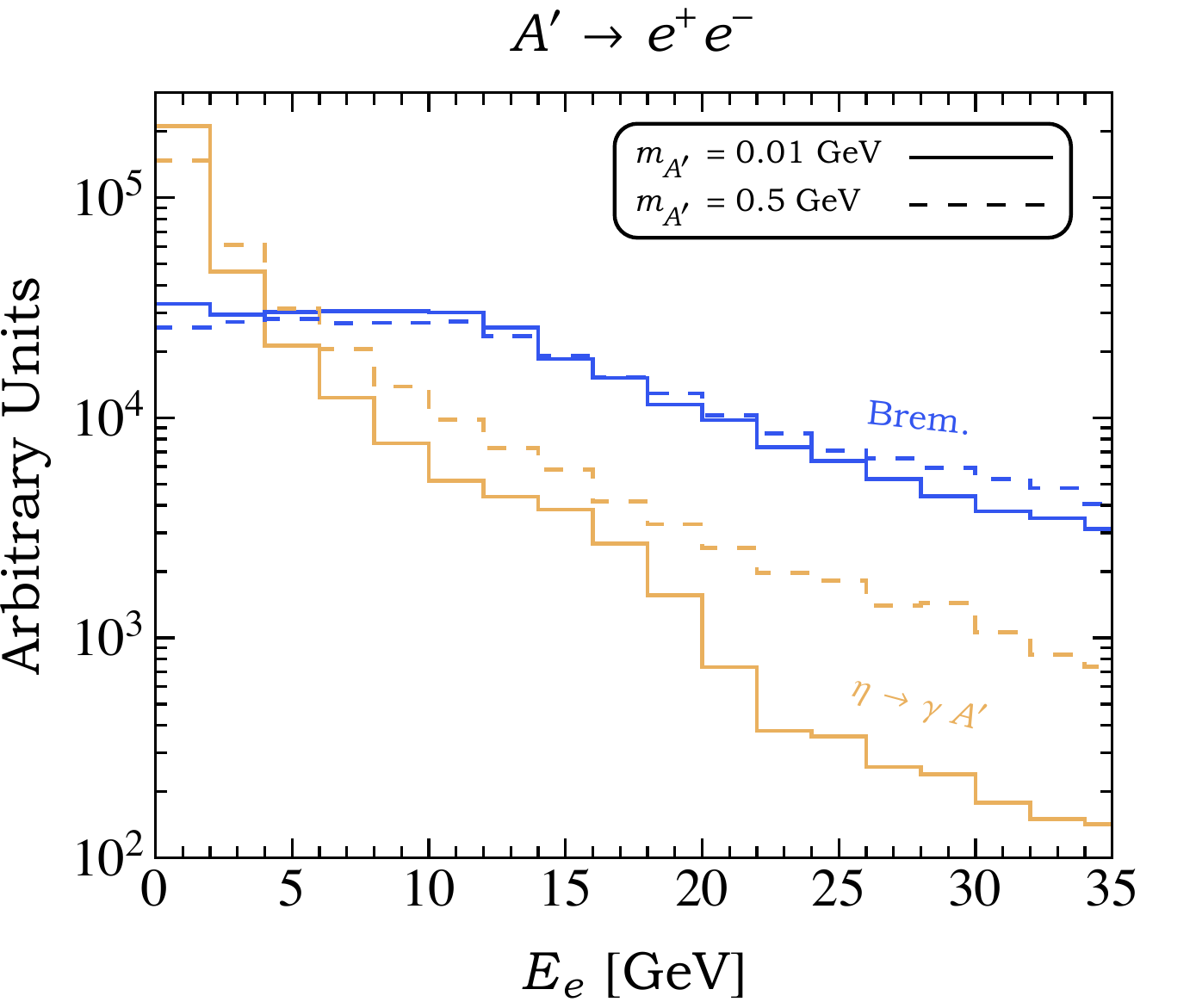}
     \includegraphics[width=0.43\textwidth]{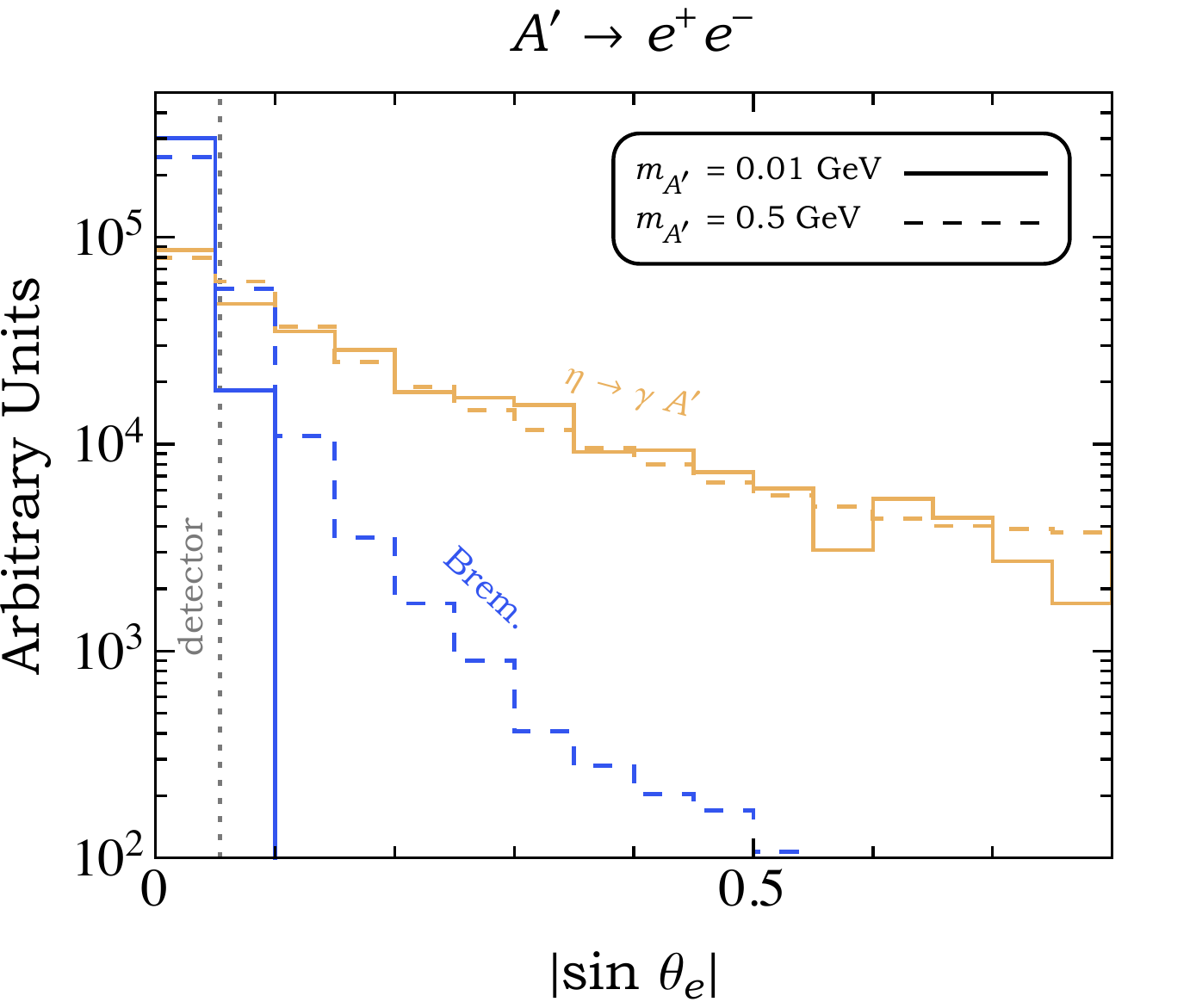}
    \caption{Signal kinematics of $A^\prime \to e^+ e^-$ for dark photons produced from eta meson decays (orange) and proton bremsstrahlung (blue). The left (right) panel displays energy (angular) distributions for electrons originating from dark photon decays before traveling through KMAG. The solid (dashed) line corresponds to $m_{A^\prime} = 0.01\; \GeV$ $(0.5  \;\GeV)$. The vertical gray dotted line in the right panel denotes the angular scale of the DarkQuest spectrometer.}
    \label{fig:KinematicsAPrime}
\end{figure}

\paragraph{Dark Photons}
Dark photons are mainly produced through the bremsstrahlung process and meson decays. Thanks to its high energy proton beam, DarkQuest will be able to produce a huge statistics of dark photons, a much larger one if compared to an electron beam dump experiment with a comparable energy. This is shown in Fig. \ref{fig:ProductionAPrime}, where the dashed gray curve indicates the number of produced dark photons assuming a 120 GeV proton beam and the same luminosity as at DarkQuest ($10^{18}$ POT). The 120 GeV proton energy also allows a large production of heavier mesons (e.g. $\eta,\omega$) enabling a copious production of dark photons well above the pion mass. Even for very small kinetic mixing parameters, $\epsilon=10^{-6}$ millions of dark photons could be produced in a broad range of masses. Once produced, the geometric acceptance of the leptons from the dark photon decay is large, particularly if the dark photon is produced from the bremsstrahlung process. In fact, the dark photons produced are typically highly boosted (see Fig. \ref{fig:KinematicsAPrime}). This, together with the compact geometry of the DarkQuest detector, leads to a large geometric acceptance of $\mathcal O(10\%)$ and above (the acceptance is close to one for $A^\prime$ with $m_{A^\prime}\lesssim 1$ GeV produced from bremsstrahlung).

\begin{figure}[tbh!]
    \centering
    \includegraphics[width=0.45\textwidth]{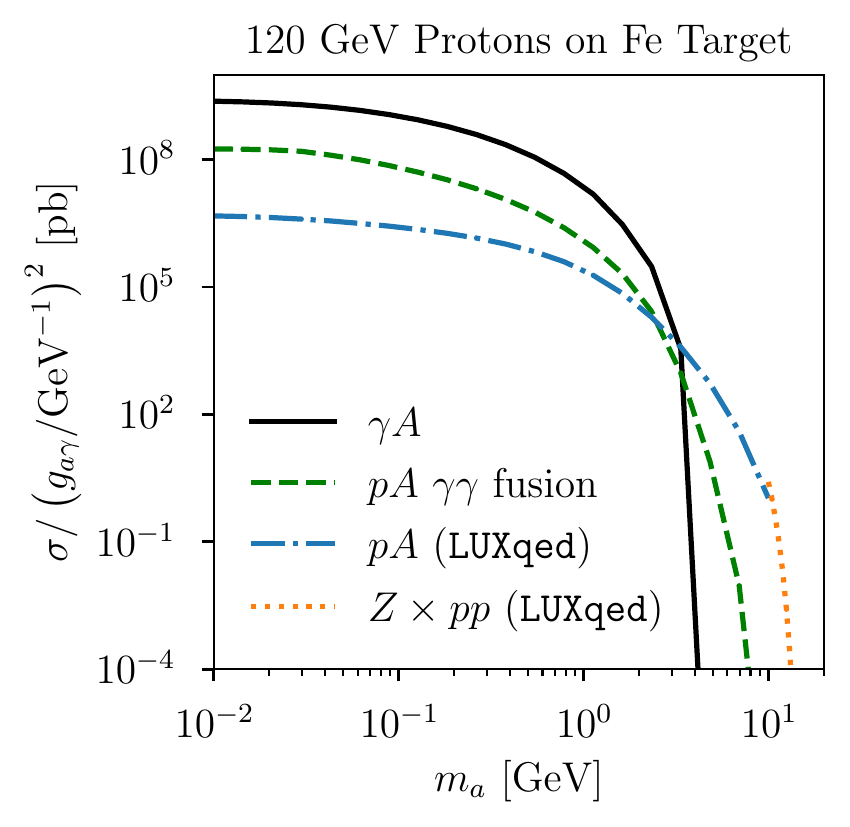}
    \caption{Cross-sections for various photon-coupled ALP production channels in collisions of a 120 GeV 
    proton beam with an iron target. Primakoff ($\gamma A$) and coherent photon fusion ($pA$) 
    processes are used to estimate the sensitivity of DarkQuest. 
    The latter channel is restricted to proton momentum transfers of $Q^2 \leq 1\;\mathrm{GeV}^2$ for 
    computational simplicity. Cross sections for processes with higher momentum 
    transfers are obtained using the \texttt{LUXqed} photon PDF, but end up being subdominant in the parameter space accessible at DarkQuest. Figure taken from Ref.~\cite{Blinov:2021say}.}
    \label{fig:photon_coupled_alp_cross_sections}
\end{figure}

\begin{figure}[tbh!]
    \centering
    \includegraphics[width=0.43\textwidth]{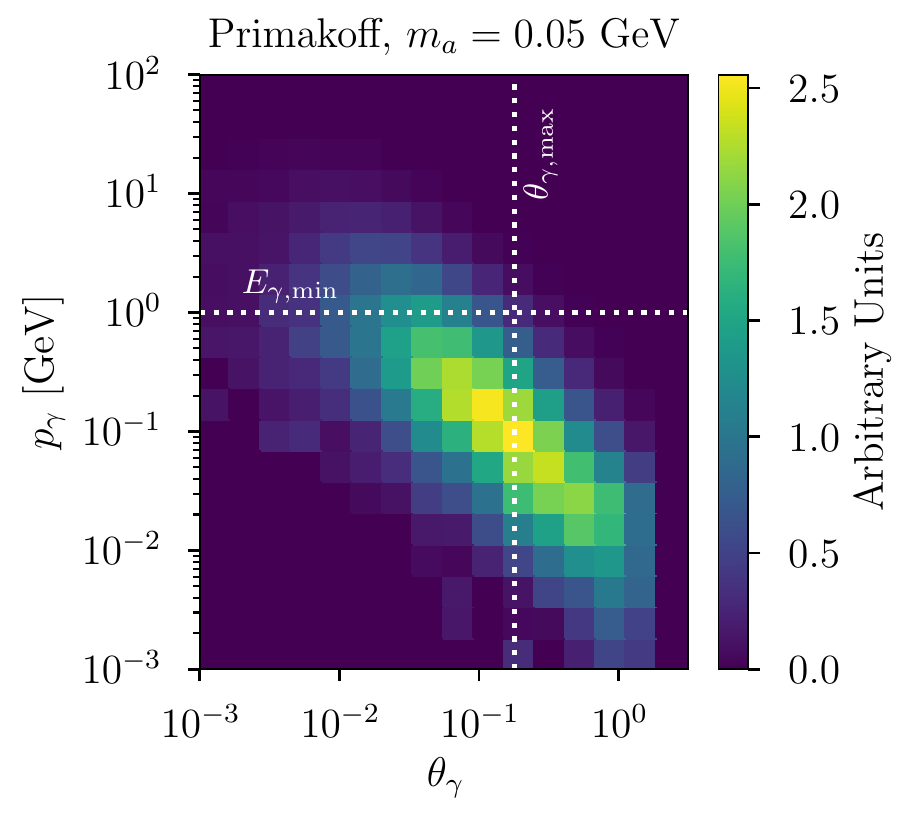}
    \includegraphics[width=0.43\textwidth]{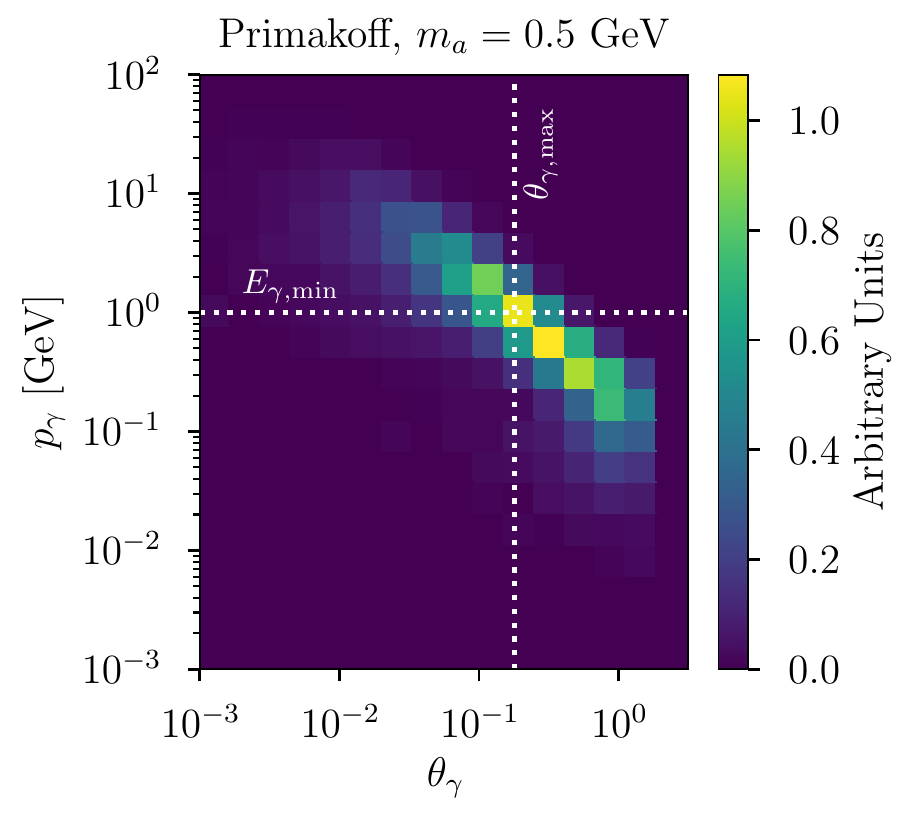}
    \caption{Distributions of ALP decay photon angles with respect to beam axis, $\theta_\gamma$, and momenta, $p_\gamma$, for two benchmark masses $m_a = 0.05$ and $0.5$ GeV in the dominantly-photon-coupled ALP model. 
    The distributions are for the dominant Primakoff production mechanism only. The dotted 
    lines indicate the approximate selections imposed in the final analysis on the photon angle (assuming the decays happen at $z=8$ m and taking a $2\; \text{m}\times 2\;\text{m}$ detector at $z=19$ m) and energy. The histograms are normalized to unity. Figure taken from Ref.~\cite{Blinov:2021say}.}
    \label{fig:alp_decay_photon_distribution}
\end{figure}

\paragraph{Axion-Like Particles}
The dominant production channel of ALPs strongly depends on whether 
they interact with hadrons and photons, or photons only~\cite{Blinov:2021say}. Even in the simpler 
photon-only case, there are several mechanisms; the cross-sections 
for these reactions in the context of DarkQuest are shown in Fig.~\ref{fig:photon_coupled_alp_cross_sections}. The dominant production channel turns out to be the Primakoff process $\gamma N \to a N$, where the initial 
state photon is a secondary from, e.g., neutral meson decays~\cite{Berlin:2018pwi, Blinov:2021say}. A Monte Carlo simulation for Primakoff production was developed in Ref.~\cite{Blinov:2021say}. The produced ALPs are then decayed into photons, whose kinematic distributions for two mass points are shown in Fig.~\ref{fig:alp_decay_photon_distribution}. The distributions are shown in the plane of photon energy $p_\gamma$ and angle with respect to the beam axis $\theta_\gamma$. These plots also show the rough acceptance criteria for the DarkQuest ECal as dotted white lines. 
Similar results for a hadronically interacting ALP are presented in Ref.~\cite{Blinov:2021say}. DarkQuest has good signal acceptance across a wide range of ALP masses, leading to the excellent sensitivity shown in the lower right panel of Fig.~\ref{fig:sesitivity_comparison}.

\begin{figure}[tbh!]
    \centering
    \includegraphics[width=0.43\textwidth]{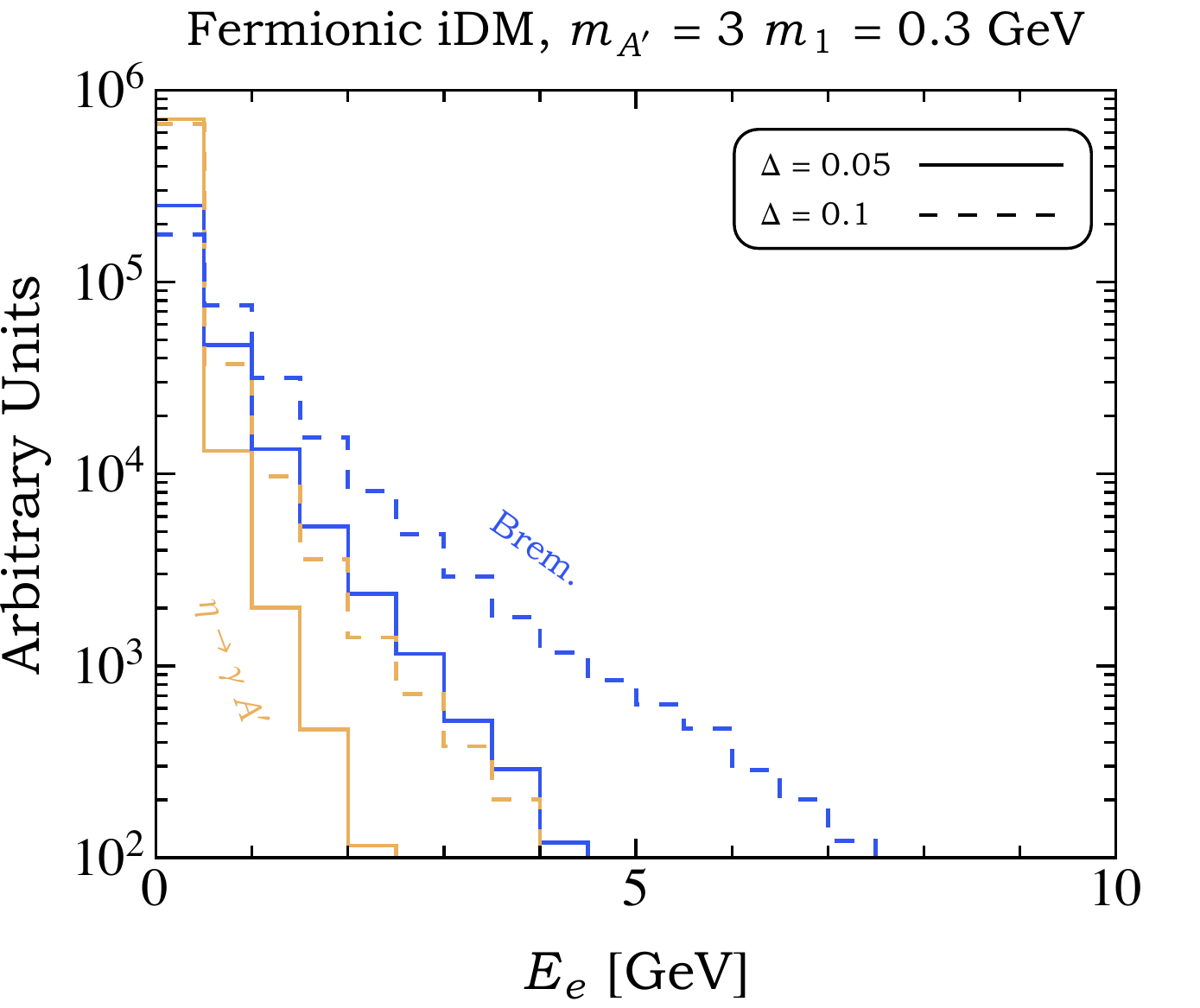}
     \includegraphics[width=0.43\textwidth]{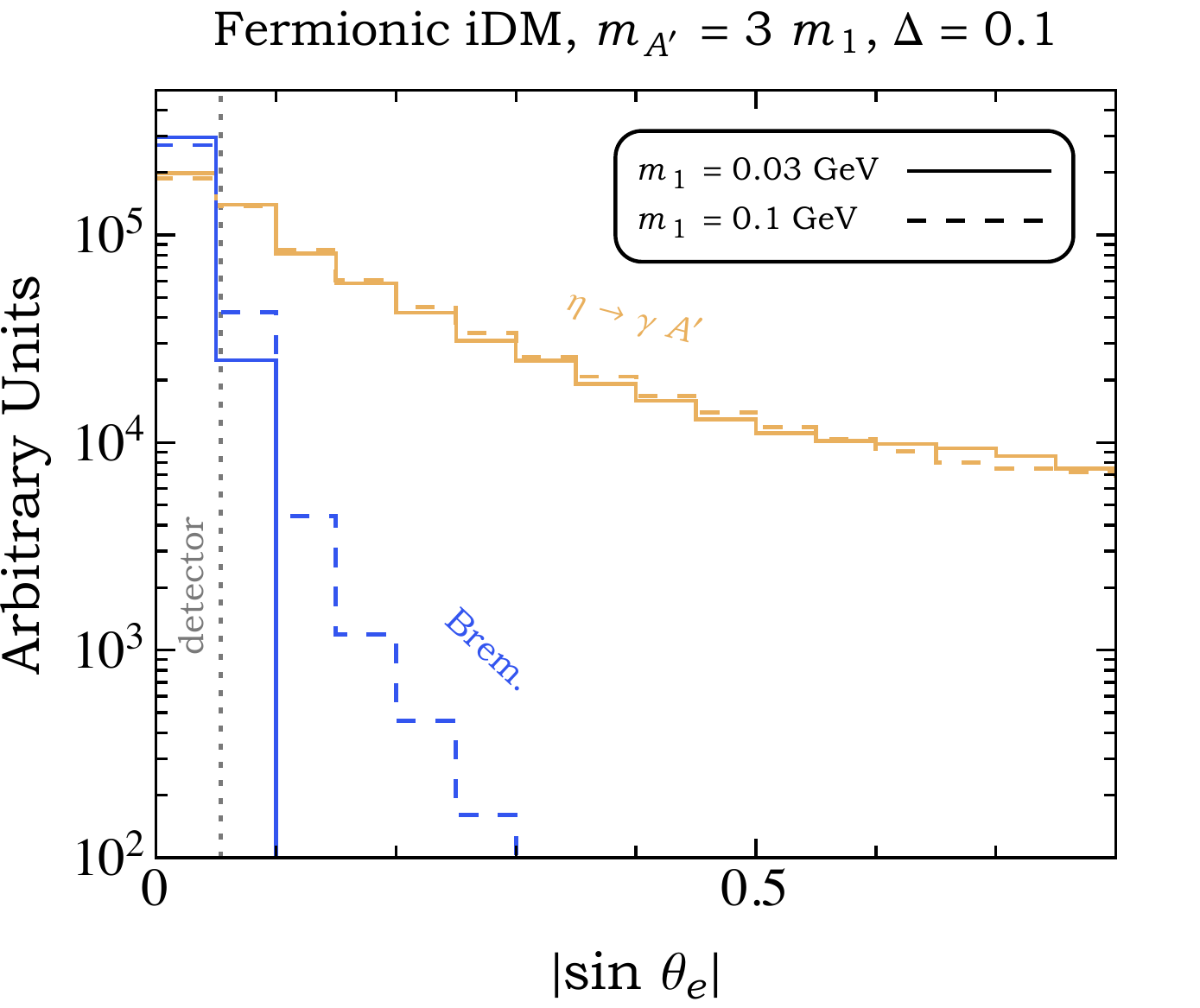}
    \caption{Signal kinematics of $A^\prime \to \chi_1\chi_2\to\chi_1 (e^+ e^-\chi_1)$. The color code is as in Fig. \ref{fig:KinematicsAPrime}.}
    \label{fig:KinematicsIDM}
\end{figure}

\paragraph{iDMs and SIMPs}
The production of DM and excited states arising from dark photon-mediated iDM models arises through the production and decay of the dark photon $A^\prime\to\chi_1\chi_2$ (see the beginning of this section for the discussion of the dark photon production at DarkQuest). In these models, the main decay mode of the dark photon is, in fact, in $\chi_1\chi_2$. DarkQuest will be able to search for the electrons produced from the $\chi_2$ decay, $\chi_2\to\chi_1 e^+e^-$. The geometric acceptance will be a bit reduced if compared to the case of the minimal dark photon model. 
In fact, as discussed in Sec. \ref{Sec:2.2}, models for iDM generically predict a small mass splitting between the DM and the excited states, $\chi_1,\chi_2$. 
For $\Delta\ll 1$, $\chi_1$ carries away a large fraction of the $\chi_2$ energy, implying that the electrons from the 3-body decay are significantly softer than those from visibly decaying dark photons. This is seen in Fig. \ref{fig:KinematicsIDM}, which shows the energy and angular distributions of electrons originating from $\chi_2$ decays. Mass splittings of the order of $\Delta=\mathcal O(0.1)$ lead to a geometric acceptance in $10^{-4}-1$, depending on the exact point where $\chi_2$ decays. As in the case of the minimal dark photon model, the production leads to a larger geometric acceptance, if compared to the production from meson decays. Larger mass splittings lead to a larger geometric acceptance.

The 3-body signatures arising from SIMP models are similar $A^\prime\to V_D\pi_D,V_D\to\pi_De^+e^-$, even if now the mass splitting $m_{V_D}-m_{\pi_D}$ is naturally larger than in iDM models. This will lead to more boosted lepton final states, and therefore to a larger geometric acceptance. The 2-body signatures arising from SIMP models, $A^\prime\to V_D\pi_D,V_D\to e^+e^-$, also have a larger geometric acceptance, if compared to the 3-body signatures.

\newpage
\section{SpinQuest spectrometer and DarkQuest detector upgrades}
\label{sec:detector}
\subsection{SpinQuest spectrometer} 
The SpinQuest experiment(E1039) is built upon the existing SeaQuest(E906) experiment and uses the SeaQuest spectrometer to reconstruct dimuon events from an upgraded target. The spectrometer is designed to detect oppositely-charged muon pairs produced in the Drell-Yan process to study anti-quark distributions in target nucleons~\cite{seaquestcollaborationSeaQuestSpectrometerFermilab2017a}. 

As shown in Fig. \ref{fig:darkquest}, the spectrometer consists of a high precision tracking system (St-1/2/3 tracking), a muon identification system (absorber and St-4 muon ID), and a high-speed vertex trigger detector (forward/backward dark photon detector). This instrument is capable of precisely measuring both the momentum and vertex of a charged particle; allowing for the identification of displaced particle identification. The 5\,m thick iron beam dump/magnet (FMag) bends and stops most of the SM particles (other than neutrinos, high energy muons, and long-lived Kaons) produced by proton-iron interactions. However, a dark particle, which interacts only weakly with normal matter, can travel a significant distance from the creation point before decaying into a pair of leptons or hadrons. This feature, combined with a high luminosity ($2.5\times 10^5$ fb$^{-1}$ per year), allows us to implement a world-leading dark sector search program while suppressing SM backgrounds to negligible levels.

Dedicated displaced dimuon dark photon detectors and triggers were successfully installed and commissioned in 2017 and will take data parasitic with the upcoming SpinQuest run. These detectors serve as a baseline for the identification and selection of displaced dimuon pair events, which are a major signature of dark photon decays, when the dark photon is heavier than twice the muon mass. 

The spectrometer has already been extensively characterized in the E906/SeaQuest experiment~\cite{seaquestcollaborationSeaQuestSpectrometerFermilab2017a}. DarkQuest can rely on the proven performance of the spectrometer. The tracking performance of the spectrometer is determined by the performance of the drift chambers. Each drift chamber consists of six planes of sense wires in the standard x/x$^\prime$, u/u$^\prime$, v/v$^\prime$ configuration. This allows each chamber to measure a track segment, with redundancy (reconstruction can recover from missed hits in up to two planes per chamber). The drift times, per-plane position resolution and efficiency were measured using muon tracks in the running period of April 2014 - June 2015 and are summarized in Table~\ref{tab:cham:performance}. Figure~\ref{fig:emcal_performance} shows a dimuon mass spectrum from the SeaQuest Drell-Yan running period which shows good agreement between data and simulation. For the SeaQuest experiment, since the primary physics signal is the high mass Drell-Yan above $4GeV$, the low mass dimuons below the $J/\Psi$ mass were rejected at the trigger level. Simulation shows that for high mass ($m > 3\,\text{GeV} $) prompt dimuons from the target, the mass resolution is $\sigma_m/m = 6\%$, mostly due to multiple scattering and energy loss of muons in the 5m thick FMag; for the low mass dimuons produced after the FMag/absorber (from dark photon decays, for e.g.), the mass resolution is much better,   $\sigma_m/m = 2.5\%$. Dark photon di-muon performance will further be discussed in Section~\ref{sec:performance}. 


\begin{table}[tbh!]
\centering
  \caption{Performance of the drift chambers in the SeaQuest experiment between April 2014 and June 2015. The position resolution and detection efficiency are the averages of the resolutions and efficiencies for all six planes in the specified chamber.  The resolutions are the average value for each chamber of the RMS of the difference between the measured position in a plane and the position calculated at the z coordinate of that plane using a fit with the plane excluded from the fit.
\label{tab:cham:performance}}
\vspace{-0.5cm}
  \begin{tabular}{lrrr}
    \\ \hline \hline
     & Max. drift  & Pos. res. & Detection eff. (\%) \\
 Chamber    & (ns)       & ($\mu$m)       & (min.-max.) \\
    \hline
    DC1.1   &     100    &   225          & 99-100 \\
    DC2     &     260    &   325          & 96-99 \\
    DC3p    &     220    &   240          & 95-98 \\
    DC3m.2  &     210    &   246          & 97-98 \\
    \hline\hline
  \end{tabular}
\end{table}

\subsection{EMCal upgrade}

To expand the physics search capability of DarkQuest, we propose an electromagnetic calorimeter we can identify electrons, photons, and pions. By including these additional detection capabilities, we will access the lower mass region below 200 MeV (di-muon mass limit), and to test many additional dark sector models.. 
The broadened dark sector search program described in Section~\ref{sec:physics} can be achieved by adding one EMCal detector behind the Station-3 (St-3) chambers.

\subsubsection{PHENIX Shashlik Calorimeter}
 A suitable EMCal detector~\cite{PHENIX:2003fvo} has been identified from the previous PHENIX experiment at Brookhaven National Laboratory (BNL). The EMCal addition would be the first stage of a long-term dedicated dark sector physics program at Fermilab.
The EMCal upgrade builds on the recent effort to add $A'\to \mu^+\mu^-$ sensitivity to SeaQuest with a displaced di-muon trigger.  A GEANT4-based simulation package has been developed to study the trigger rate and background level in the di-muon channel. The data collected during a commissioning run in 2017 agrees very well with the simulation. This simulation has been extended to include EMCal. 

The proposed EMCal for DarkQuest would cover a 2\,m~$\times$~4\,m area perpendicular to the beamline.  This corresponds to 3$\times$6 EMCal super modules where:
\begin{itemize}
    \item 1 EMCal super module = 6$\times$6 = 36 EMCal modules
    \item 1 EMCal module is 11\,cm$\times$11\,cm and includes 2$\times$2 = 4 towers with a granularity of approximately 5.5\,cm.  
\end{itemize}
In total, this is 2592 towers or channels to read out of the EMCal.  
Adding an EMCal to the SeaQuest spectrometer serves two essential purposes. First, the EMCal provides a generic trigger for non-minimum-ionizing particles. A simple energy threshold will reject the muons that penetrate the beam dump, and trigger only on pions, electrons, and photons produced after the beam dump. Second, matching cluster energies in the EMCal to track momenta allows particle identification, which is necessary for rejecting backgrounds. In particular, the track-calorimeter cluster linking can allow for the separation of electrons from photons, pions, and muons by requiring the track momentum and electron energy to be similar. Furthermore, photons can be identified by looking for the absence of a track when a large electromagnetic deposit is placed. Details about the available particle identification are discussed in section~\ref{sec:performance}. 

The EMCal's ability to distinguish pions from electrons is important for rejecting background processes such as kaon decays. Misidentification of $K^0_L\to\pi^\pm e^\mp\nu_e$ as $A'\to e^+e^-$ is expected to be the dominant physics background, with $O(1000)$ such decays in the $(5-6)$ m) fiducial region, after having accumulated $1.44\times 10^{18}$ POT. An EMCal pion rejection factor of 1\% or better reduces this background to $O(10)$ events, which can be further reduced using mass resolution and pointing resolution \cite{berlinDarkSectorsFermilab2018}. 

\begin{figure}
\centering
\includegraphics[width=0.53\textwidth]{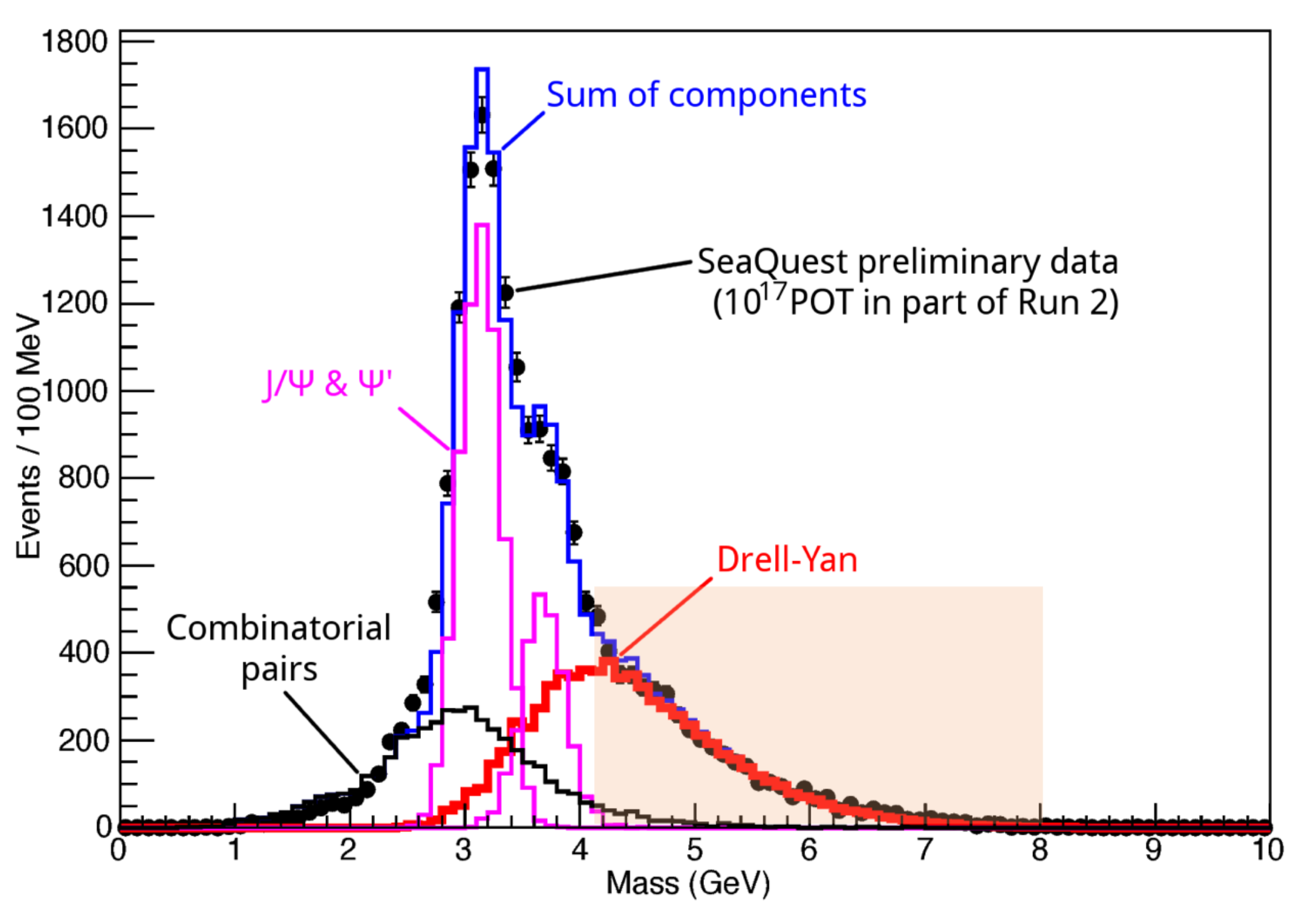}~~
\includegraphics[width=0.4\textwidth]{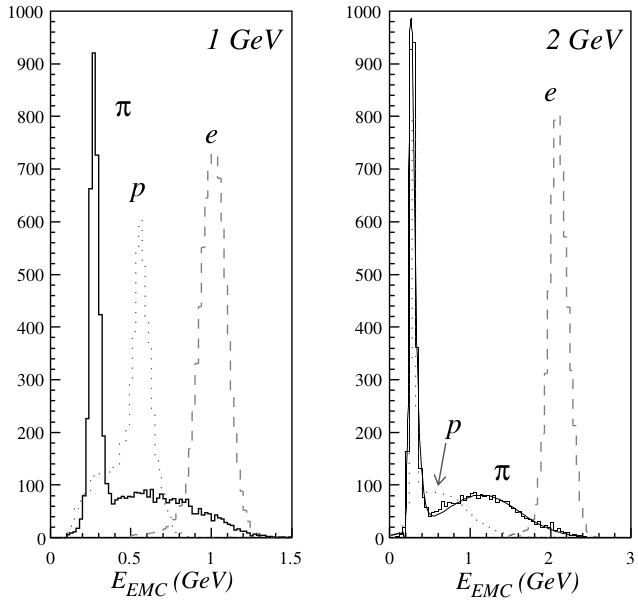}
\caption{Left: SeaQuest dimuon mass spectrum from Drell-Yan running comparing simulation and data, Right: the measured energy spectra when the EMCal is exposed to electrons, pions and protons at 1 GeV (left) and 2 GeV (right).}
\label{fig:emcal_performance}
\end{figure}

The energy resolution for electrons has been measured to be $8.1\%/\sqrt{E\mathrm{(GeV)}}+2.1\%$ by the PHENIX experiment at BNL. A typical electron from a potential dark photon decay has a momentum range of 2--20 GeV, which corresponds to 4--8\% energy resolution. Fig. \ref{fig:emcal_performance} shows the different energy deposition spectra when the EMCal is exposed to electrons, pions, and protons. The MIP peak at 0.25 GeV is well separated from the electron peak, which demonstrates clean triggering on electrons. The separation of the peaks from 2 GeV pions and electrons is large enough to allow pions to be identified with a veto from the St-4 muon ID system, and separation improves at higher energies. We, therefore, expect that the EMCal will meet or exceed the 1\% benchmark. More discussion of this performance is shown in section~\ref{sec:performance}.

The thickness of the beam dump determines both the sensitivity and the backgrounds for DarkQuest. Dark Photon di-muon decays can be reconstructed and separated from original collision point even when dark photon decays occur within the beam dump, provided they are sufficiently well separated. However, when considering dark photon to di-electron decays, DarkQuest is only sensitive to electron or photon production downstream of the beam dump, so a thinner dump increases the experiment's acceptance and allows us to access shorter decay lengths. A thicker dump absorbs a larger fraction of the kaons produced when the beam hits the beam dump, and therefore a thicker dump reduces the background. This interplay is critical towards understanding the ideal dark photon sensitivity. 
This current dump thickness of 5m (or 25 $\pi$ interaction lengths) has been used in the initial studies of DarkQuest performance, and we will use simulation studies to understand if further optimization is necessary.

Lastly, placement of the magnet relative to the tracking chambers is another critical component to achieving optimized sensitivity. Mass reconstruction, moreover muon identification is only achieved through the use of the magnet. As a consequence, track identification both before and after the magnet are essential to identifying muons. This limits the allowed possible lifetimes that the muon final state can probe to be before station 1 (or roughly 6m). An additional station or a re-positioning of station 1 and the Magnet further down the beam line would potentially increase the possible lifetimes that can be probed. Electrons, on the other hand, do not need the magnet to allow for energy, and subsequently mass reconstruction. As a consequence, electron identification may be possible for decays up to 12m at station 2 of the detector or even further. 

\subsubsection{Readout electronics and triggering}

In order to integrate the PHENIX EMCal into the existing SpinQuest data acquisition system, new readout electronics will need to be developed. The readout scheme must be able to digitize the light pulses from the 2592 EMCal towers at the 53\,MHz RF frequency and perform an energy sum calculation to provide a trigger signal that is efficient for dark sector signatures. The overall philosophy will be to build upon related electronics developments to minimize design costs. It is currently planned to factorize the readout scheme into two components: a front-end board to digitize the light pules and a back-end board to perform ADC, energy sums, and trigger algorithm computations. The front-end boards will use silicon photomultipliers (SiPMs) as photodetectors, replacing the photomultiplier tubes previously used by PHENIX.  The advantage of SiPMs are that they do not require high voltage, are radiation hard, and have a small form factor. Therefore the integration into the SpinQuest spectrometer and NM4 cavern will be greatly simplified. The cost of SiPMs from industry (e.g. Hamamatsu MPPCs) are relatively low and their performance is well understood. The STAR collaboration also plans to use repurposed PHENIX EMCal sectors for their Forward Calorimeter System, and they have verified that Hamamatsu MPPCs are suitable photodetectors for performing the front-end readout~\cite{STARFCS}. The back-end board will build upon a similar design being developed for the EMPHATIC experiment’s Aerogel Cherenkov Detector~\cite{emphatic}. The analog signals from the front-end boards will be aggregated and processed by 32-channel CITIROC 1A ASICs from WeeROC. These chips will shape the signal pulse and apply a discriminating threshold to produce a digital signal. These signals will then be processed by FPGAs located on the same back-end board to execute the trigger algorithms (e.g. total calorimeter energy).

\subsection{Additional tracking layers}


The SpinQuest spectrometer depicted in Fig.~\ref{fig:darkquest} shows two tracking layers between FMag and KMag, collectively referred to as Station 1. However, the second of those high-precision tracking stations is currently not operating due to difficulties in operation. Reviving additional tracking layers between the FMag and KMag would provide substantial improvement in the physics performance of the experiment.  Additional pattern recognition before KMag, where detector occupancy is the greatest would improve pattern recognition capability, rejecting background particles more efficiently, and improve vertexing resolution.  Furthermore, the additional tracking layer would extend the acceptance of the detector to charged displaced particles to roughly 7\,m (the position is configurable, to a degree).  This would then increase the overall sensitivity of the experiment to dark sector particles.  We leave it to future studies to quantify the impact of this additional tracking layer before the KMag, but anticipate that it will bring important additional capabilities to DarkQuest.  All simulation studies show below do not include the performance of this proposed detector.

We have identified a modest-cost technology which would be well-suited to the DarkQuest detector.  
Existing proportional chambers from the HyperCP experiment~\cite{HyperCP:2004kbv} stored at the Fermilab site would be a good candidate detector for this proposed additional tracking layer.  We would propose, in particular, to use chambers C5 \& C6.  These two chambers each have an aperture of 1.212$\times$0.405\,$m^2$ with a wire pitch of 1.5\,mm.  Their installation would provide additional points for tracks upstream of the analysis magnet, KMag.  The chambers are in an $X,V,U,X'$ configuratoin where the $X,X'$ planes each measure in the magnet bend plane (global $\hat{x}$ direction and the $V,U$ stereo planes are inclined at $\pm$26.57\,degrees to provide 3 dimensional reconstruction.  There are 320 wires for the $X,X'$ planes and 384 wires for $V,U$ planes.  We plan for the readout through the SeaQuest/SpinQuest ASDQ, Level-shifter board,.  The TDC electronics require 176 ASDQ cards and 22 Level shifter boards and 44 TDCs.  These chambers previously ran on fast gas mixture of argon/ethane bubbled through isopropyl alcohol at -0.7\,C in a 50/50 mixture and later on a fast gas mixture of CF4-isobutane (50-50 mixture).

\newpage
\section{Detector simulation and reconstruction}
\label{sec:performance}
In this section, we present a number of full simulation studies performed to detail the expected performance of the DarkQuest experiment to search for dark sector signatures.   

\subsection{Signal and background generation and simulation} 
\label{sec:sigsim}

For the signal simulation, we take the dark photon model as a benchmark and consider two production modes: via proton Bremsstrahlung and via the decay of $\eta$ SM mesons. For proton Bremsstrahlung, we manually generate $p p \rightarrow A'$ events that are weighted by the relevant differential cross-section $d^2\sigma/(dz\ d p_T^2)$, where $p_T$ and $z$ are the transverse momentum and the fraction of beam momentum carried by the outgoing dark photon, respectively. This procedure is outlined in \cite{Blumlein:2013cua,Gorbunov:2014wqa,deNiverville:2016rqh}. For meson decays, we model the production of SM $\eta$ mesons via $p p \rightarrow \eta$ with Pythia 8.2~\cite{Sjostrand:2014zea}. From the meson spectra generated in Pythia, we manually decay the $\eta$ meson to the dominant process $\eta \rightarrow \rho^0 A'$. Details on the branching ratio of this process and the  validation of the meson spectra are given in Ref.~\cite{berlinDarkSectorsFermilab2018}. 

\begin{figure}[tbh]
    \centering
    \includegraphics[width=0.32\textwidth]{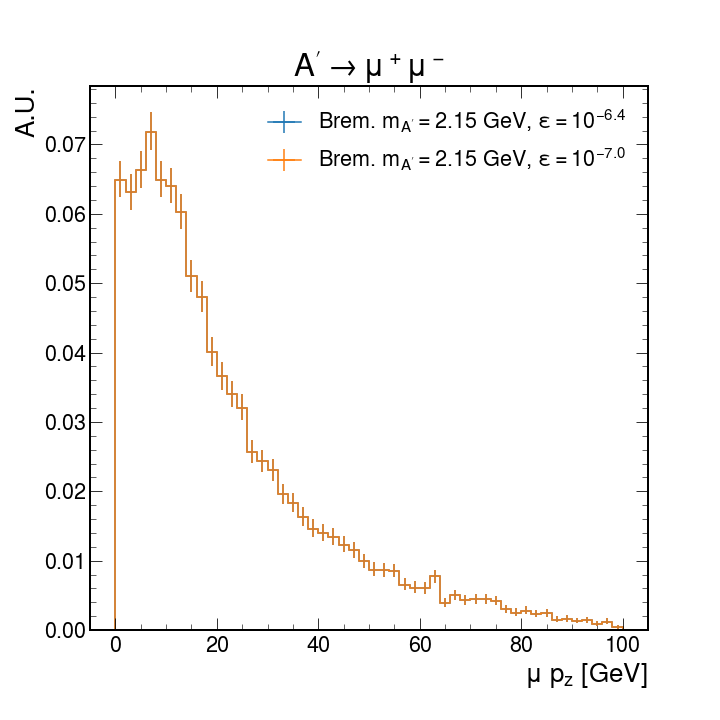}
    \includegraphics[width=0.32\textwidth]{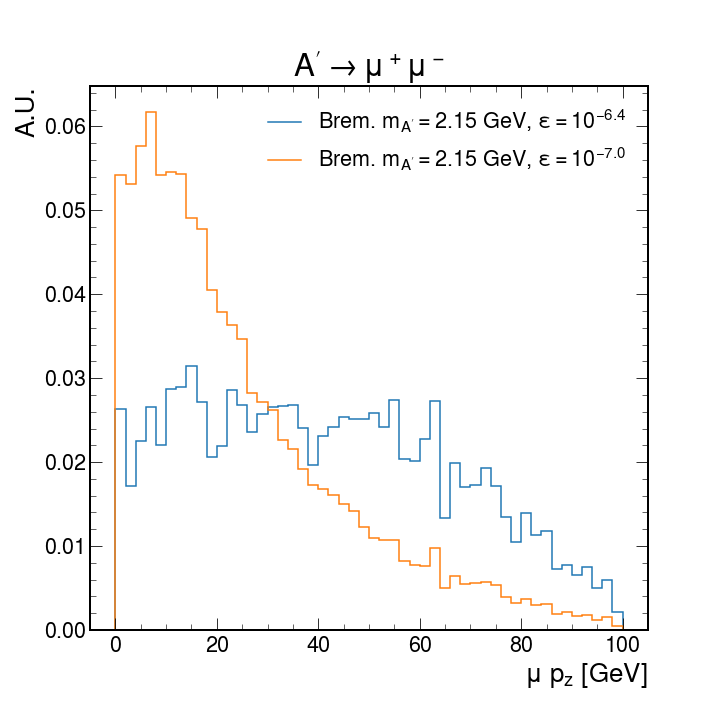}
    \includegraphics[width=0.32\textwidth]{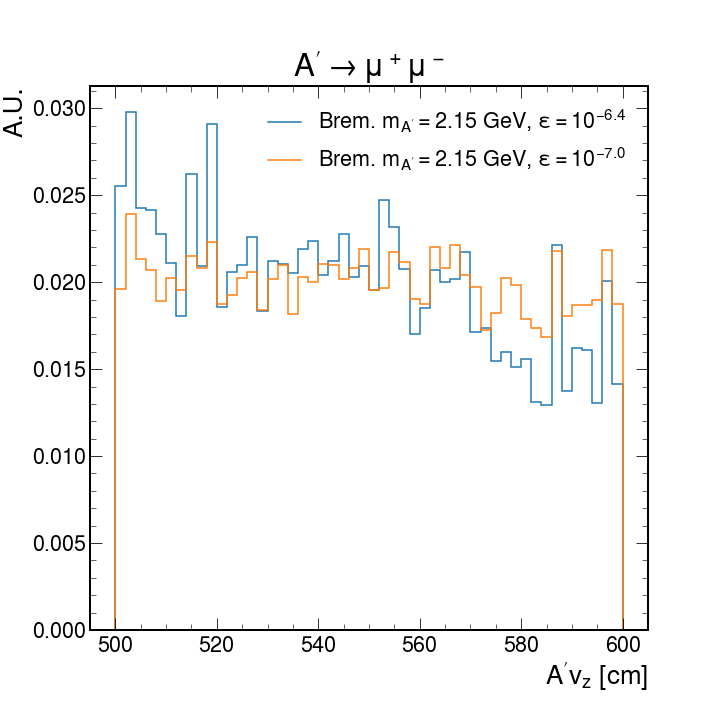}
    \caption{\textbf{Left and center}: $A'$ $p_z$ for two different values of $\epsilon$ before and after the re-weighting procedure, respectively, that accounts for a displacement in $z$. \textbf{Right}: $A'$ vertex position in $z$ after re-weighting. Here, the events are generated via proton Bremssthrahlung, they correspond to a fixed $m_{A'} = 2.15$ GeV and are displaced a fixed distance from the target that varies between 5 and 6 meters.}
    \label{fig:sigkinematics}
\end{figure}

We then consider the decays of the long-lived dark photons into SM leptons, $A' \rightarrow \ell^+\ell^-$. Such decays are controlled by kinetic mixing, $\epsilon$. For $m_\ell < m_{A'} < m_Z$, the partial width for decays to a pair of SM leptons is approximately: 
\begin{equation}
\centering
    \Gamma(A' \rightarrow \ell^+\ell^-) \approx \frac{\alpha_{\text em}\epsilon^2}{3} m_{A'}
\end{equation}
Thus, for a fixed $m_{A'}$, we sample the generated $A'$ events and manually displace the decay vertex $v_z$. We assign an event weight according to the $A'$ lifetime and its probability to decay in a fixed z range given its kinematics and $\epsilon$. For our studies we consider $v_z$ to be within 5 and 6 meters and $\epsilon$ to vary within $10^{-7.0}$ and $10^{-4.0}$. Figure~\ref{fig:sigkinematics} shows the $p_z$ and $v_z$ distributions of the generator-level muon before and after this re-weighting procedure for $A'\rightarrow \mu^+\mu^-$. Dark photons with lower couplings are generally more boosted.

To model the response of the spectrometer we use the official E1039 software simulation. This simulation package uses GEANT4~\cite{AGOSTINELLI2003250} to model each of the spectrometer stations, including the EMCal, and simulate the digitization, hit, track and vertex reconstruction. The simulation studies detailed in the next sections use events with reconstructed-level information for signal and background.

To simulate background events we generate a large sample of proton-on-target interactions. These result in a large number of primary and secondary particles aside from leptons, such as pions, kaons or neutrons. On average, the simulation and reconstruction steps for so many protons interacting on target requires large computing time and resources. Moreover, it is even more difficult to generate a full spill of $5 \times 10^{12}$ protons per spill. For this reason, we take a different approach for our background simulation and divide it in three steps:
\begin{enumerate}
    \item We generate one large dataset with $10^8$ events at generator-level, referred to as the proton-gun dataset. Each proton-gun event corresponds to a single 120 GeV proton that traverses the detector. The proton vertex is along in the beam line: the $z$ position is centered at the target position ($z=-7.$m) and the $x$ and $y$ position are sampled from the beam profile distribution. 
    \item We take $N$ events from the proton-gun dataset and embed them together to generate one background event. Here, $N$ is derived from a random number sampled from the beam intensity profile, which is equivalent to the distribution of number of protons in a bunch.
    \item We run the reconstruction steps including hit efficiency, tracking and vertexing, for each background event.
\end{enumerate}
These steps result in a sample of background events with an intensity that mimics the beam intensity in a run. 

\begin{figure}[htp]
    \centering
    \includegraphics[width=0.4\textwidth]{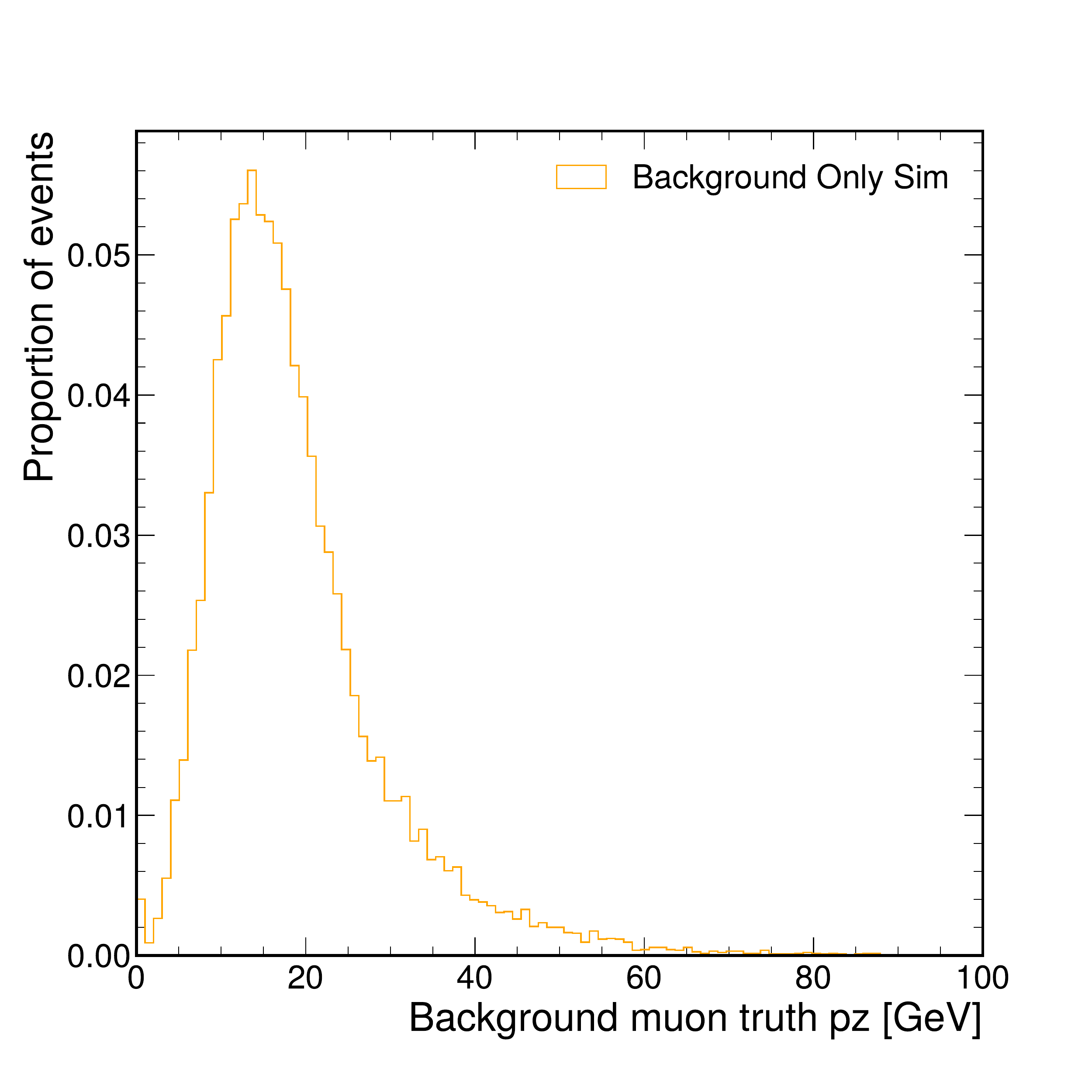}
    \includegraphics[width=0.4\textwidth]{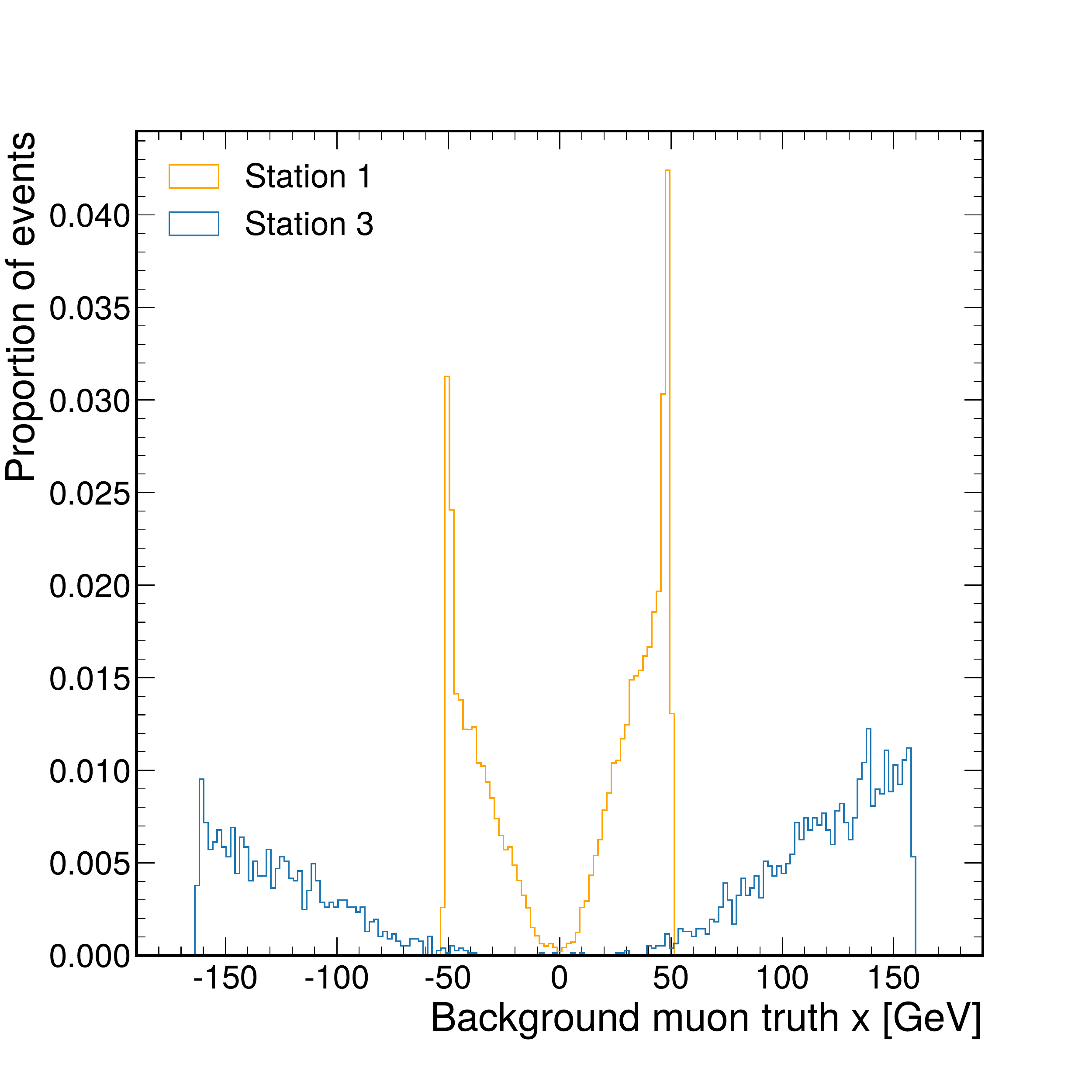}
    \caption{(Left) Distribution of $p_z$ for particles that pass through the FMag in the background simulation. (Right) Distribution of $x$ location in drift chamber station 1 and station 3 for these particles.  Particles that leave hits in station 1 but no hits in station 3 \textit{are} included in the station 1 distribution, but are left out of the station 3 distribution.}
    \label{fig:bkgPz}
\end{figure}

For our studies shown below, we used this procedure to generate a sample of 30k background-only events. We filter muons to pass through the FMag with a $p_z$ distribution shown in Fig.~\ref{fig:bkgPz}. After passing through the FMag, these particles are often are swept out to the edges of the detector - they tend to have high $x$-values, as also shown in Fig.~\ref{fig:bkgPz}. Nearly 85\% of the muons in the background simulation that leave a hit in station 1 do not leave enough hits in station 3 to be reconstructed. At the event level, there are about 1.9 generator-level particles that leave hits in station 1, though only about 25\% of events have generator-level particle that leaves hits in station 3.

\subsection{Trigger}

The beam repetition rate for the main injector is 53\,MHz and the data acquisition rate for SpinQuest is approximately 4\,kHz.  Therefore, events needs to be filtered, or triggered, by 4 orders of magnitude before being saved for later analysis.  For dark sector particles decaying to minimum ionizing particles, such as in the muon case, the hodoscope tracking layers provide information for triggering.  In a DarkQuest upgrade, we propose to use the EMCal as a triggering detector as well.  We explore the performance of both in this section.  

\subsubsection{Hodoscope triggering}
\begin{figure}[htpb!]
    \centering
    \includegraphics[width=0.42\textwidth]{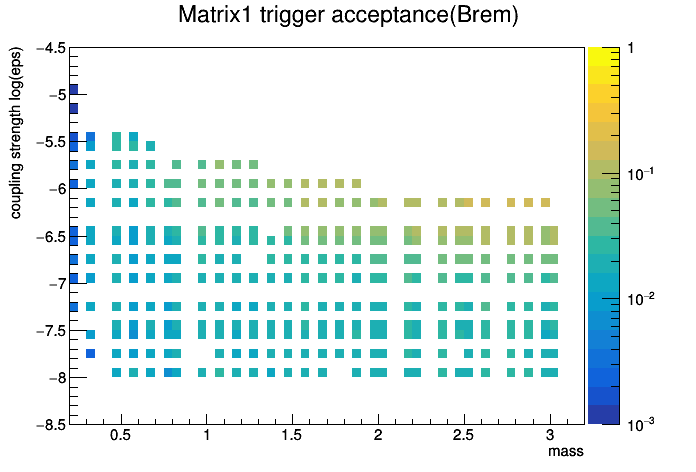}
    \includegraphics[width=0.42\textwidth]{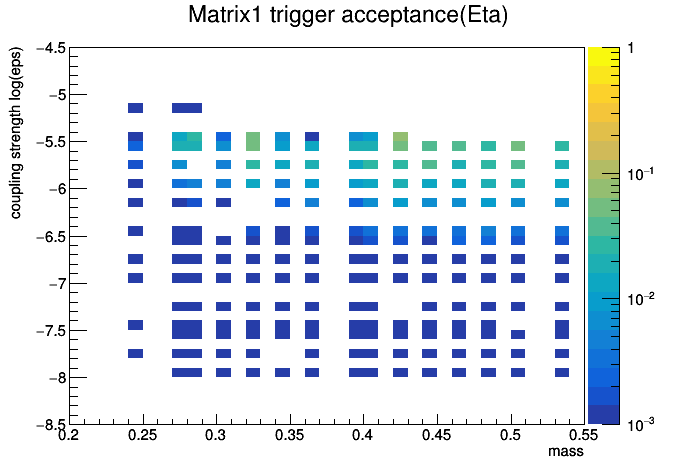}
\caption{Trigger efficiency for A' (displaced 5-6m) to dimuon events from Monte Carlo simulation. Each small box corresponding to a mass and coupling constant combination for A'. 10,000 events were simulated for each combination. The efficiency of the original SeaQuest Trigger system is shown for dimuon events produced by proton bremsstrahlung (left) and eta decay (right) respectively.}
\label{fig:OldTriggerEff}
\end{figure}

The current SpinQuest trigger system contains three separate levels of CAEN V1495 VME modules which include an Altera EP1C20F400C6 Field-Programmable Gate Array~(FPGA). The trigger system reads the hits on the fast-responded hodoscopes and distinguishes the signal dimuon final states from a Drell-Yan process from background events that include accidental single muon hits and dimuon events generated in the beam dump. Although the overall performance of this FPGA-based trigger system works quite well for the SpinQuest experiment, especially for the high-mass (4-10 GeV) dimuons, the current trigger-roads are significantly suppressing the efficiency for low-mass dimuons because of background events generated by charmonium decays \cite{seaquestcollaborationSeaQuestSpectrometerFermilab2017a}. While it may not be a big issue for previous nuclear experiments focusing on the Drell-Yan process only, it affects the efficiency for detecting dark sector particles in MeV-GeV mass region. Figure~\ref{fig:OldTriggerEff} shows the calculated trigger efficiency for simulated dark photons (A') to dimuons events to pass the current main physics trigger. The efficiency, which defined as number of events identified by the main physics trigger over total event number simulated, is relatively low for both production channels. For this region where new physics potentially exists, a new complementary trigger road is proposed to cover our physics interests.

\begin{figure}[htpb!]
    \centering
    \includegraphics[width=0.85\textwidth]{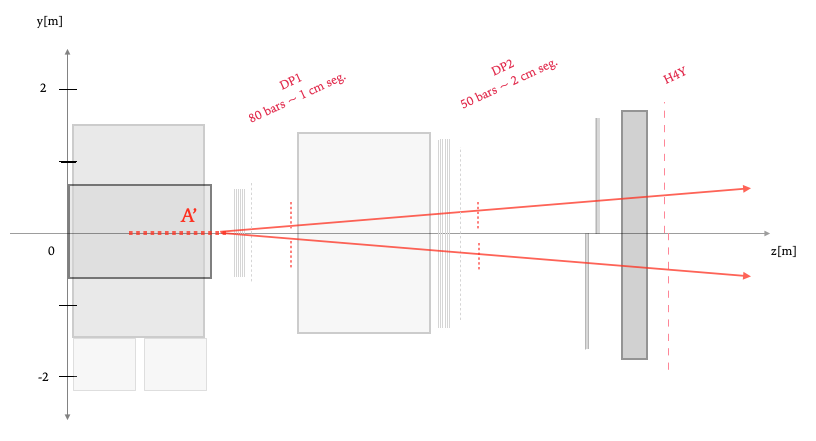}
\caption{The vertex reconstruction logic using two DP fiber hodoscopes. Unbent y position can be measured relatively accurately for each hit on the hodoscopes and two tracks of dimuon events will then be reconstructed. We find the intersection point to determine the z-vertex position, and it can distinguish displaced dimuon events. The H4Y detector with lower resolution works as an additional check.}
\label{fig:TriggerLogic}
\end{figure}

In order to detect the dark sector particles produced by proton bremsstrahlung and meson decay in low mass region, two hodoscopes performing a hit measurement in y were installed in 2017. 
Fig.~\ref{fig:TriggerLogic} shows the trigger logic. The trigger fires if there are hits in the same quadrant for DP1 and DP2 detectors and for the dimuon events. We can use their unbent y position to draw two straight lines to find the crossing point in z. In our simulation we observe that the z-vertex position resolution can reach 30 cm. If this reconstructed z-vertex position is relatively far away from the target, it can be a unique signal to identify the possible dark sector particles. 

\begin{figure}[htb!]
    \centering
    \includegraphics[width=0.85\textwidth]{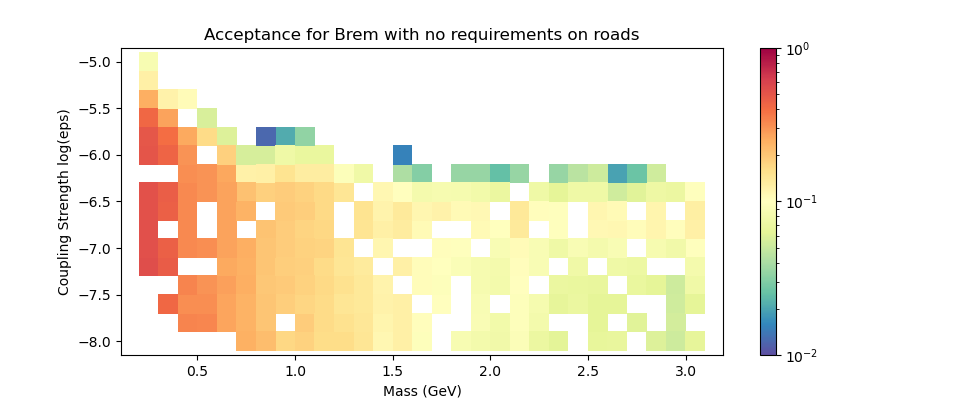}
\caption{Trigger acceptance from displaced vertex reconstruction using DP fiber hodoscopes for proton bremsstrahlung. }
\label{fig:NewTriggerEff}
\end{figure}

Following this new trigger logic, we scan the trigger acceptance for signal $A'$ dimuon decays, produced via bremsstrahlung, in Fig.~\ref{fig:NewTriggerEff}. The numerator is the number of events with reconstructable hits on two DP hodoscopes while the denominator is the total number of $A'$ to dimuon events generated through simulation. The new trigger path complements the original one and increases acceptance for lower mass dimuons.

\subsubsection{Calorimeter triggering} 
\label{sec:calTrig}


The addition of the EMCal detector creates the possibility of a calorimeter-based trigger. The calorimeter is designed to capture $e$/$\gamma$ showers, so this trigger would primarily target events with $e^+ e^-$ decays, although it could also trigger events with displaced photons or even $\pi$/$K$ that decay to leptons or photons.

\begin{figure}[htp]
    \centering
    \includegraphics[width=0.4\textwidth]{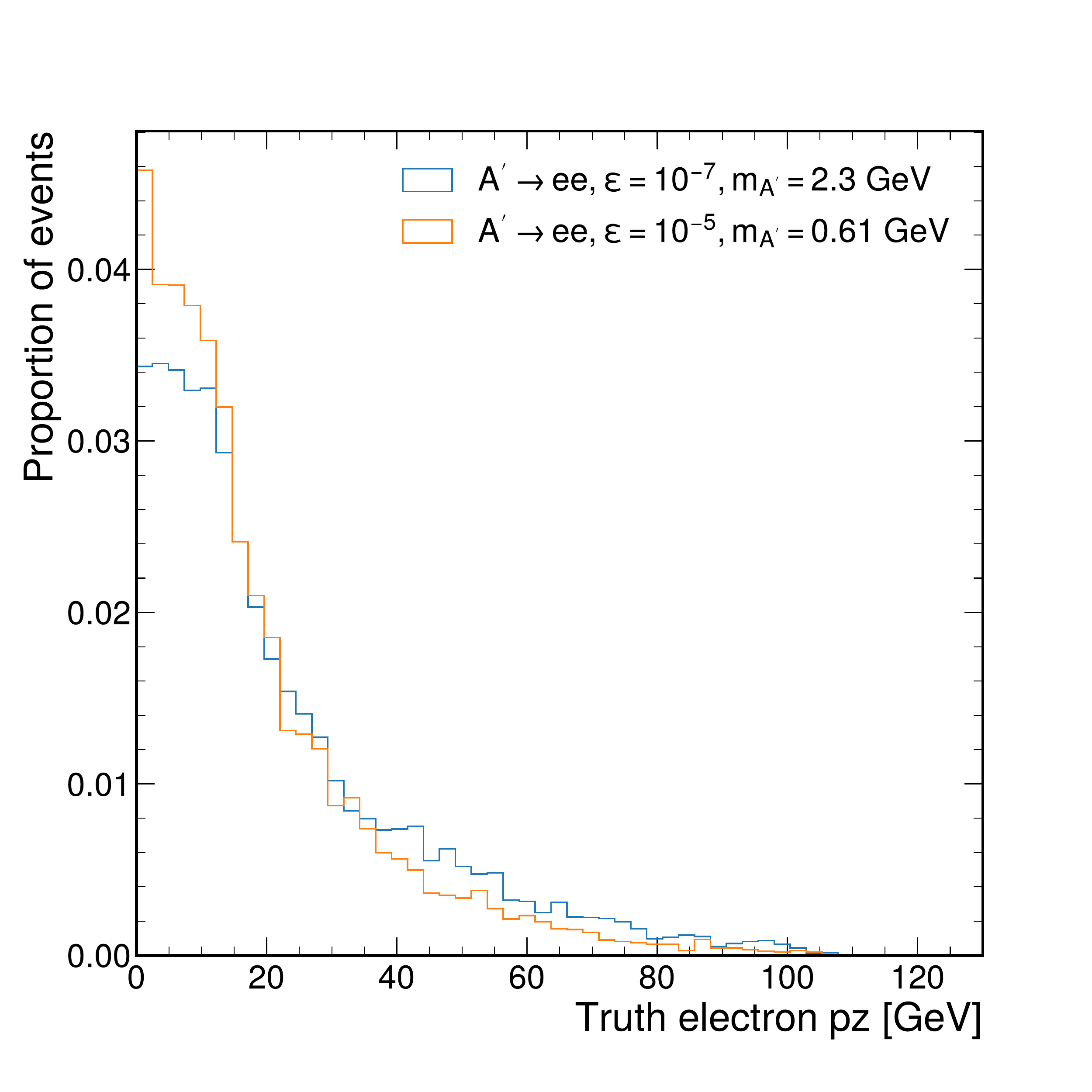}
    \includegraphics[width=0.4\textwidth]{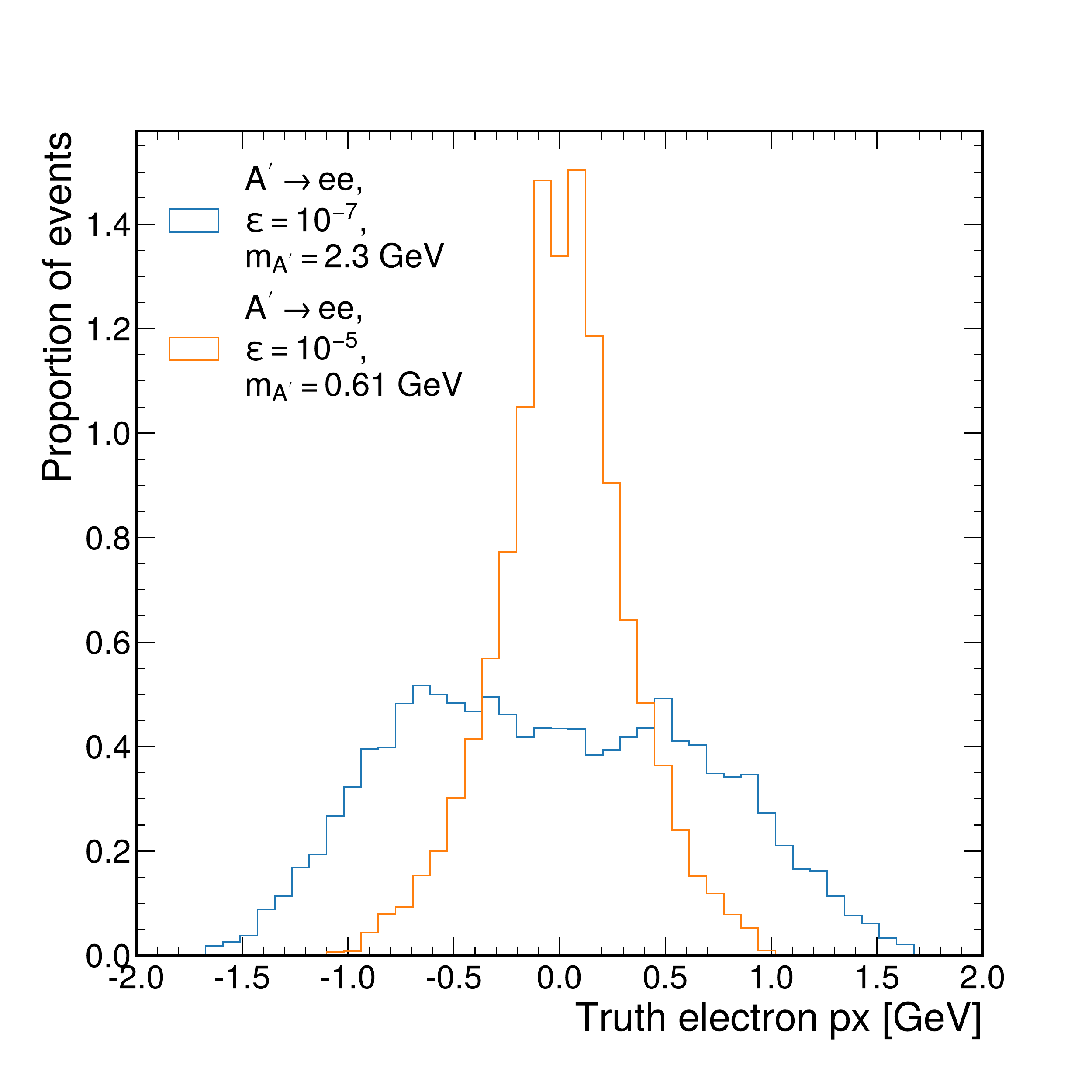}
    \caption{(Left) Distribution of $p_z$ for the electrons in two of the $A' \rightarrow e^+ e^-$ signal samples discussed in Sec.~\ref{sec:calTrig}.  (Right) Distribution of $p_x$ for these same two samples.}
    \label{fig:EEPZ}
\end{figure}

To design the trigger, we simulate the energy deposits of signal and background events in the calorimeter. For the $A' \rightarrow e^+ e^-$ signal, we choose four points in the $\epsilon$-$m_{A'}$ space: $\epsilon = 10^{-7}, 10^{-5}$ and $m_{A'} = 0.614, 2.303$~GeV and generate 10k events. Figure~\ref{fig:EEPZ} shows the $p_z$ and $p_x$ distributions for electrons in two of the $A' \rightarrow e^+ e^-$ signal samples. For the background we use the 30k background event dataset described in Sec.~\ref{sec:sigsim}.

\begin{figure}[htp]
    \centering
    \includegraphics[width=0.5\textwidth]{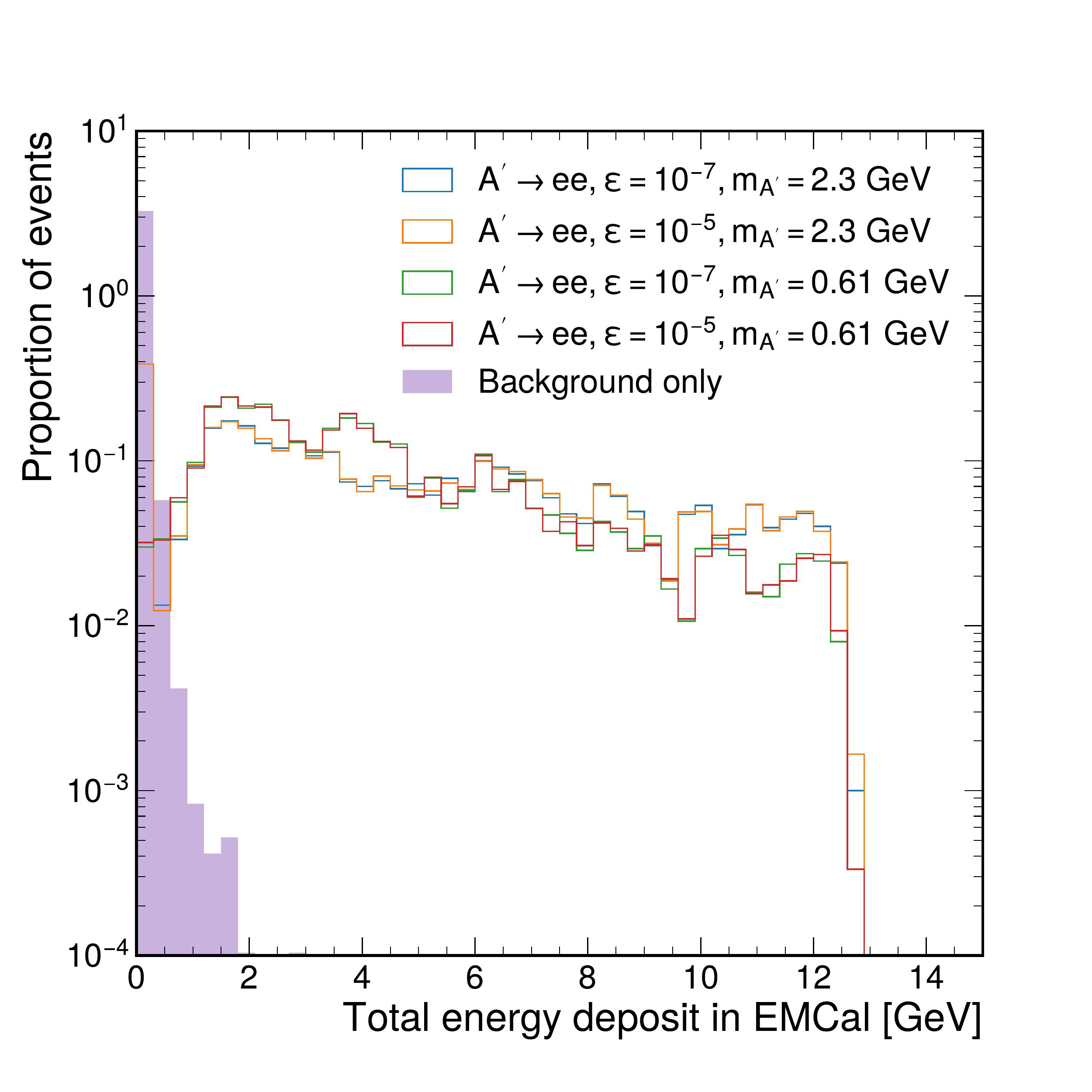}
    \caption{Total energy deposition in the EMCal for background events (shaded purple) and four $A' \rightarrow e^+ e^-$ signal points (unfilled line histograms).  The energy deposition is quoted without applying the sampling fraction of the EMCal.}
    \label{fig:calTrig}
\end{figure}

The distributions of the total energy deposited in the EMCal for the background and signal samples are shown in Fig.~\ref{fig:calTrig}.  Background events typically have very little energy deposited in the EMCal, with fewer than 0.06\% of events having a total energy deposit greater than 1 GeV, and about 0.006\% have an energy sum greater than 2 GeV.  Conversely, the $A' \rightarrow e^+ e^-$ signal events typically have relatively large energy depositions.  For example, 86.3\% of events in the $\epsilon = 10^{-7}$,~$m_{A'} = 2.303$~GeV signal point sample had a total EMCal energy deposition greater than 1 GeV, and 71.2\% had a total greater than 2 GeV.  By applying a cut of 3 GeV on the total EMCal energy deposit, it is likely that a sufficiently low trigger rate could be achieved to allow for a more complete processing of events.  Further studies with more statistics in real data will ultimately help us to determine the final EMCal trigger threshold, though, the final threshold will be able accept a large fraction of signal events.  

\subsection{Displaced tracking and vertexing} 
A critical component of the DarkQuest reconstruction scheme is reconstructing the trajectories (``tracking'') of ``displaced'' particles using the existing drift chambers.  Here ``displaced'' indicates the following:
\begin{itemize}
    \item For muon final states: we require that the particle was not created in the proton-on-target interaction. Because muons can traverse the FMag, a full trajectory can be reconstructed for muons created anywhere between $z = 0$~cm and the beginning of the first drift chamber.
    \item For electron final states:  we require that the particle was created after the FMag.  The radiation length of Fe is about 1.8 cm, electrons produced within the FMag effectively cannot create reconstructable trajectories in the drift chambers.  As such, electron tracks can be fully reconstructed as long as the particle is produced after the end of the FMag, at approximately $z=500$~cm and the beginning of the first drift chamber.  Similar arguments apply for charged pions and kaons, although these particles are slightly more penetrating in iron.
\end{itemize}
In both cases, it is possible to reconstruct a \textit{linear} trajectory using hits in the second and third drift chambers, which are downstream of the KMag.  While a momentum measurement cannot be made in this way, this information \textit{can} be used for particle identification, and in conjunction with an EMCal measurement, electron energy can be recovered.

The SpinQuest tracking algorithm is highly specialized for reconstructing prompt muon tracks, making assumptions about the charge and momentum of a trajectory based on the location in the detector as it is being reconstructed~\cite{Sanftl:2014}.  These assumptions can be valid for particles that pass through the FMag, but they are \textit{not} necessarily valid for displaced particles, rendering the default algorithm largely ineffective for DarkQuest purposes.

Because of these issues, a new tracking algorithm is being developed which has the flexibility to reconstruct both displaced and prompt particle trajectories.  This section describes the current performance of this algorithm in terms of efficiency, fake rate, single-particle resolution, and displaced vertex reconstruction. The results shown here demonstrate the capability of the detector and will continue to improve, which we leave to future work.


\subsubsection{Displaced tracking efficiency}
\label{sec:trackEff}

To study the tracking efficiency for displaced particles a dataset of single muons was generated with the following kinematic information:
\begin{itemize}
    \item The $z_0$ of the muons was set at $z = 520$~cm, to ensure that the particle was generated between the end of the FMAG and in front of the first drift chamber station.
    \item The $x_0$ and $y_0$ of the muon were independently sampled from flat distributions between [-10, 10] cm.
    \item The $p_z$ of the muon was sampled from a flat distribution between [0,100] GeV.
    \item The $p_x$ and $p_y$ of the muon was sampled from a flat distribution between [-1.5,1.5] GeV.
\end{itemize}
We will refer to this type of simulated event as a ``muon-gun'' event.  The studies in the following section were performed using a muon-gun sample of 32k events.

The displaced tracking algorithm was optimized for reconstructing trajectories with $p_z > 10$~GeV.  This was done primarily to improve algorithm speed in events with large numbers of hits in the drift chambers, but the algorithm's reach can easily be extended to lower momenta with straightforward adjustments. 

After the generation of the truth muons and subsequent simulation of detector response, an extra layer of drift chamber emulation is applied to replicate wire inefficiencies and resolution effects. In particular, every wire is assumed to have a hit efficiency of 94\%, conservatively, and the resolution is approximated by smearing the drift distance with a value drawn from a Gaussian with width parameter of 400 $\mu$m.

\begin{figure}[htp]
    \centering
    \includegraphics[width=0.5\textwidth]{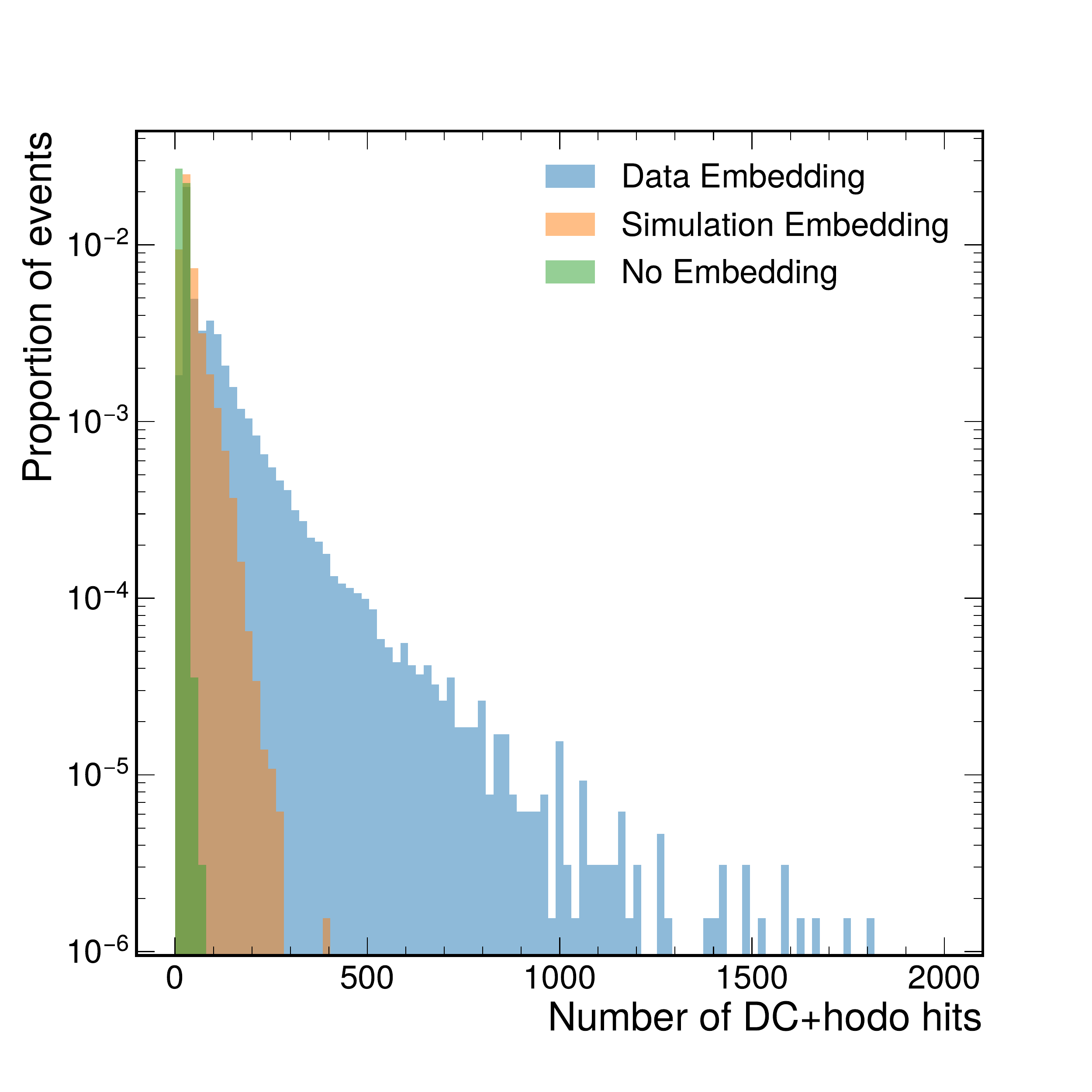}
    \caption{Distributions of the number drift chamber and hodoscope hits before and after applying simulated background and data event overlay to muon-gun events.}
    \label{fig:nHitsDCH}
\end{figure}

Pileup can be emulated by ``overlaying'' simulated background events or data events from the E906 experiment.  Here, an ``overlay'' just means that the hit collection from the background or data event is combined with the hit collection from the muon gun event.  Drift chamber efficiency and resolution emulation \textit{is} applied to simulated background events but \textit{is not} applied to data events.  The background simulation is the same as that discussed in Sec.~\ref{sec:sigsim}. After applying the event overlay, the distributions of the number of drift chamber and hodoscope hits - which are the hits relevant to tracking - are shown in Fig.~\ref{fig:nHitsDCH}. The number of hits added in the data overlay tends to be larger than that in the simulated background overlay.

As the number of tracking-relevant hits in the detector increases, the processing time of the reconstruction algorithm tends to increase exponentially.  While the displaced tracking algorithm actually has better timing performance than the original E906 algorithm, pileup mitigation techniques are still required.  As an example, the simplest mitigation technique would be a cut on the number of drift-chamber hits in station 1.  A benchmark cut at 300 hits would result in about 17\% of data-overlaid events being ignored.  Regardless of the mitigation scheme employed, some loss in reconstruction efficiency at high-pileup is expected.

In the following studies, a reconstructed trajectory is considered to be ``matched'' to a generator-level particle if the trajectory's $x$-value is within 3 cm of the generator-level particle's $x$-position in station 1 and station 3, and if the trajectory's $y$-value is within 4 cm of the generator-level particle's $y$-position in station 1 and station 3.  More refined truth-matching schemes are possible based on hit-level information, but they are still under development for the new displaced tracking algorithm.

\begin{figure}[htp]
    \centering
    \includegraphics[width=0.4\textwidth]{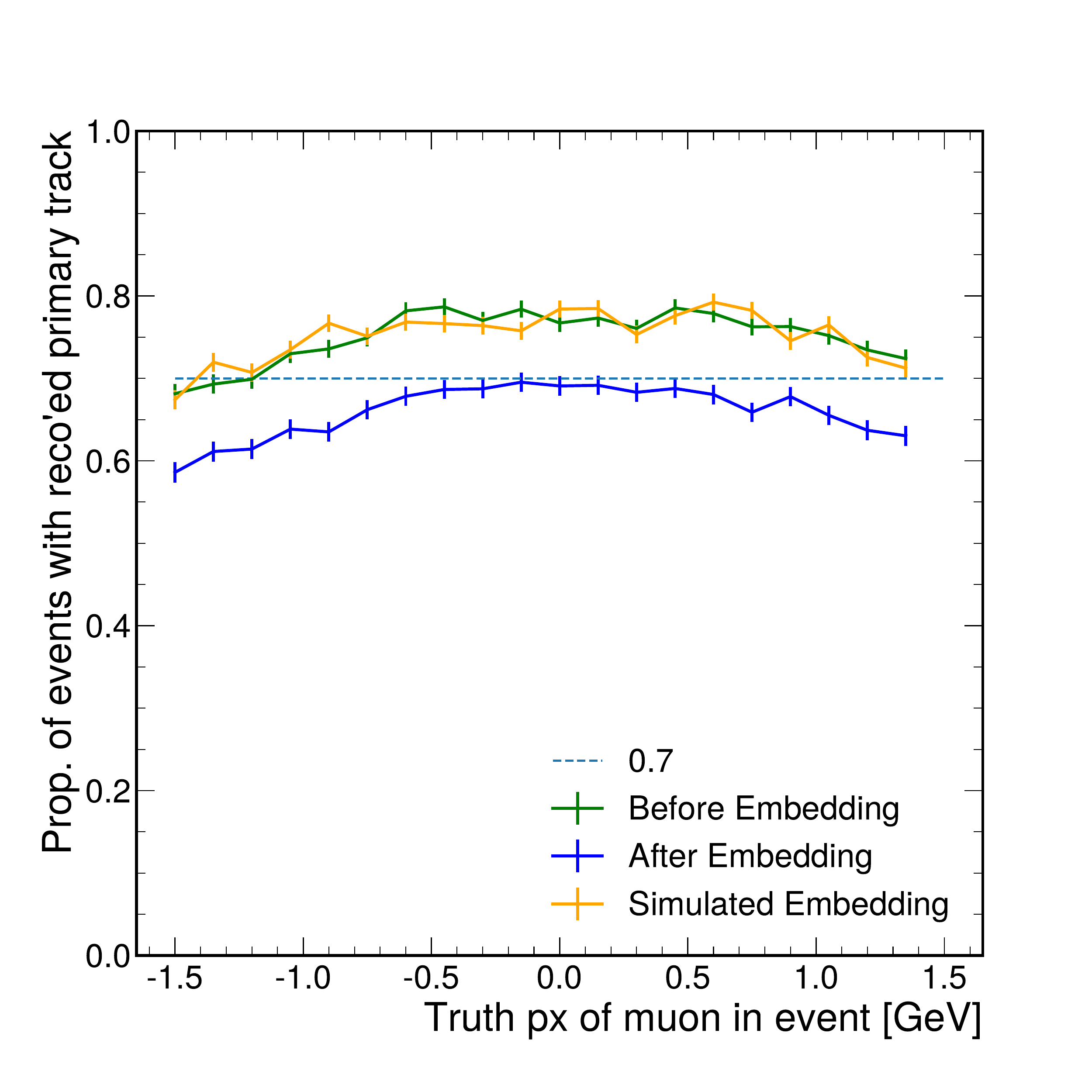}
    \includegraphics[width=0.4\textwidth]{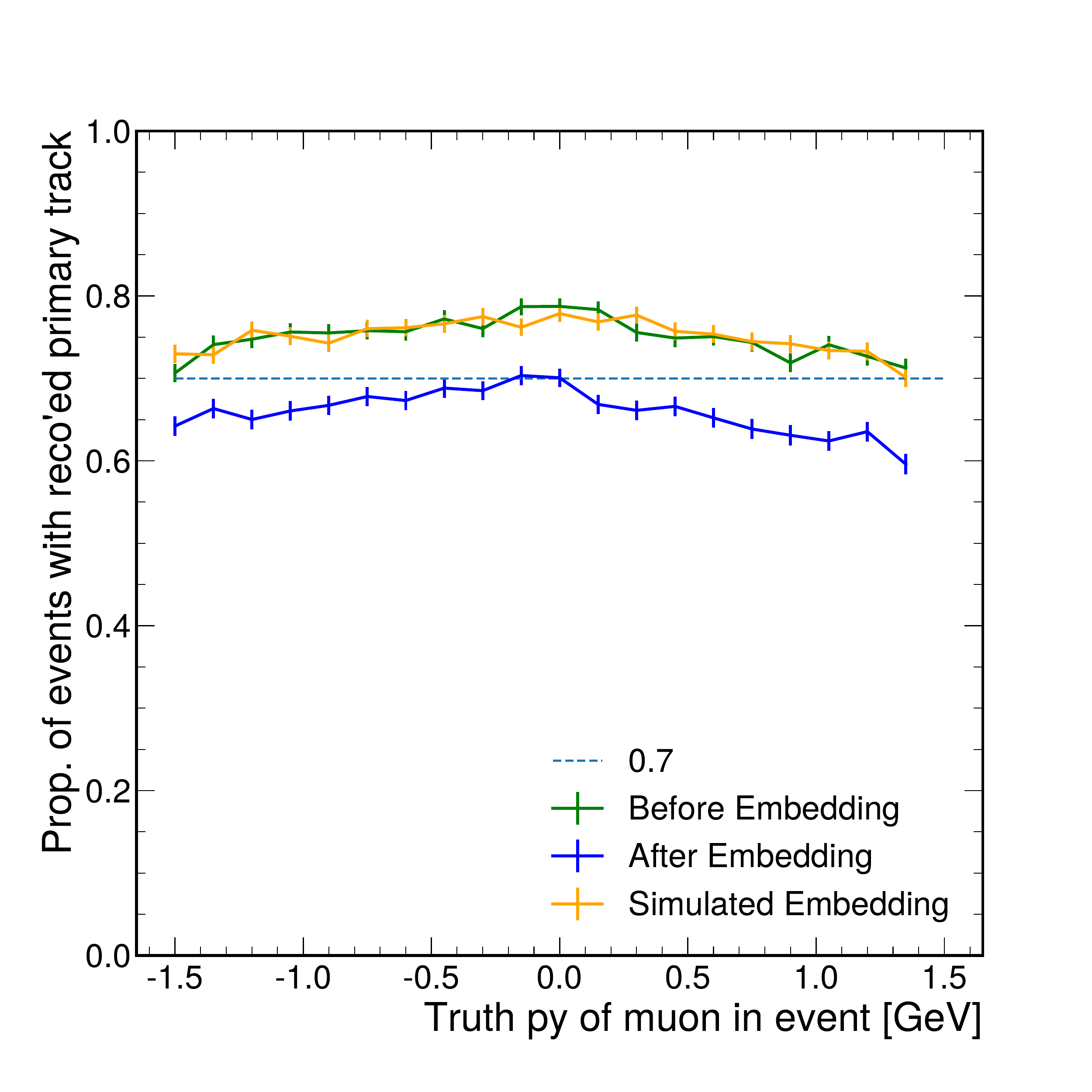}
    \caption{(Left) Primary particle trajectory reconstruction efficiency as a function of $p_x$ for the muon-gun sample described at the beginning of Sec.~\ref{sec:trackEff}.  (Right) Primary particle trajectory reconstruction efficiency as a function of $p_y$ for the same sample.  Efficiency is shown both before any overlay, after simulated background overlay, and after data event overlay.}
    \label{fig:pxpyEff}
\end{figure}

The primary particle reconstruction efficiency is shown as a function of $p_x$ and $p_y$ for the muon-gun events in Fig.~\ref{fig:pxpyEff} before any overlay, after simulated background overlay, and after data event overlay.  The efficiencies are relatively flat against both $p_x$ and $p_y$, and while the overlay of simulated background events has little effect on the efficiency (i.e. it rarely triggers the algorithm's pileup mitigation scheme), there is about a 10\% efficiency loss after embedding E906 data events. From this plot, the efficiency after data embedding peaks around 70\%, but it should be noted that this includes an implicit integration over the region $0<p_z<10$~GeV, where the algorithm is known to be inefficient. 

\begin{figure}[htpb!]
    \centering
    \includegraphics[width=0.42\textwidth]{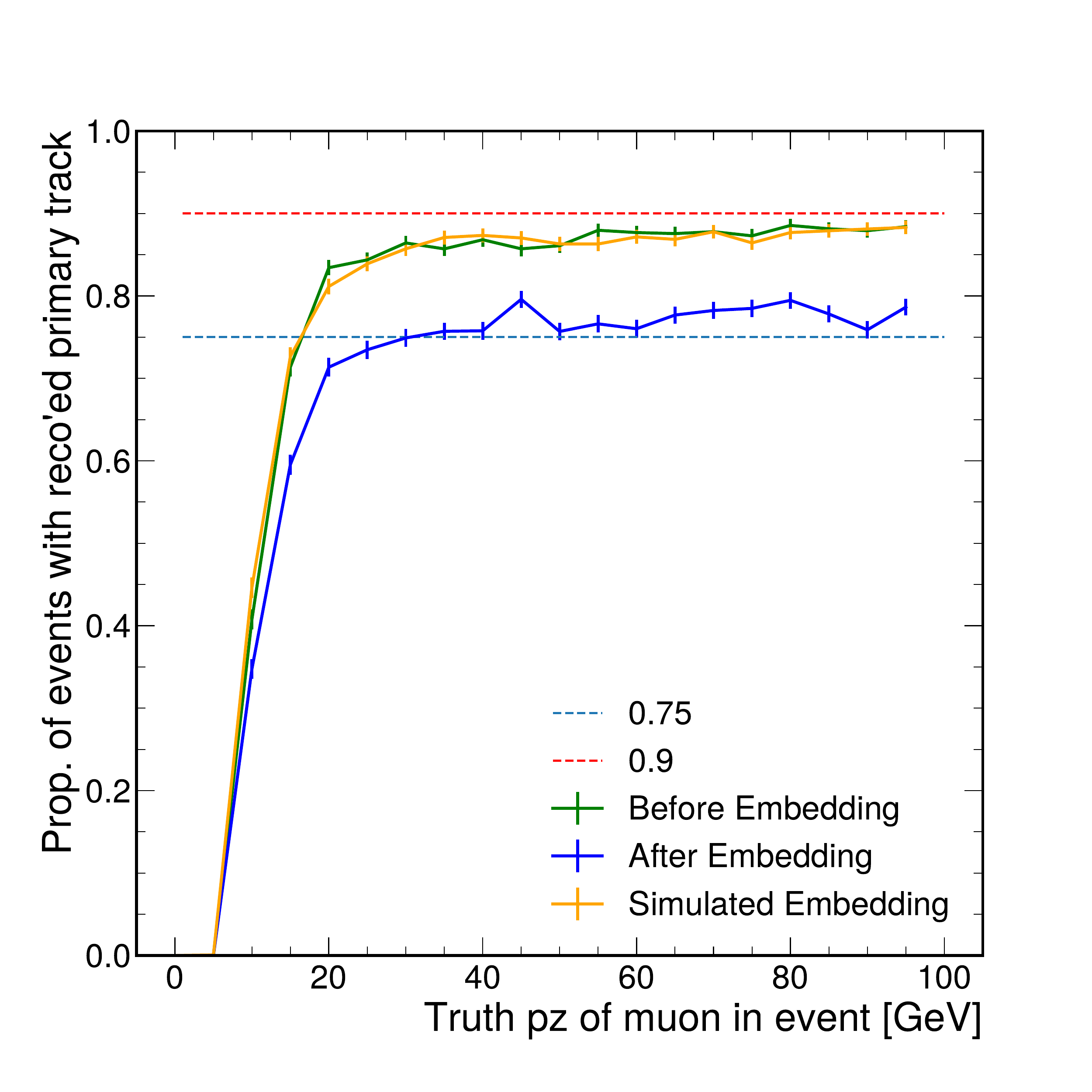}
    \includegraphics[width=0.42\textwidth]{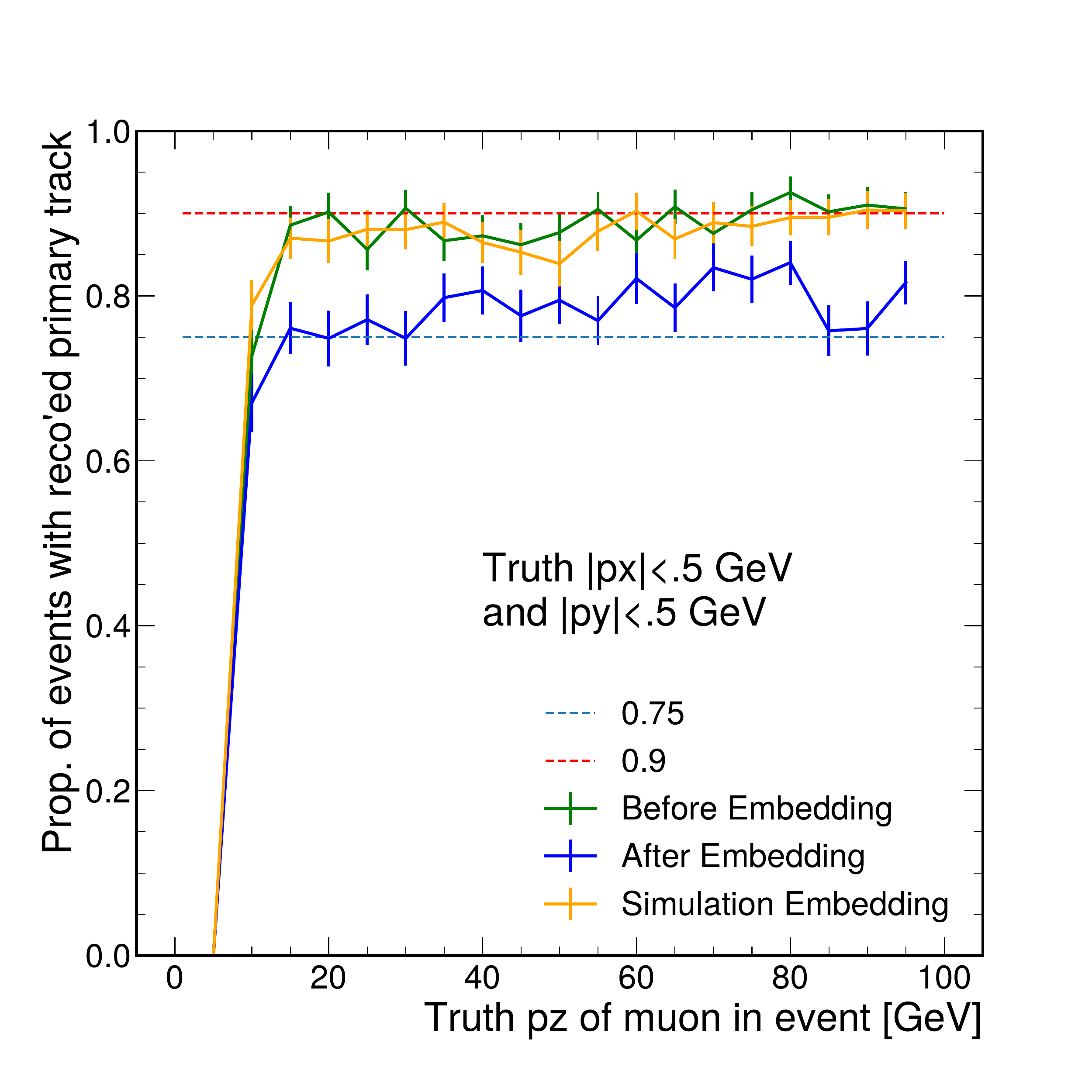}
    \caption{(Left) Primary particle trajectory reconstruction efficiency as a function of $p_z$ for the muon-gun sample described at the beginning of Sec.~\ref{sec:trackEff}.  (Right) Primary particle trajectory reconstruction efficiency as a function of $p_z$ for the same sample after restricting the $|p_x|$ and $|p_y|$ to be less than 0.5.  This largely keeps the particle in detector acceptance.  Efficiency is shown both before any overlay, after simulated background overlay, and after data event overlay.}
    \label{fig:pzEff}
\end{figure}

The efficiency as a function of $p_z$ is shown in Fig.~\ref{fig:pzEff}, both with unrestricted $p_x$ and $p_y$ and with a cut on $p_x$ and $p_y$ to keep the particle within detector acceptance (the acceptance in $x$ at the final drift chamber station is about [-100, 100] cm).  From these plots it is clear that after data embedding, the muon reconstruction efficiency is slightly better than 75\% down to about $p_z = 10$~GeV.  Prior to the data-overlay, the reconstruction efficiency for particles with $p_z > 10$~GeV is just shy of 90\%.  If we take the per-lepton efficiency to be 85\% and the pileup efficiency loss to be 10\%, then for di-lepton events where both leptons are in acceptance with $p_z > 10$~GeV, the expected efficiency is about $0.85*0.85 - .1 \approx 0.62$.  The efficiency loss due to pileup is applied at the event-level, rather than per-event level.

\subsubsection{Tracking momentum resolution}
\begin{figure}[htpb!]
    \centering
    \includegraphics[width=0.32\textwidth]{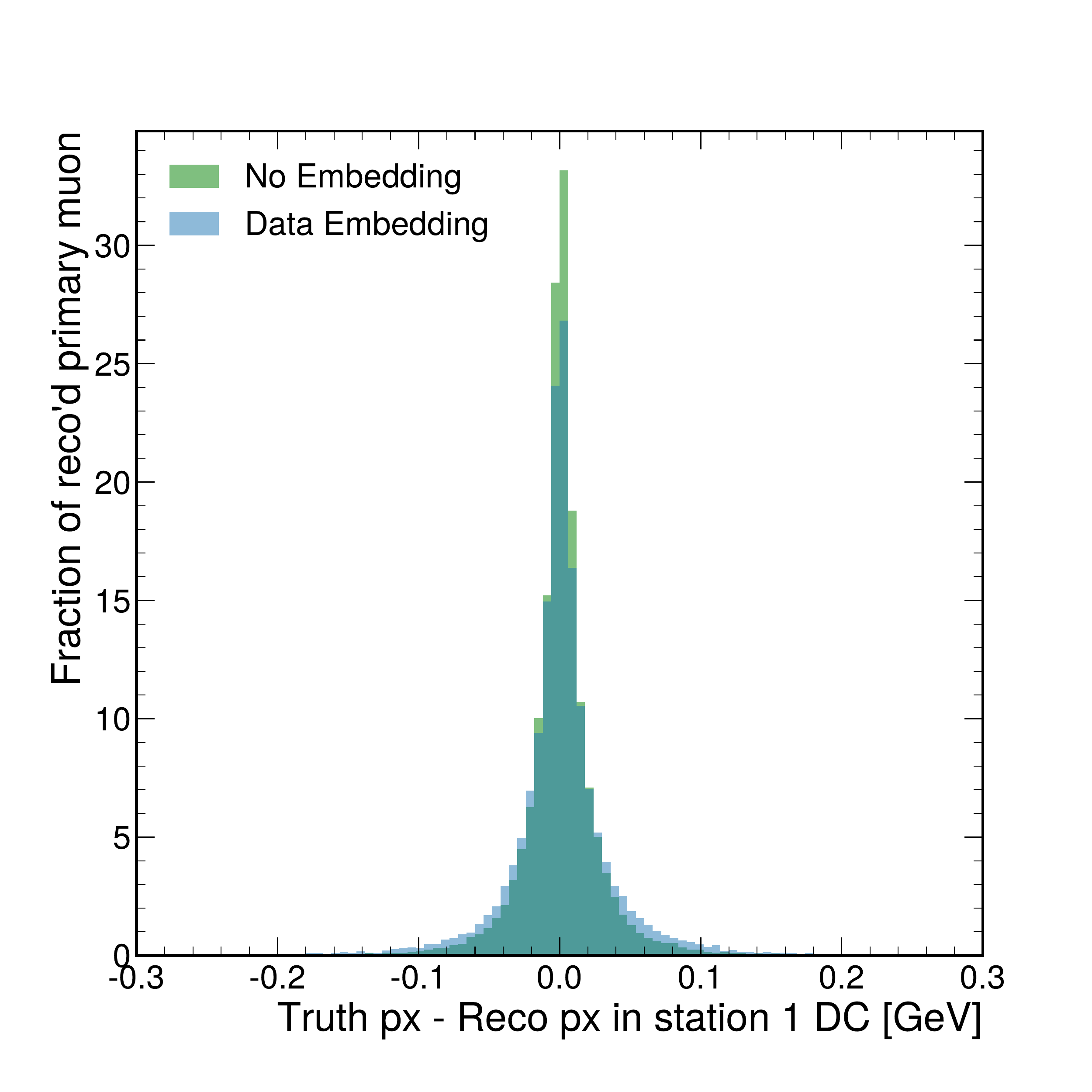}
    \includegraphics[width=0.32\textwidth]{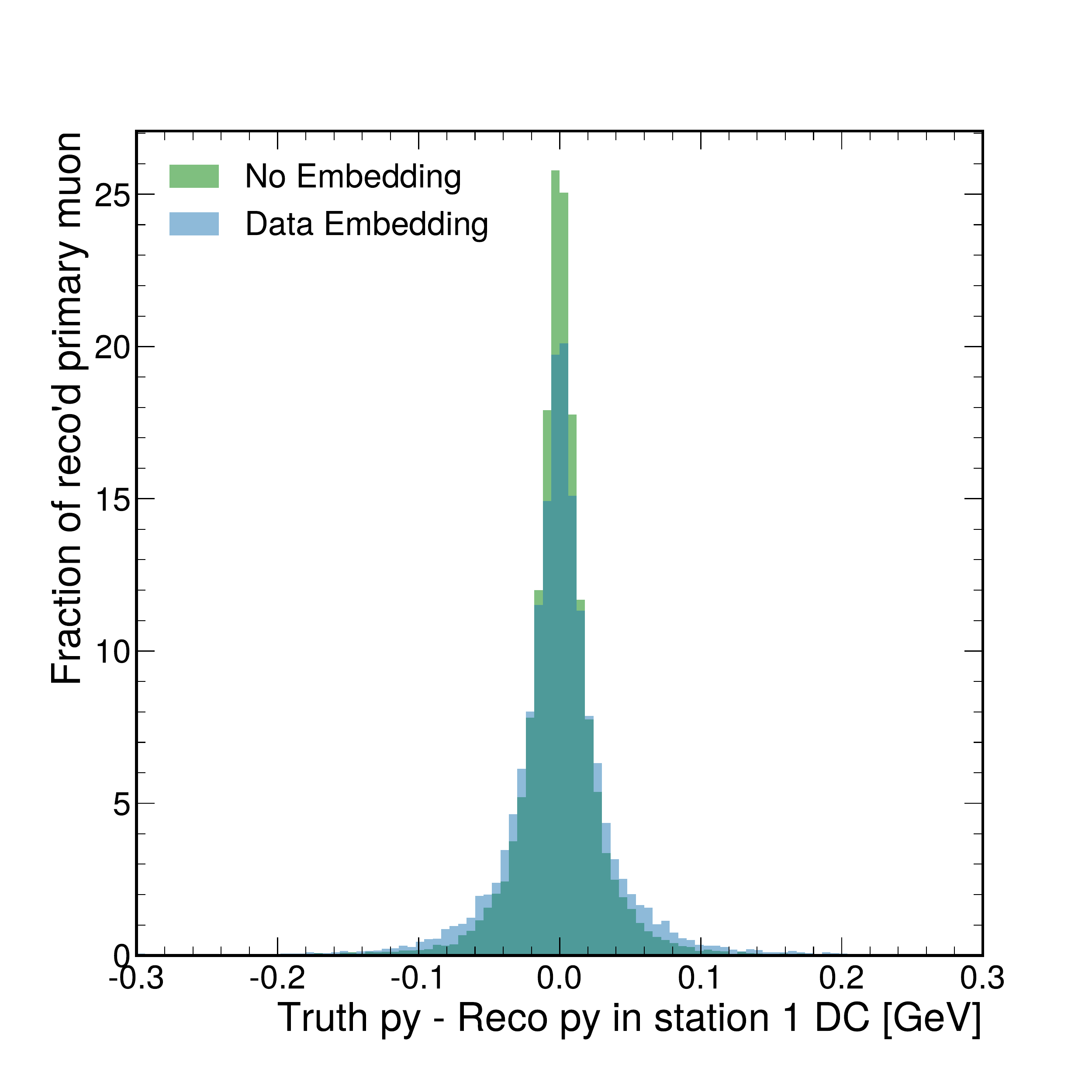}
    \includegraphics[width=0.32\textwidth]{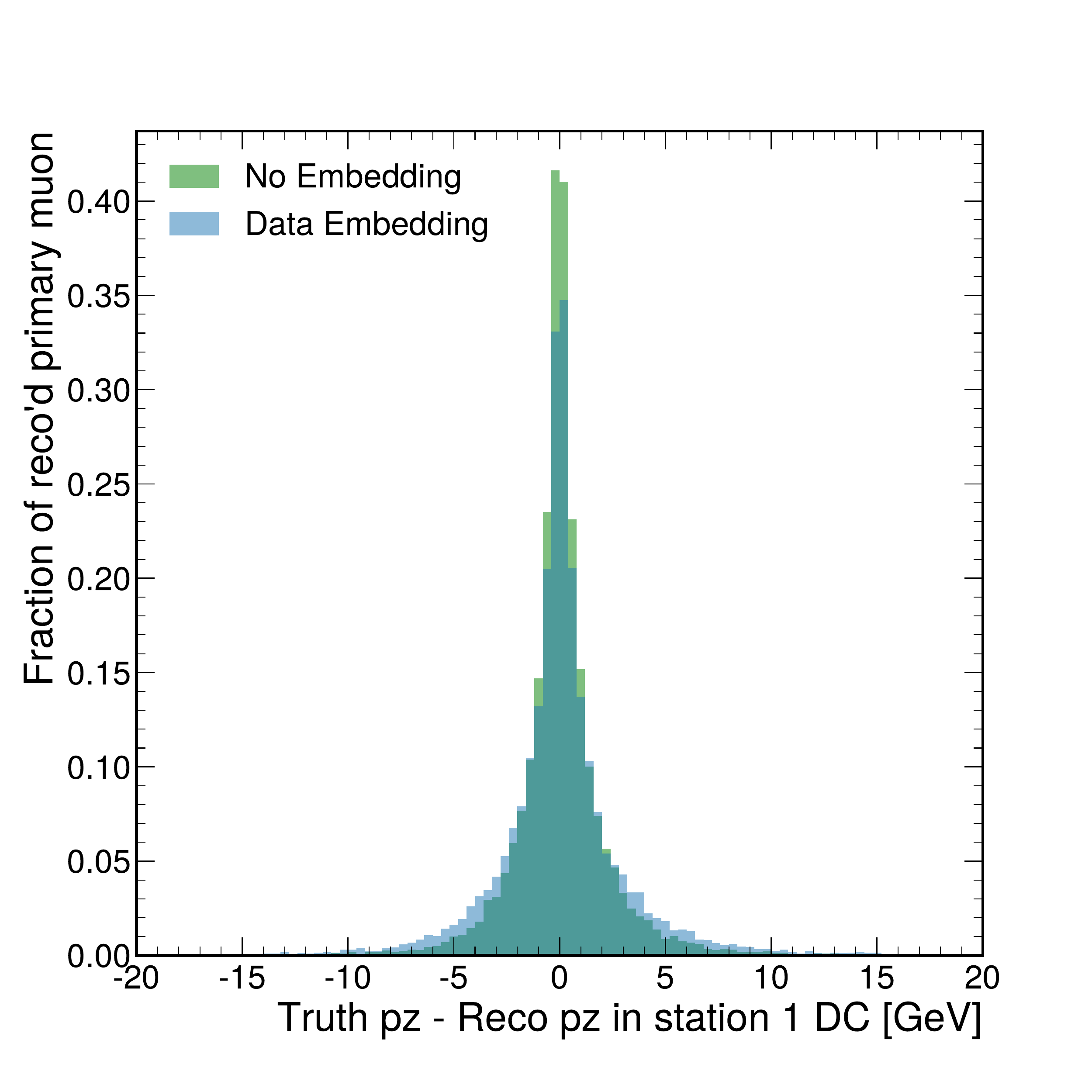}
    \caption{(Left) Resolution of $p_x$ reconstruction.  (Center) Resolution of $p_y$ reconstruction.  (Right) Resolution of $p_z$ reconstruction.  Resolution in all cases is shown both before and after data event embedding.}
    \label{fig:Res}
\end{figure}

Having investigated the \textit{likelihood} that the trajectory of a particle will be found, we can investigate the \textit{quality} of the reconstruction.  In particular, we can look at the $p_x$, $p_y$, and $p_z$ resolution.  These resolution distributions are shown in Fig.~\ref{fig:Res}.  The momentum measurement is generally good to within a few percent, with tails that extend to about 10\%.  Embedding data events has a slightly negative impact on the momentum resolution, although the effect is manageable.  

\subsubsection{Tracking algorithm fake rate}
A final natural area of investigation for tracking algorithms is the rate of creating combinations of hits that do not correspond to the true trajectory of a particle, or the ``fake rate''.  Because data does not have a truth record, it is easiest to do a true ``fake rate'' study in fully simulated events.  In particular we will look at the fake rate in the simulation-overlaid muon-gun events.

In simulation, each trajectory can be sorted into one of the following three categories:
\begin{enumerate}
    \item \textit{Primary track}: this is a reconstructed trajectory that is matched to the muon-gun primary particle.
    \item \textit{Background track}: this is a reconstructed trajectory that is matched to a generator-level particle in the overlaid background sample.
    \item \textit{Fake track}: this is a reconstructed trajectory that \textit{is not} matched to any generator-level particle.
\end{enumerate}
The matching mentioned above uses the same scheme discussed in Sec.~\ref{sec:trackEff}.

\begin{figure}[bhtp!]
    \centering
    \includegraphics[width=0.3\textwidth]{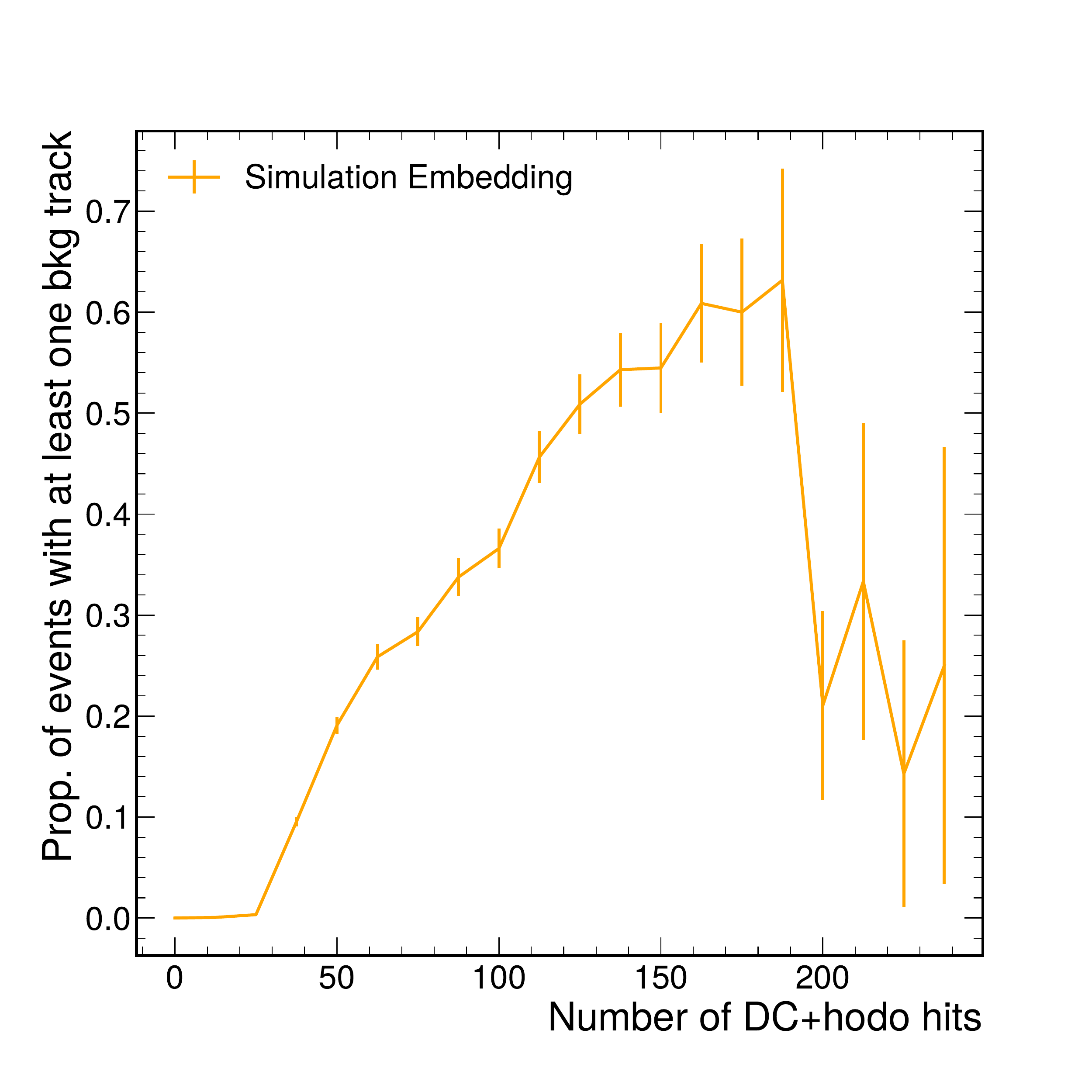}
    \includegraphics[width=0.3\textwidth]{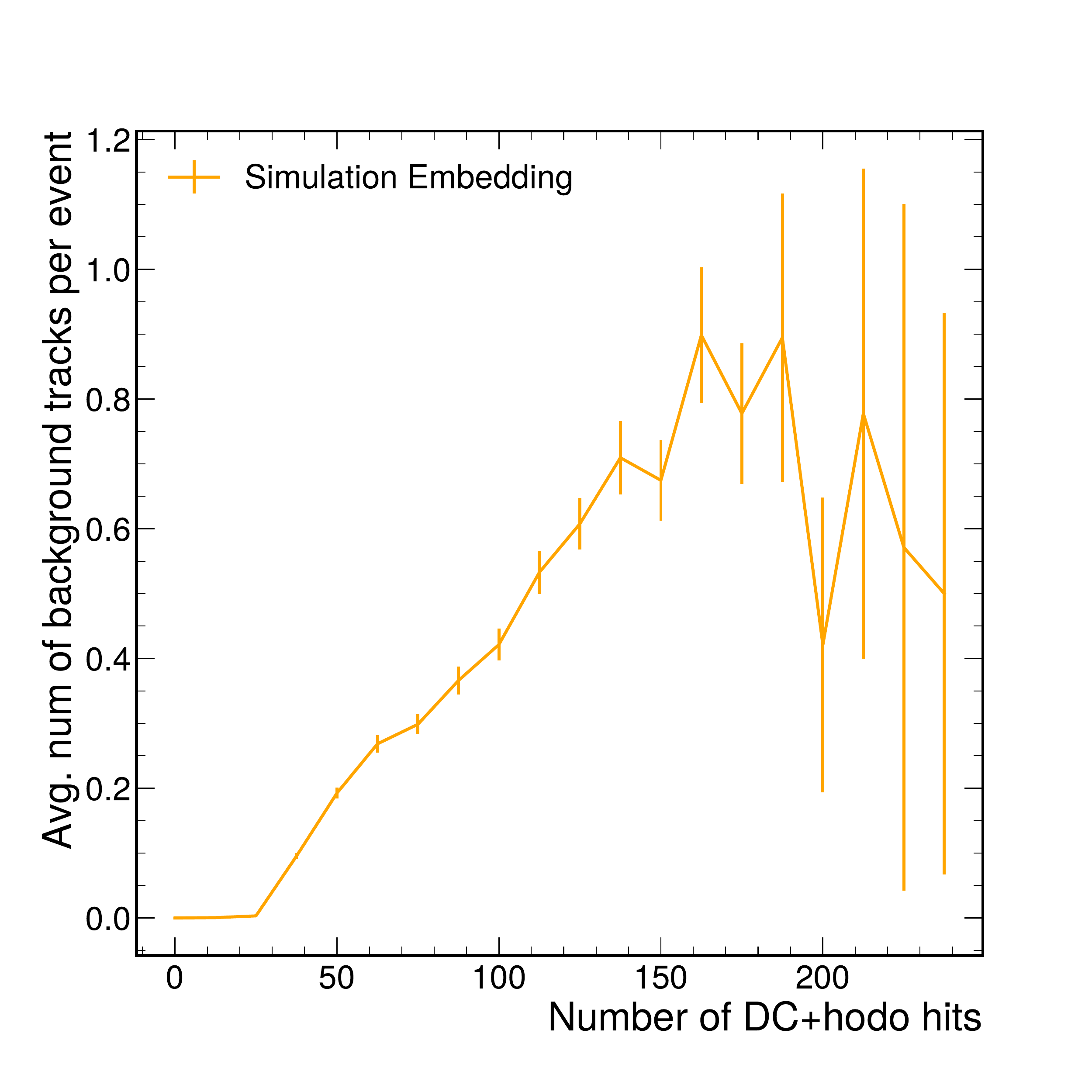}
    \caption{(Left) Proportion of events with at least one reconstructed background track as a function of the number of tracking-relevant hits.  (Right) Average number of reconstructed background tracks as a function of the number of tracking-relevant hits.  Events have overlaid background simulation.}
    \label{fig:simBkg}
\end{figure}

First, we can look at the background-track rate in simulation-overlaid events.  This is shown in Fig.~\ref{fig:simBkg}.  Keeping in mind that a particle will generate 18 hits if it passes by and leaves a hit in all six wires in each of the three drift chamber stations, it can be seen that almost 50\% of events with over 100 tracking-relevant hits have at least one background track.  At high numbers of hits, it seems that the algorithm's pileup mitigation techniques might restrict the reconstruction of background tracks, though the relatively small drop-off in the average \textit{number} of background tracks suggests that events that are fully processed often have more than one background track in them.

\begin{figure}[bhtp!]
    \centering
    \includegraphics[width=0.35\textwidth]{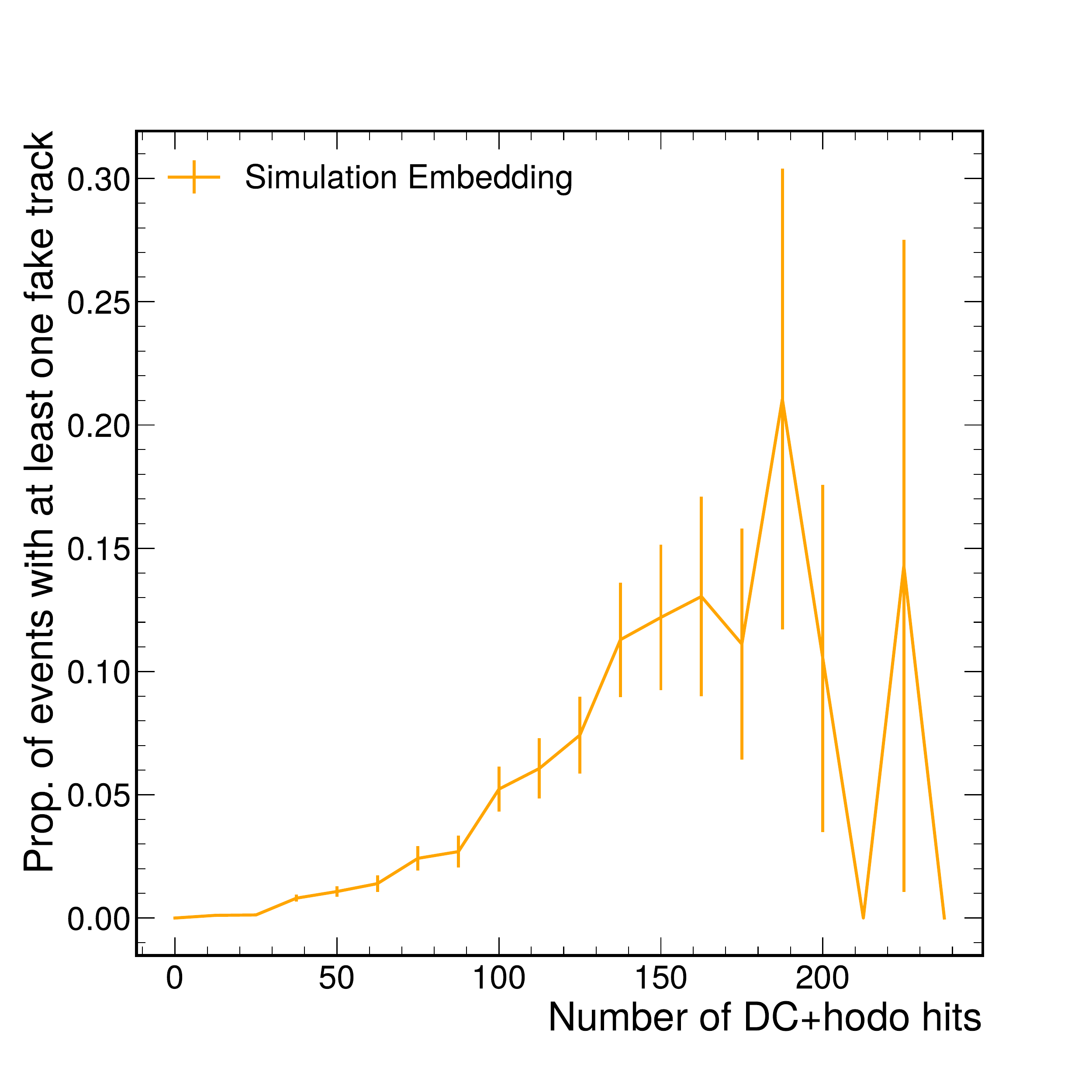}
    \includegraphics[width=0.35\textwidth]{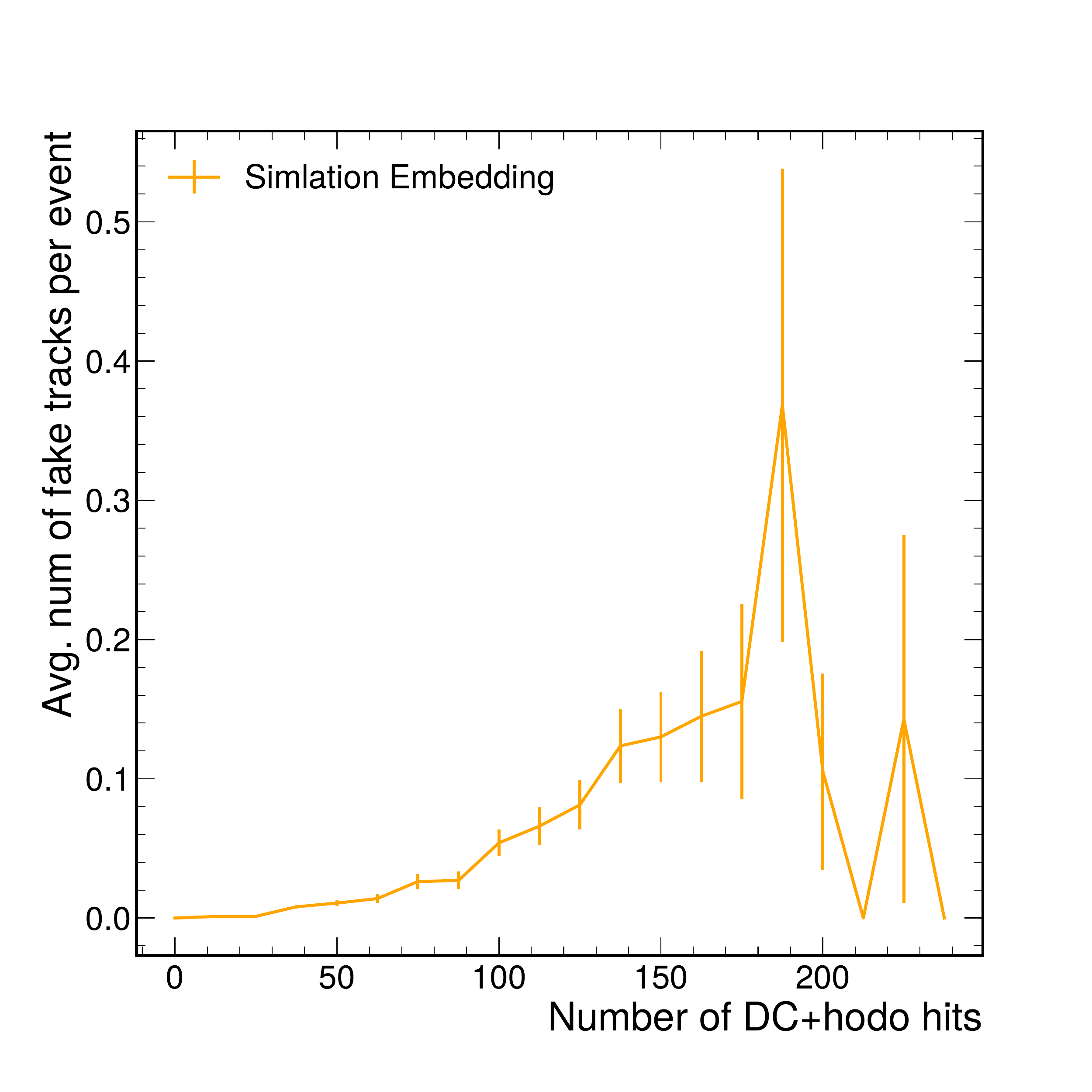}
    \caption{(Left) Proportion of events with at least one fake track as a function of the number of tracking-relevant hits.  (Right) Average number of fake tracks as a function of the number of tracking-relevant hits.  Events have overlaid background simulation.}
    \label{fig:simFake}
\end{figure}

Next, we can look at the true fake rate, which is presented in Fig.~\ref{fig:simFake}.  The fake rate increases with the number of tracking-relevant hits, which is not unexpected, and up to about 130 hits, fewer than 10\% of events have a fake track.  The average number of fake tracks is typically higher than the proportion with at least one due to the fact that some events have multiple fake tracks, but this does not seem like a very large effect.

Lastly, while we cannot perform a true fake track analysis in data, we find the fraction of events that have tracks that aren't matched to the primary particle.  In data, each trajectory can be sorted into one of the following two categories:
\begin{enumerate}
    \item \textit{Primary track}: this is a reconstructed trajectory that is matched to the muon-gun primary particle.
    \item \textit{Non-primary track}: this is a reconstructed trajectory that is \textit{not} matched to the muon-gun primary particle.  This is effectively a combination of the background and fake track categories of the simulation-overlaid study.
\end{enumerate}

\begin{figure}[bhtp!]
    \centering
    \includegraphics[width=0.35\textwidth]{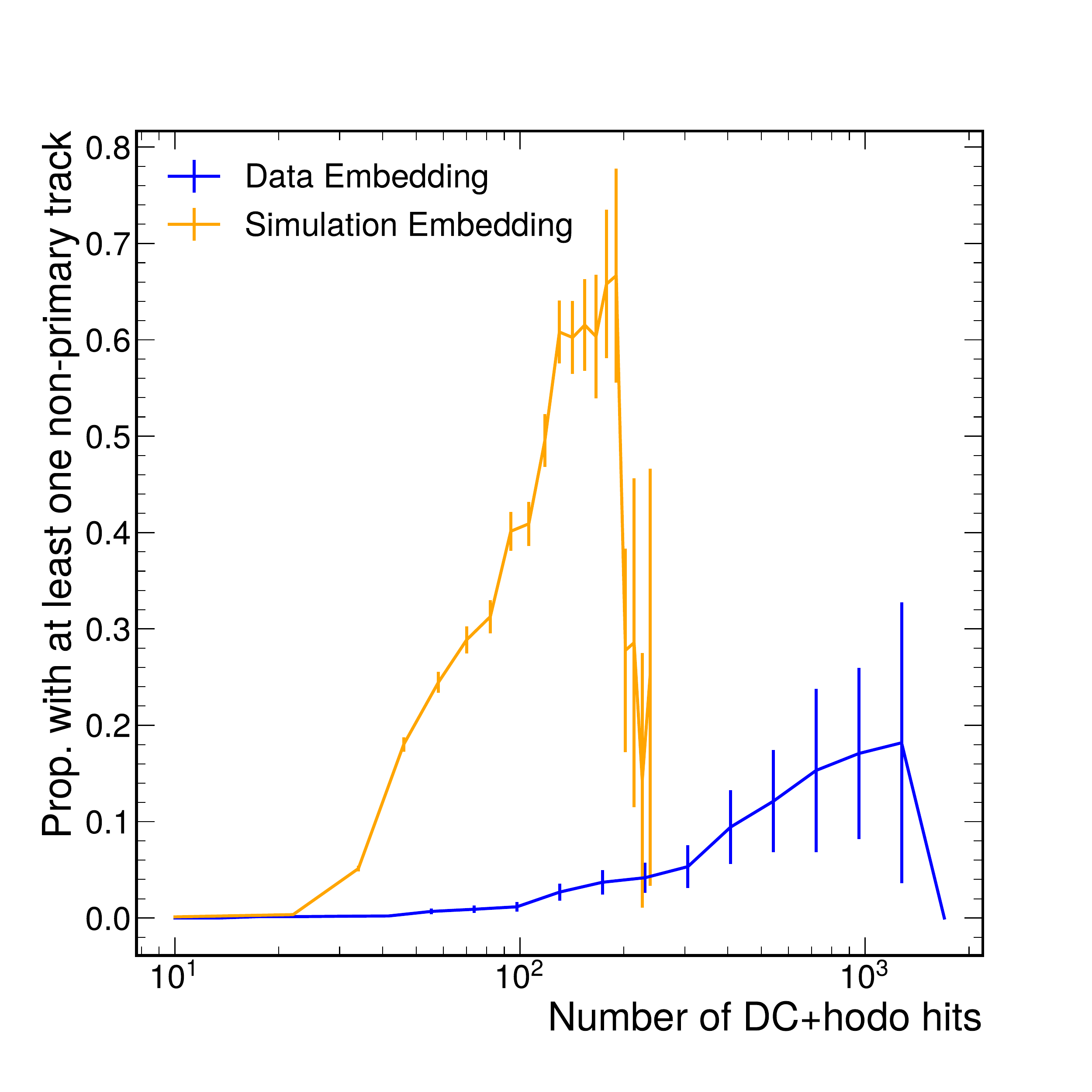}
    \includegraphics[width=0.35\textwidth]{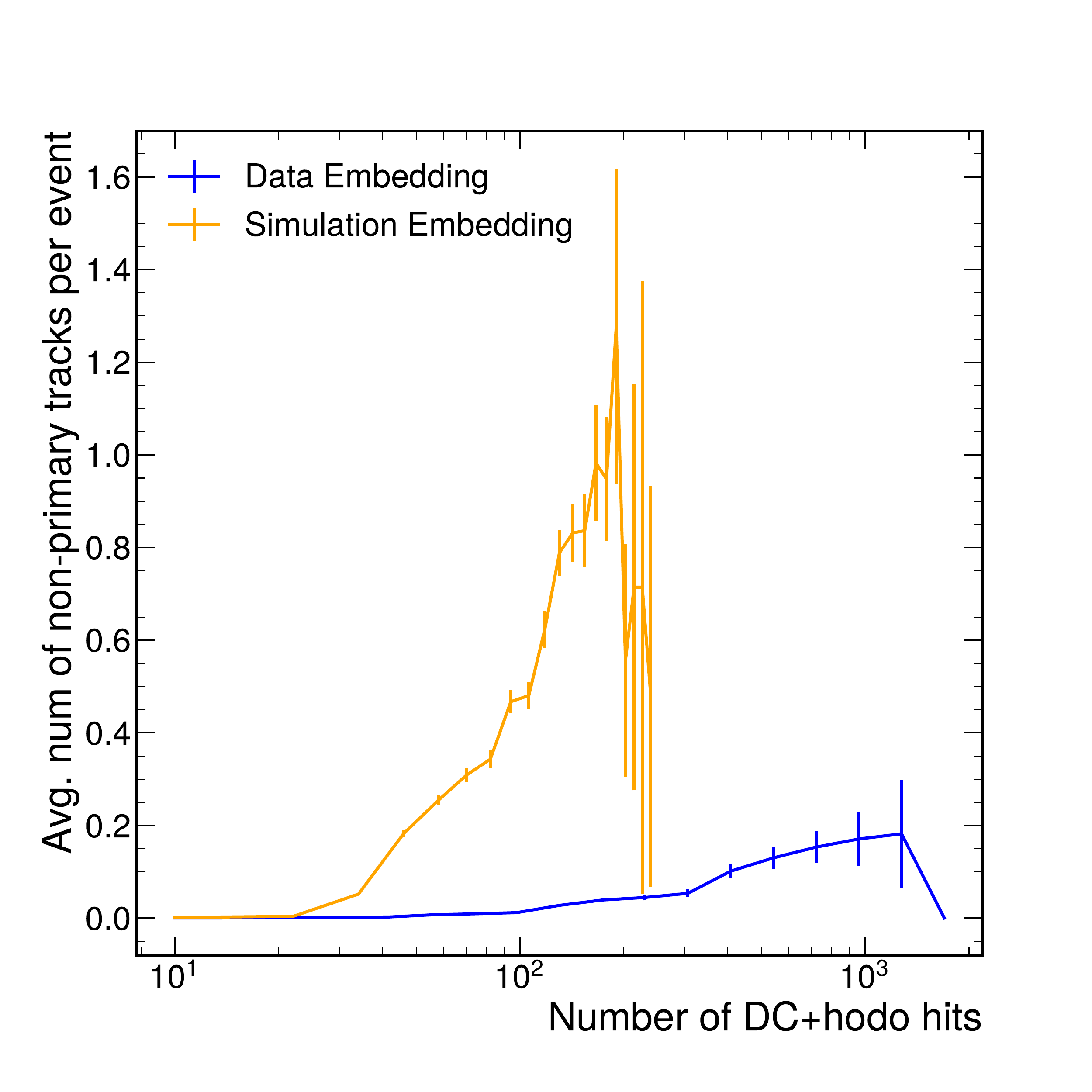}
    \caption{(Left) Proportion of events with at least one non-primary track as a function of the number of tracking-relevant hits.  (Right) Average number of non-primary tracks as a function of the number of tracking-relevant hits.  Events with data embedding are plotted in blue and events with overlaid background simulation are shown in orange.}
    \label{fig:dataNP}
\end{figure}

The event rate and average number of non-primary tracks as a function of the number of tracking-relevant hits is shown in Fig.~\ref{fig:dataNP}.  In data-overlaid events, the proportion of events with a non-primary track is significantly lower than the rate found in simulation-overlaid events.  For example, when there are about 100 hits, the rate of events with a non-primary track is about 1\% for data-overlaid events and about 45\% for simulation-overlaid events.  A potential difference between these two setups is the presence of drift chamber noise.  The data events have spurious hits in the drift chambers, while all hits in the simulation-overlaid events are created by a generation-level particle.  Additionally, the muons in the background only have a relatively hard $p_z$ spectrum after passing through the FMag, as shown in Fig.~\ref{fig:bkgPz}, leading to relatively easily reconstructable trajectories.  
Continuing to reduce the fake rate is another component of the displaced tracking algorithm that will be investigated in more detail in the near future, particularly using the vertexing information and track quality ($\chi^2$, for example) to reduce fake and background tracks.

\subsubsection{Vertex reconstruction}
The displaced vertex reconstruction is modified from the SpinQuest E-1039 vertex reconstruction code, which is based on the Kalman Filter. The detailed steps are explained below:

\begin{enumerate}
    \item \emph{Preparing all possible dimuon candidates}. \emph{Valid} reconstructed tracks with positive and negative charges are looped separately in this step, and all possible pairs of tracks are formed and saved in the dimuon candidate collection.
    \item \emph{Initial estimation of the vertex position}. After the track reconstruction, the point of the closest approach to the beam pipe (z-axis) is calculated, and this is referred to as the \emph{vertex} of one track. For one dimuon candidate, the average of the two track vertices is used as one initial estimation of the dimuon vertex position. In addition, each track is \emph{swum} from station 1 to upstream. In the gap between the downstream side of FMag and station 1, the swimming is done by projecting the track linearly, since the magnet effect in this region is small. The point with the smallest distance between the two tracks during this swimming process serves as the 2nd estimation of the dimuon vertex position. The swimming is then done in the region between the dump and FMag, taking into account the FMag effects. The point with smallest distance between the two tracks here serves as the 3rd estimation of the dimuon vertex position.
    \item \emph{Vertex fit}. For one initial estimation of one vertex pair candidate, the Kalman Filter is running with the two associated tracks. The resulting vertex with smallest $\chi^2$ among all the initial estimations will be the vertex of the corresponding dimuon pair. 
\end{enumerate}

\begin{figure}[htp]
    \centering
    \includegraphics[width=0.95\textwidth]{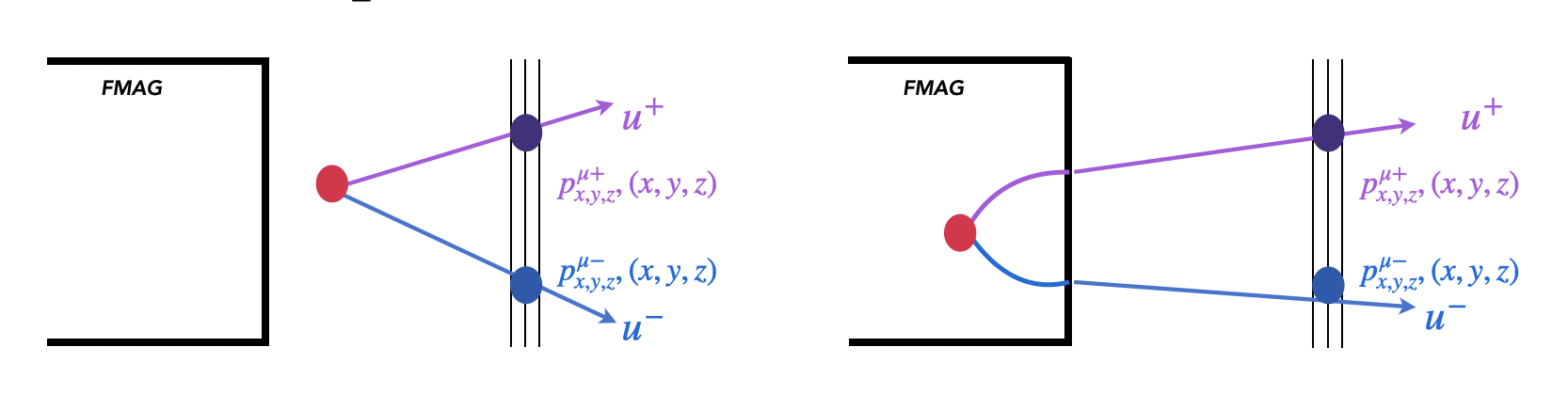}
    \caption{Illustration of the vertex reconstruction scenario of dimuon pairs produced after FMag (left) and inside FMag (right).}
    \label{fig:vertexrecodemo}
\end{figure}

Figure~\ref{fig:vertexrecodemo} shows the schematic of the two scenarios where the $A^{\prime}\to\ell^{+}\ell^{-}$ decay happens after the FMag (left) and inside the FMag (right). If $A^{\prime}$ decays after the FMag, the effects of FMag is small, the track upstream propagation is linear, and the track and vertex momentum resolutions are better; if $A^{\prime}$ decays before the FMag, the effects of FMag is large, the track upstream propagation needs to take the FMag effects into account, and therefore the track and vertex reconstructions are worse. In the scope of this paper, we are focusing on $A^{\prime}$ decays between 5-6m, so most of these events will decay after the FMag.

\begin{figure}[htp]
    \centering
    \includegraphics[width=0.394\textwidth]{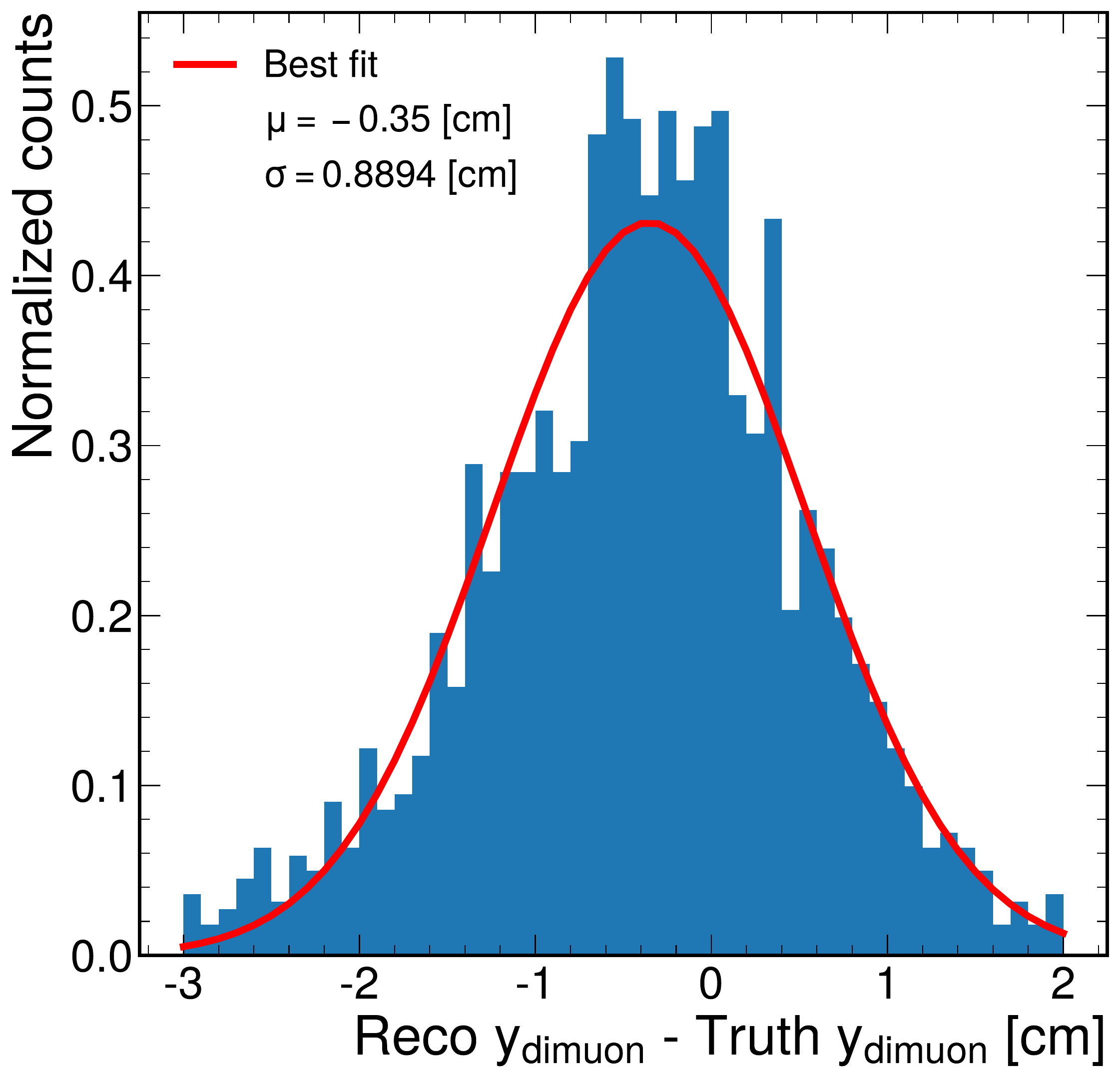}
    \includegraphics[width=0.40\textwidth]{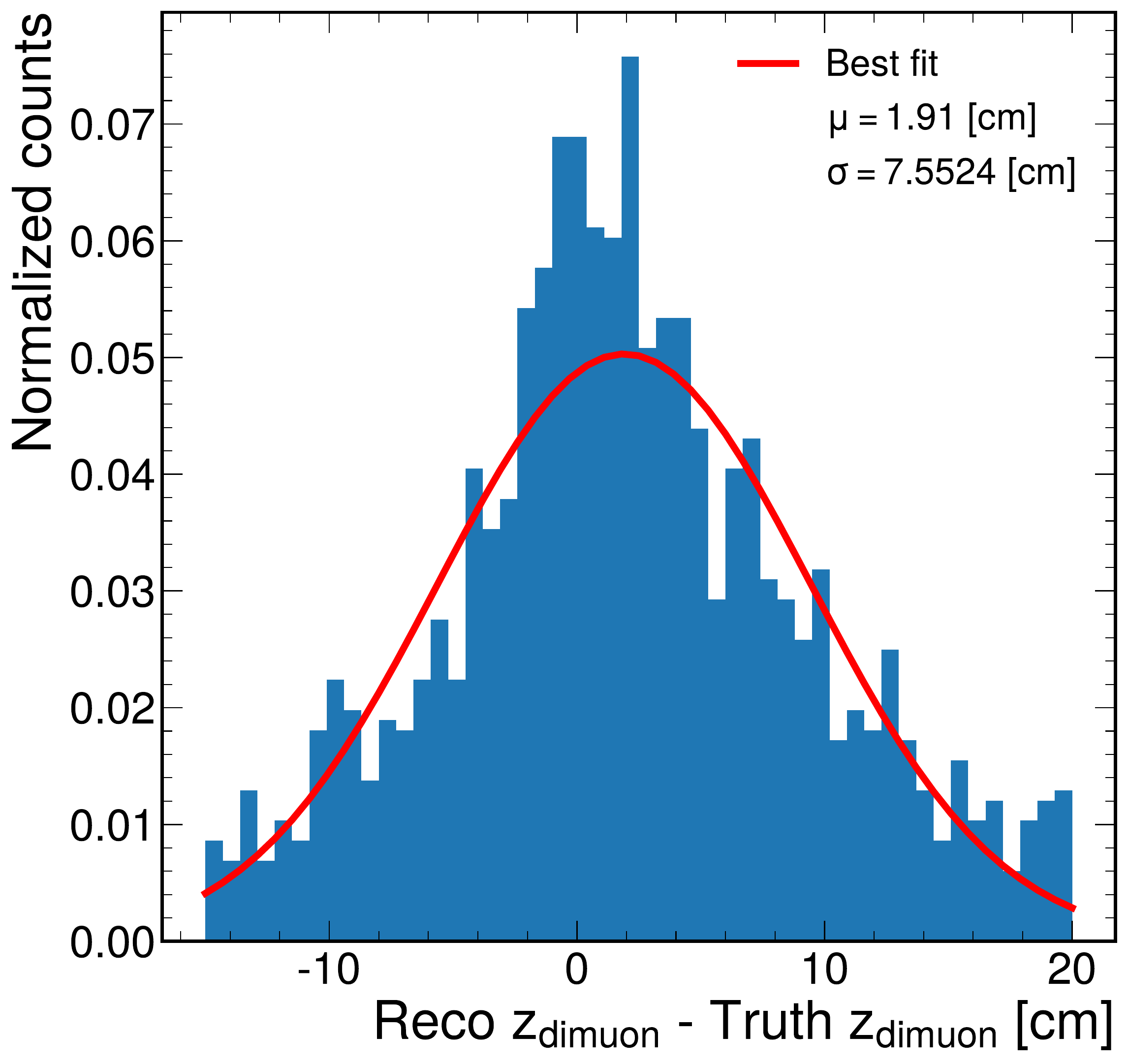}
    \caption{Vertex resolutions along $y$-axis (left) and $z$-axis (right) of $A^{\prime}\to\mu\mu$ decays between 5-6m. The signal sample in this plot is with $m(A^\prime)=0.95$\,GeV and the coupling $\epsilon=10^{-4.6}$.}
    \label{fig:vtxresol_afterFMag}
\end{figure}

Figure~\ref{fig:vtxresol_afterFMag} shows the vertex resolutions in $y$-axis and $z$-axis in the case where $A^{\prime}$ decays after the FMag. The signal sample in this plot is with $m(A^\prime)=0.95$\,GeV and the coupling $\epsilon=10^{-4.6}$. In such scenarios, the vertex resolution in $x$ and $y$ axis is around 0.9\,cm, and in $z$ axis the resolution is around 8\,cm. The vertex position resolutions of other samples with $A^{\prime}$ decaying in 5-6\,m are found out to be similar.

\begin{figure}[htp]
    \centering
    \includegraphics[width=0.40\textwidth]{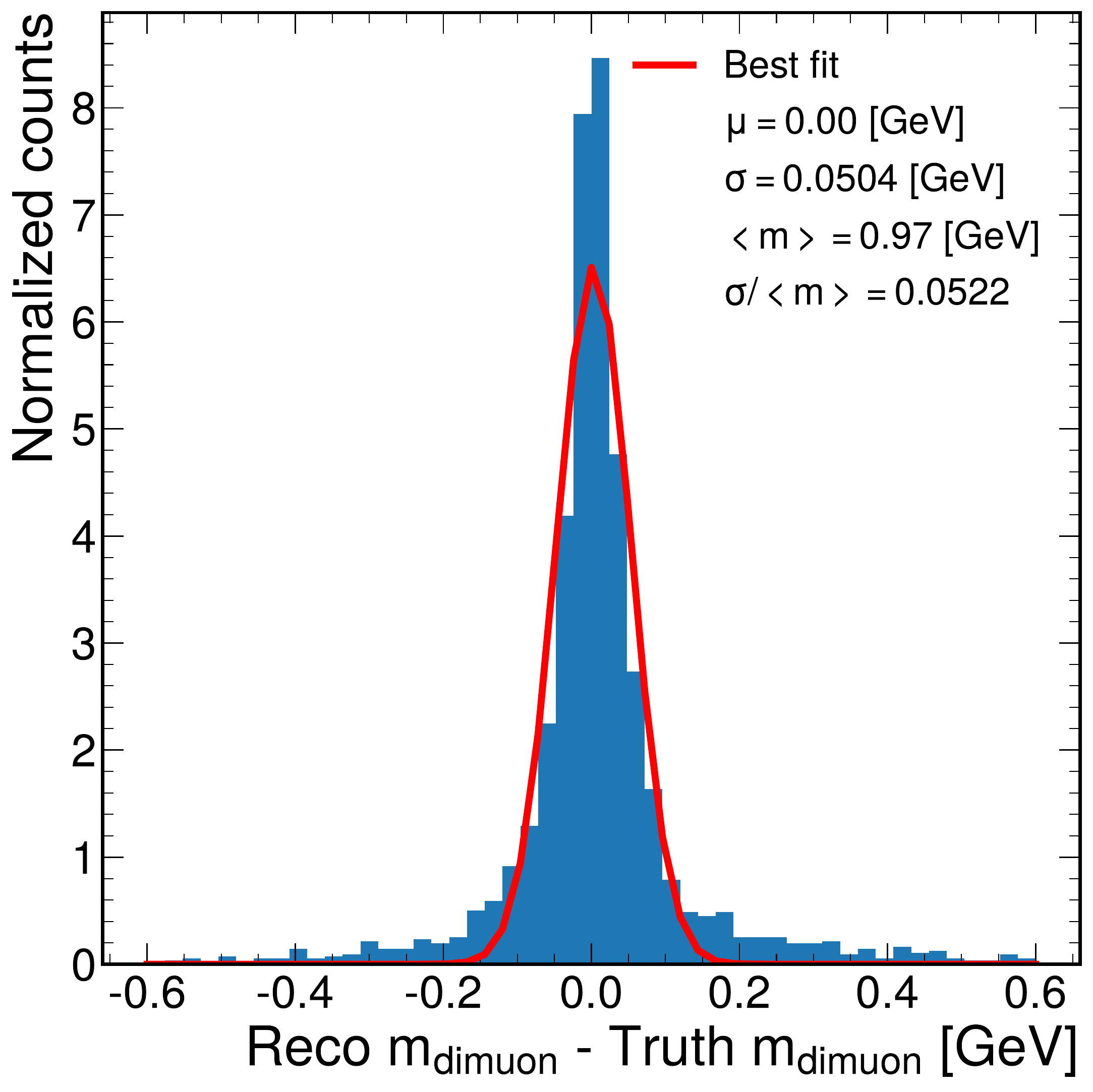}
    \caption{Vertex mass resolutions for $A^{\prime}\to\mu\mu$ decays between 5-6m. The signal sample in this plot is with $m(A^\prime)=0.95$\,GeV and the coupling $\epsilon=10^{-4.6}$.}
    \label{fig:vtxmassresol}
\end{figure}

Figure~\ref{fig:vtxmassresol} shows the $A^{\prime}$ mass resolution from the two reconstructed tracks, with $m(A^\prime)=0.95$\,GeV and the coupling $\epsilon=10^{-4.6}$. In this case, the mass resolution is around 0.05\,GeV, about 5\% of the $A^{\prime}$ mass.

\subsection{Particle ID and reconstruction}
In this section, we study higher particle-level identification including electromagnetic particle reconstruction in the EMCal and how we can potentially separate electrons, pions, and muons using EMCal information.  

\subsubsection{Electron/photon shower reconstruction and identification}
\label{sec:showers}

\begin{figure}[tbh!]
    \centering
    \includegraphics[width=0.45\textwidth]{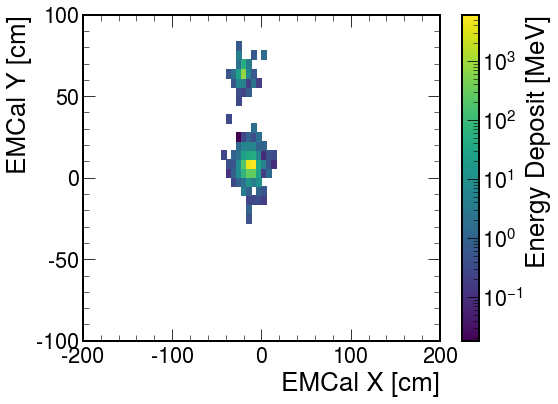}
    \hspace{0.5cm}
    \includegraphics[width=0.40\textwidth]{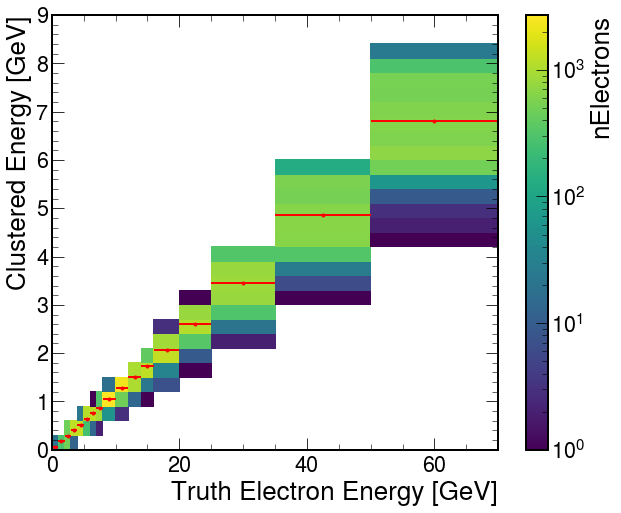}
    \caption{(Left) One event display example of the two EMCal shower clusters from the $A^{\prime}\to e^{+}e^{-}$ decay. (Right) EMCal clustered energy with respect to the truth energy of electrons from the $A^{\prime}$ decay. }
    \label{fig:emcluster}
\end{figure}

\begin{figure}[tbh!]
    \centering
    \includegraphics[width=0.45\textwidth]{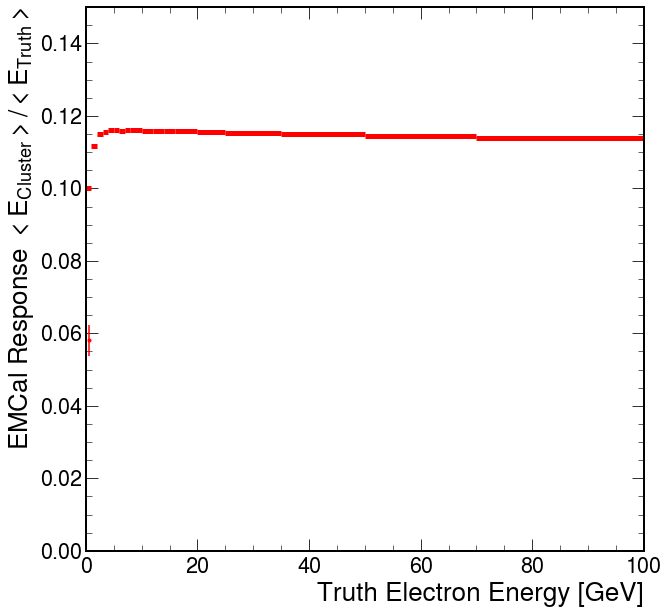}
    \includegraphics[width=0.45\textwidth]{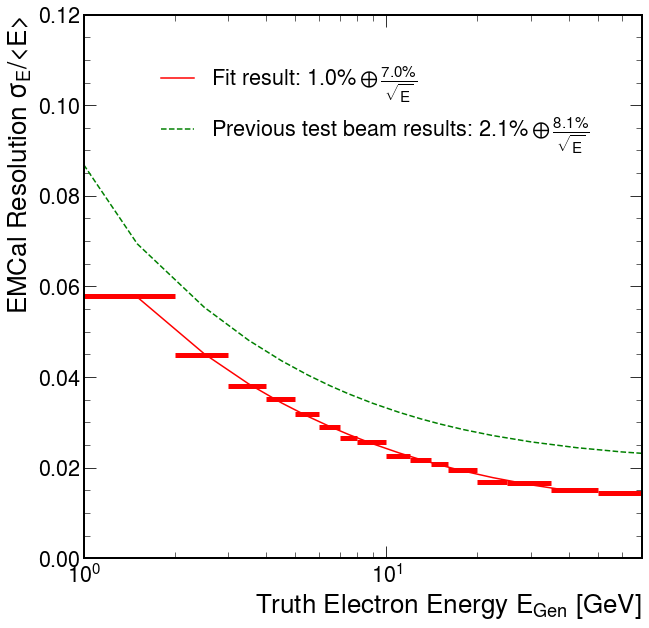}
    \caption{Simulated response (left) and resolutions (right) of the electron showers in the EMCal, as a function of the generator-level electron energy.}
    \label{fig:emperf}
\end{figure}

For the $A^{\prime}$ signals that decay into electron pairs, the two electrons will produce some track hits in station 1, 2, and 3, and level some energy deposits in the EMCal. Figure~\ref{fig:emcluster} (left) is an example of the EMCal showers from the $A^{\prime}\to e^{+}e^{-}$ decay, where two well-separated showers can be found.  Figure~\ref{fig:emcluster} (right) shows the distribution of the EMCal clustered energy with respect to the gen level electron energy. They align very well linearly. Figure~\ref{fig:emperf} shows the simulated response and resolutions of the electron showers in the EMCal. The response, defined as the ratio between $<E_\mathrm{cluster}>$ and $<E_\mathrm{truth}>$, is a flat distribution around 11\% with respect to the truth electron energy, similar to the expected sampling fraction of the EMCal (about 11\%). The simulated resolutions also agree well with the previous EMCal test beam results~\cite{PHENIX:2003fvo}.

%

\subsubsection{Particle identification scheme}


To study the overall detector response to a variety of particle types, simulated particle gun samples were produced with the same ``vertex'' and kinematic information discussed in Sec.~\ref{sec:trackEff}.
In particular, 100k events were generated for each of the following particle types: $e^-$, $\gamma$, $\mu^-$, $K^0_L$, and $\pi^+$.
To focus solely on the particle type under study, no background hit information was overlaid on these samples.

\begin{figure}[htp]
    \centering
    \includegraphics[width=0.4\textwidth]{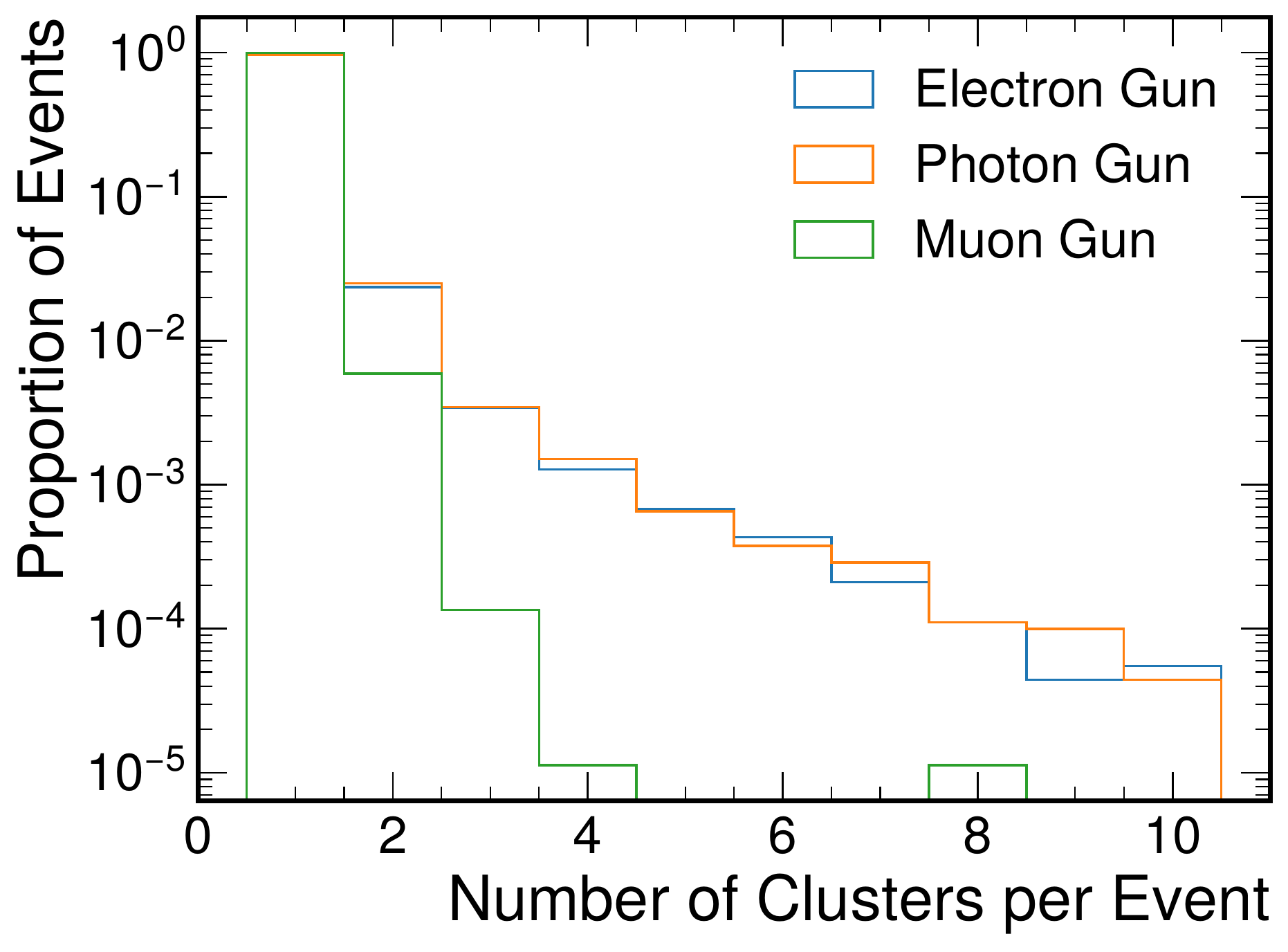}
    \includegraphics[width=0.4\textwidth]{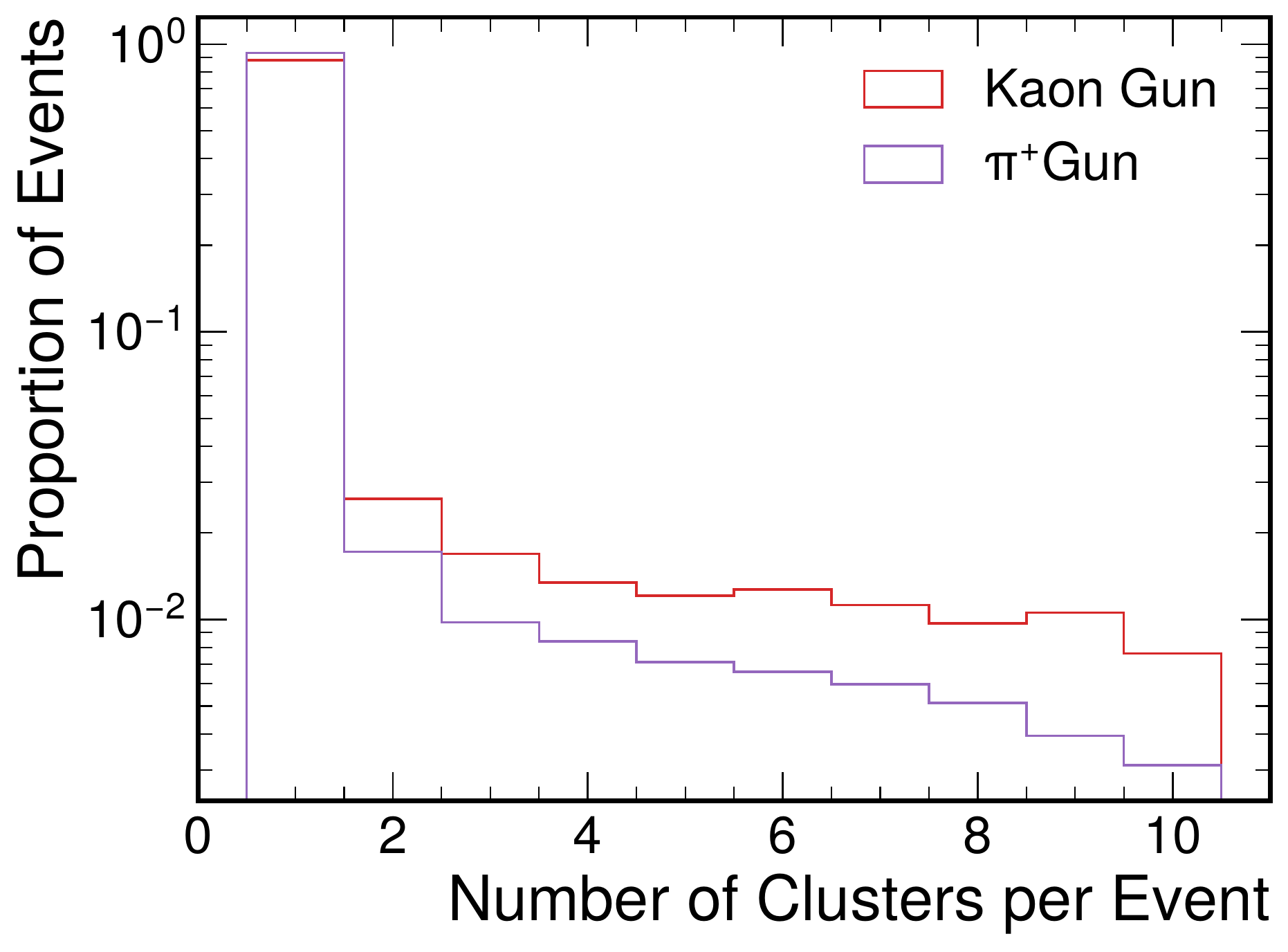}
    \caption{(Left) Distribution of the numbers of EMCal clusters for displaced $e^-$, $\gamma$, and $\mu^-$ particle guns. (Right) Distributions of number of clusters for $\pi^+$ and $K^0_L$ particle guns.}
    \label{fig:nClusEGM}
\end{figure}

First, we can examine the number of EMCal clusters created by the different particle types, the distributions of which are shown in Fig.~\ref{fig:nClusEGM}.
Clustering was performed on individual hits in the EMCal using the K-Means clustering algorithm implemented in the \texttt{sklearn.cluster} Python package~\cite{scikit-learn}.
The $e^-$ and $\gamma$ distributions are very similar, where multiple clusters can be created via bremsstrahlung or pair production, and $\mu^-$, as the most MIP-like particle tends to have the fewest clusters.
The $\pi^+$ and $K^0_L$ samples have the most pronounced tails, as decays into multiple particle types are expected.

\begin{figure}[htp]
    \centering
    \includegraphics[width=0.4\textwidth]{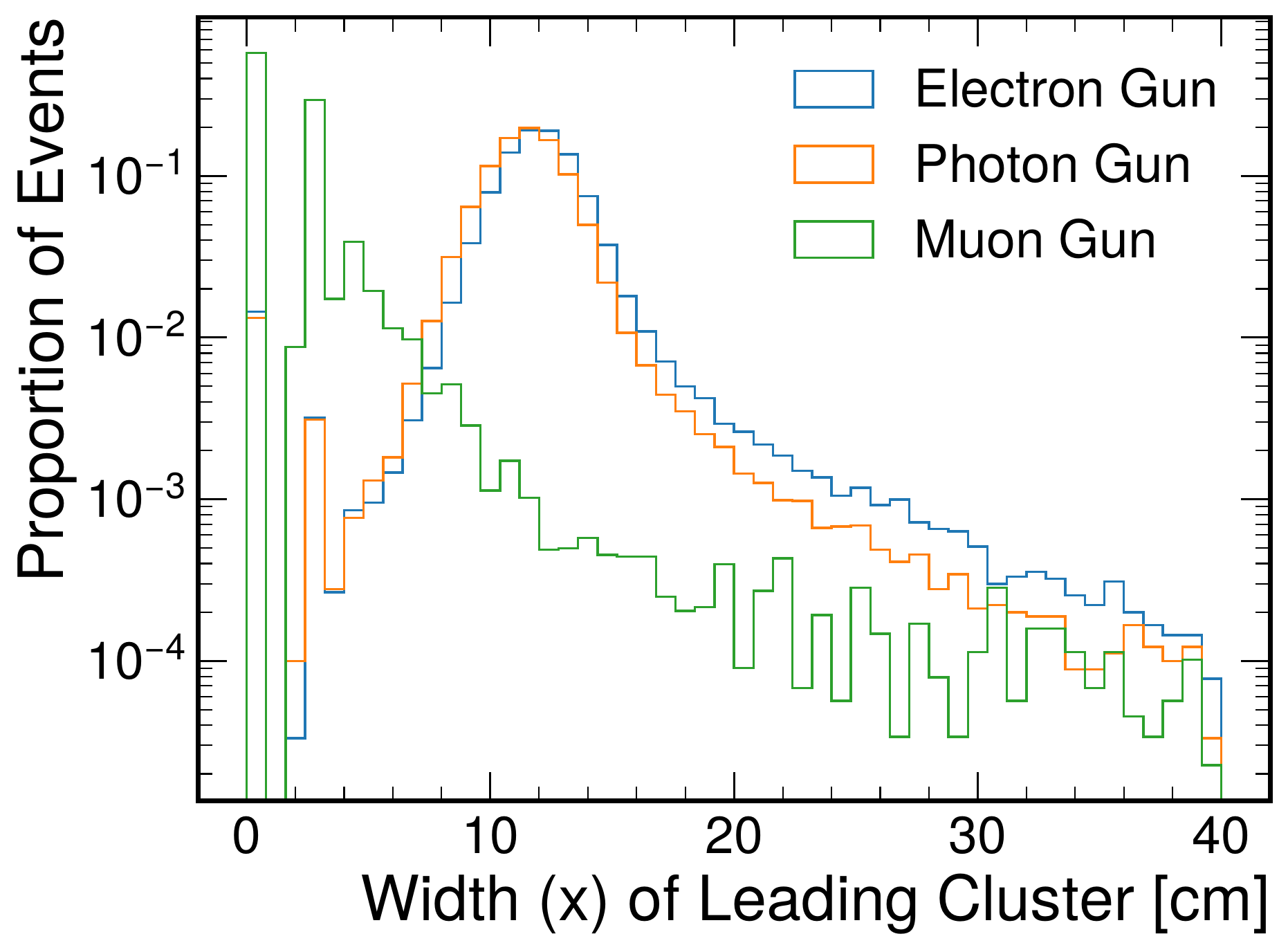}
    \includegraphics[width=0.4\textwidth]{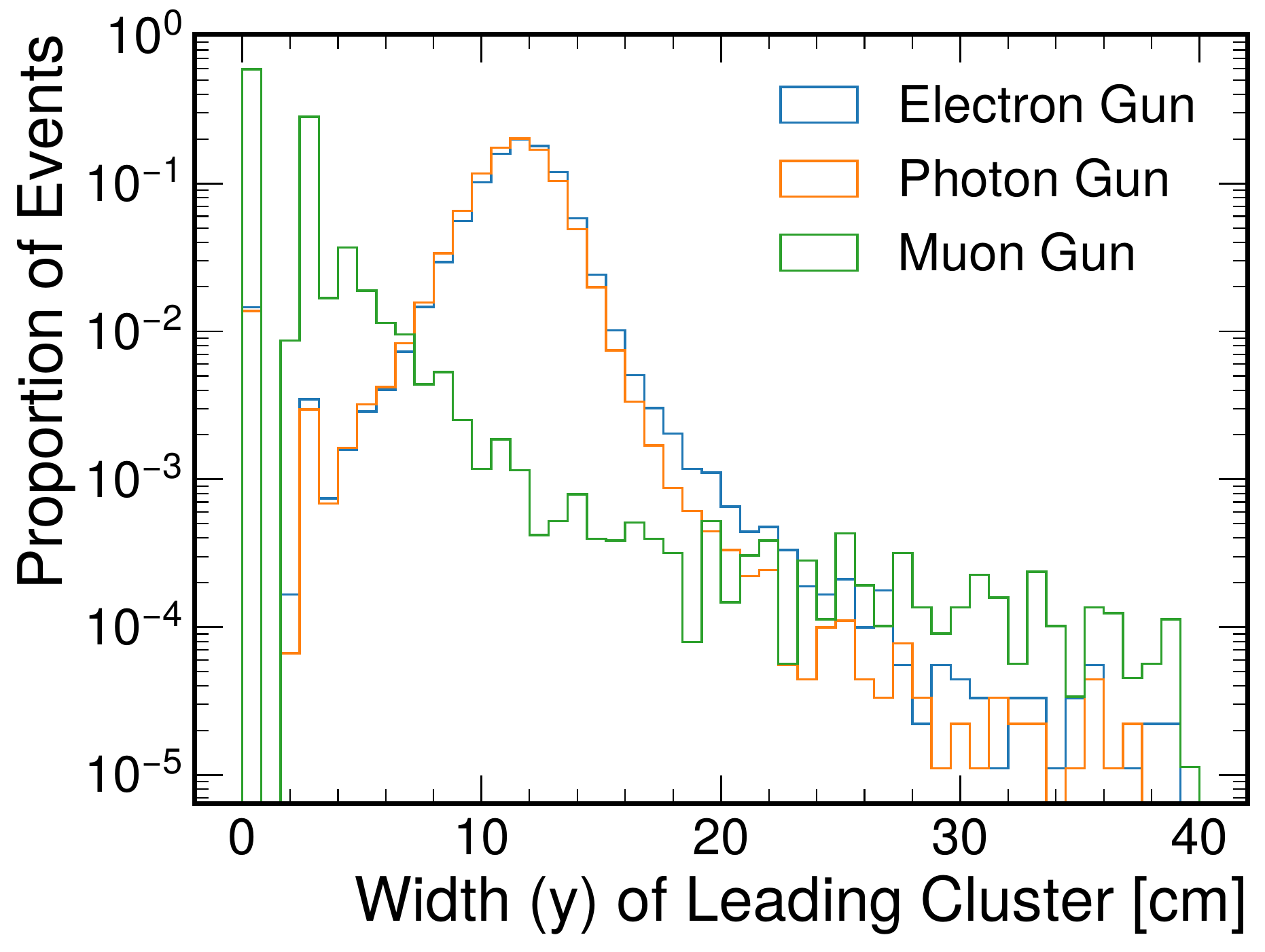}
    \includegraphics[width=0.4\textwidth]{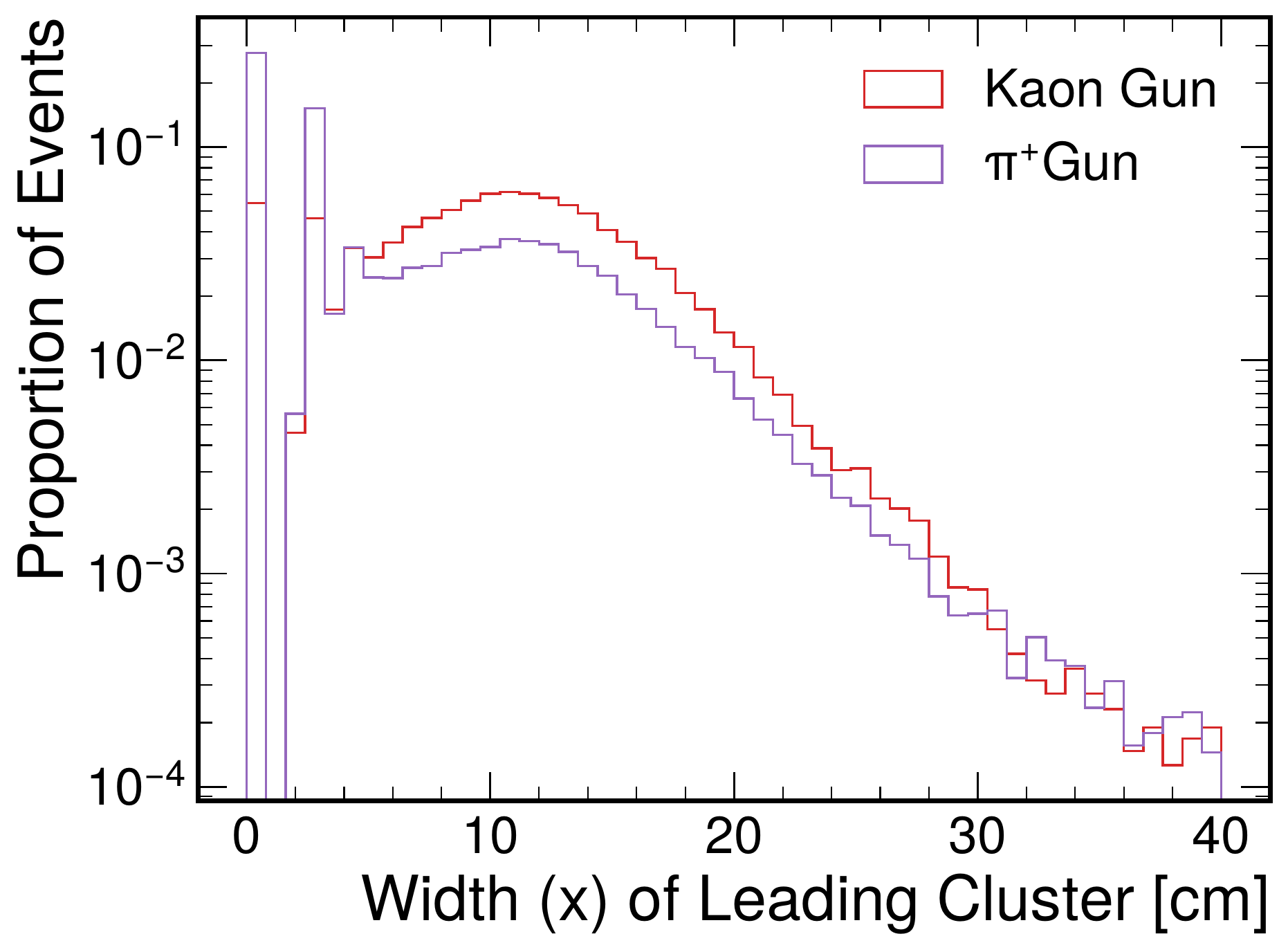}
    \includegraphics[width=0.4\textwidth]{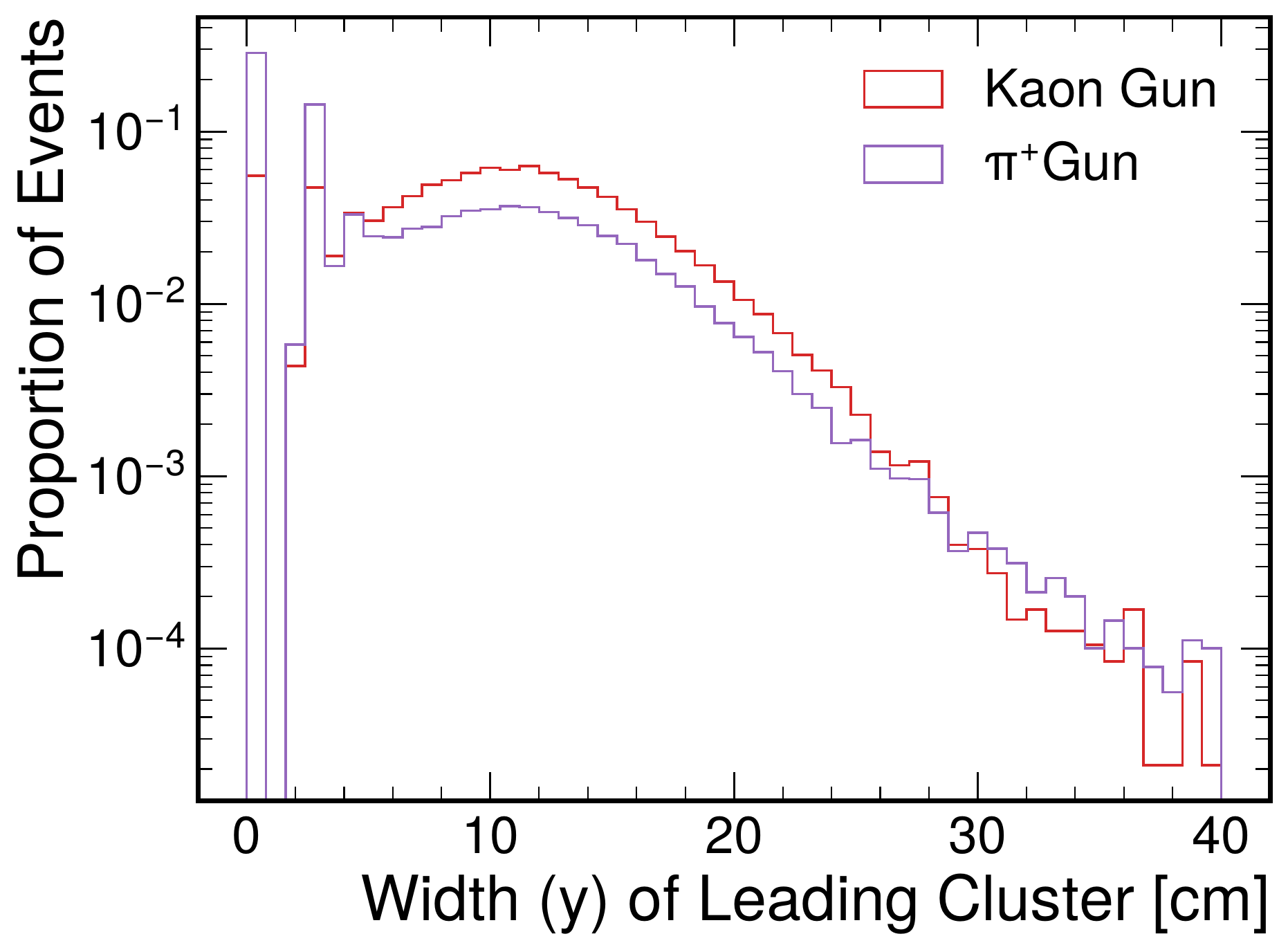}
    \caption{(Top) Distributions of width of EMCal energy deposits in $x$- (left) and $y$- (right) directions for $e^-$, $\gamma$, and $\mu^-$ guns. (Bottom) Distributions of width of EMCal energy deposits in $x$- (left) and $y$- (right) directions for $\pi^+$ and $K^0_L$ guns.  The KMag magnetic field is oriented in the $\hat{z}$ direction, resulting in slightly greater width in the $x$-direction.}
    \label{fig:widths}
\end{figure}

\begin{figure}[htp]
    \centering
    \includegraphics[width=0.4\textwidth]{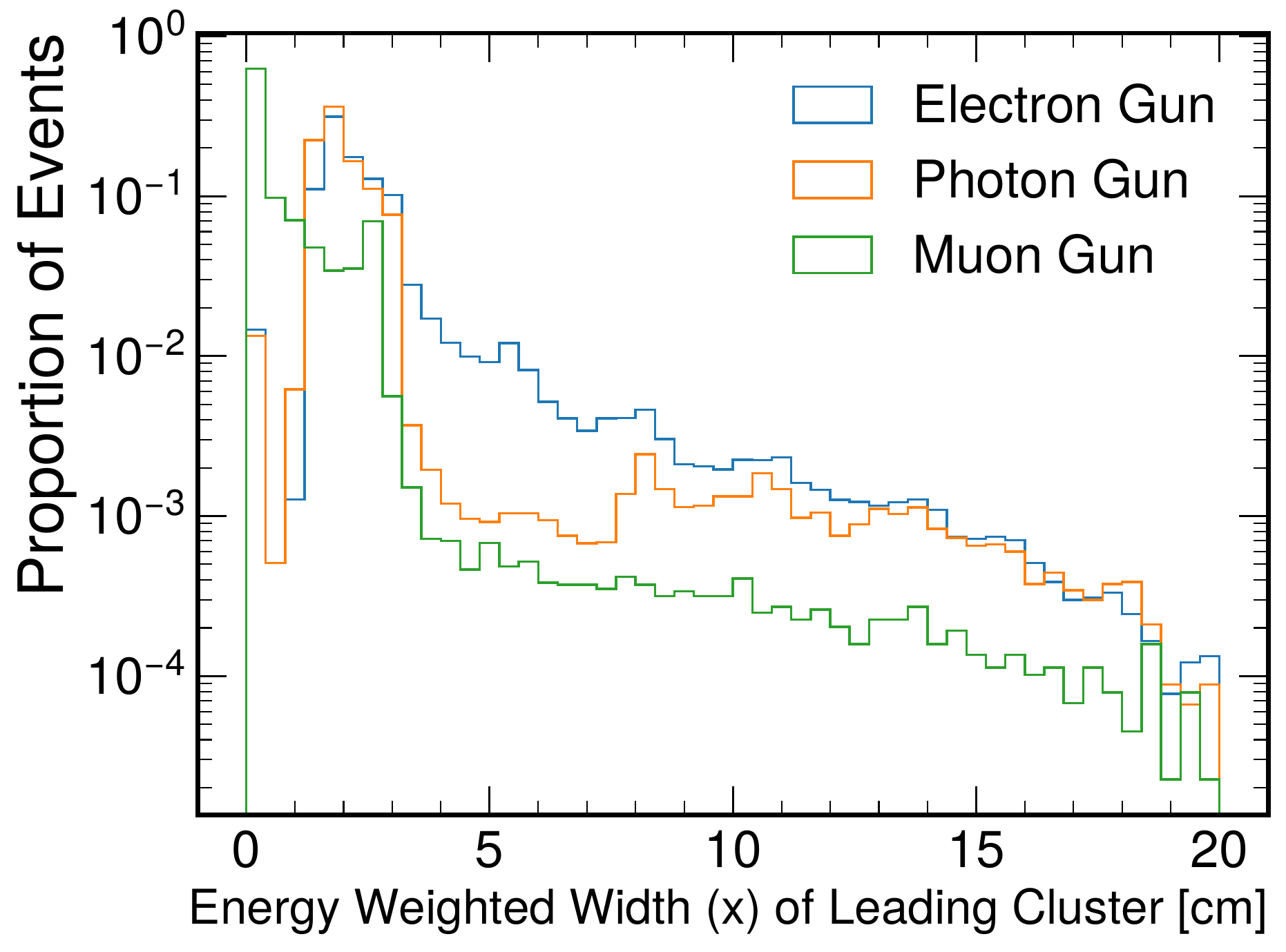}
    \includegraphics[width=0.4\textwidth]{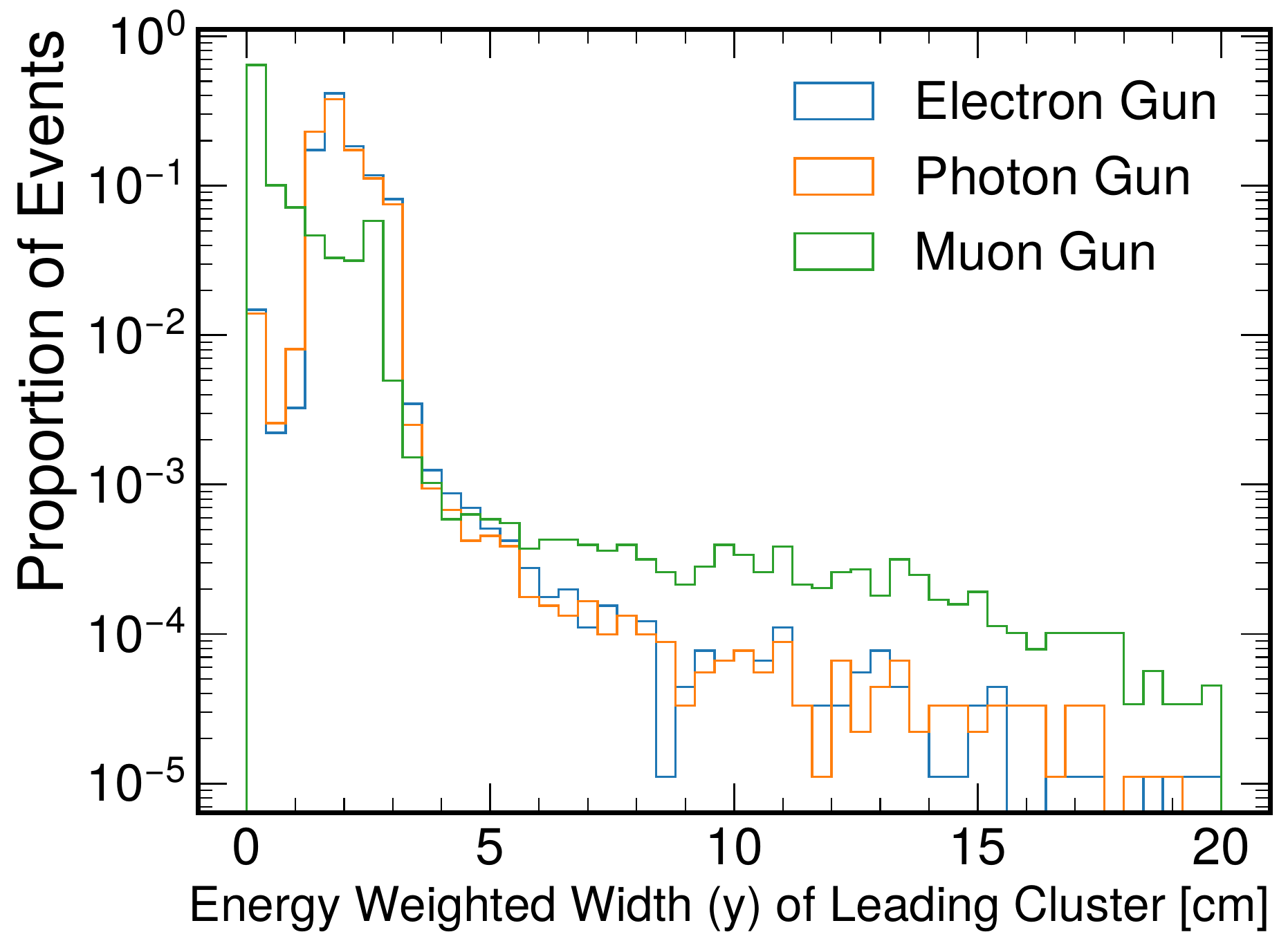}
    \includegraphics[width=0.4\textwidth]{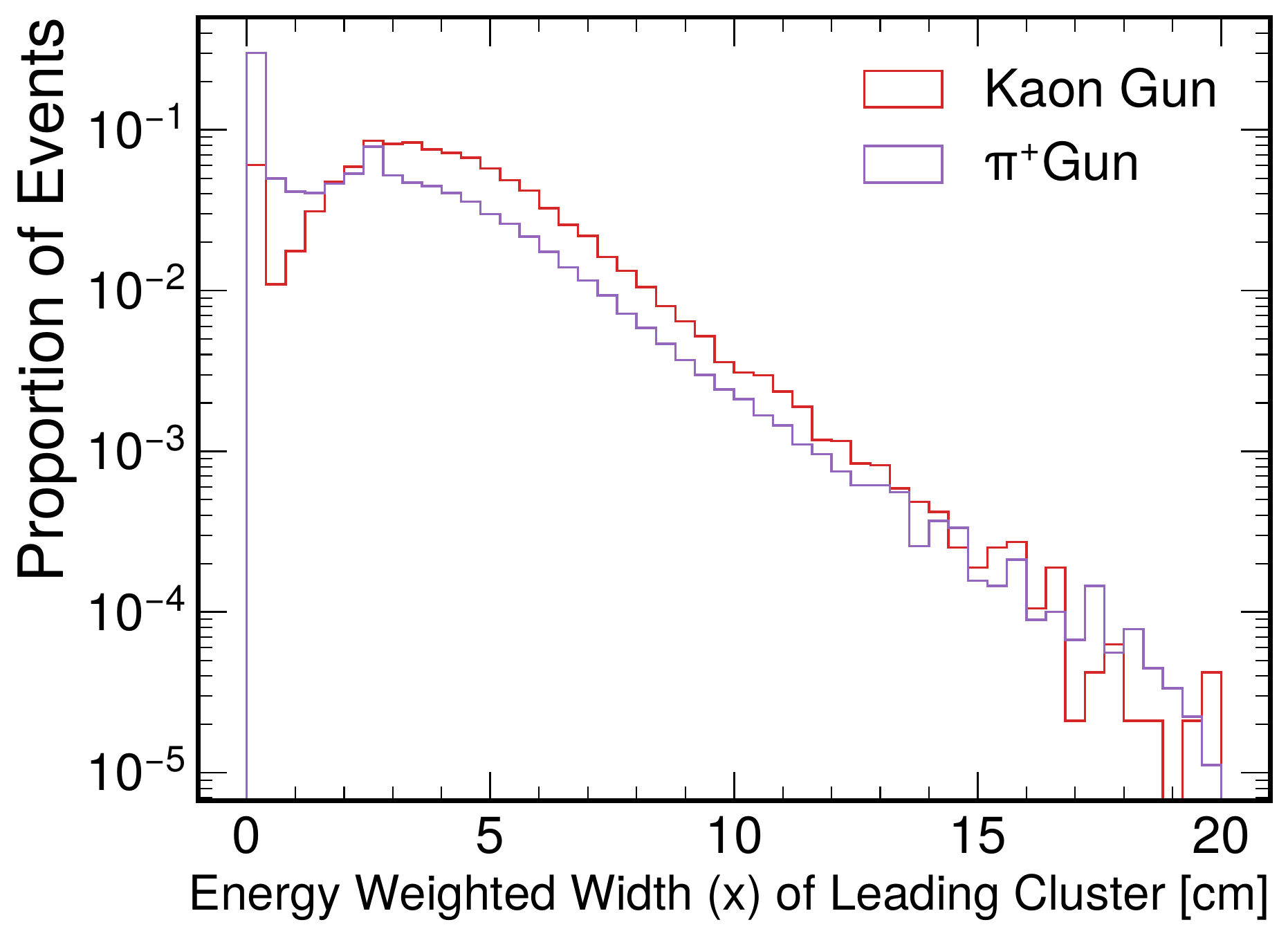}
    \includegraphics[width=0.4\textwidth]{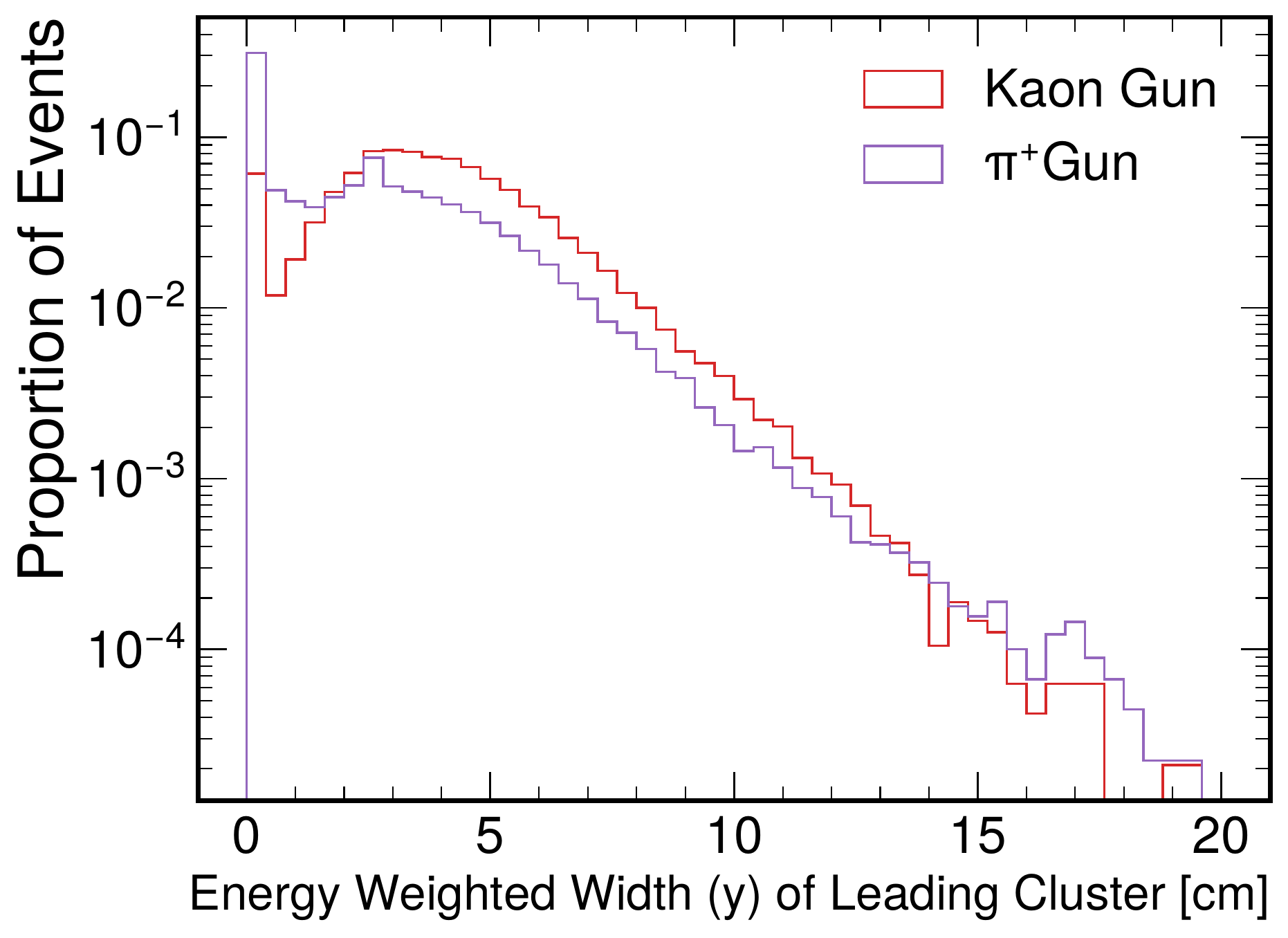}
    \caption{(Top) Distributions of energy-weighted width in $x$- (left) and $y$- (right) directions for $e^-$, $\gamma$, and $\mu^-$ guns. (Bottom) Distributions of energy-weighted width in $x$- (left) and $y$- (right) directions for $\pi^+$ and $K^0_L$ guns.  The KMag magnetic field is oriented in the $\hat{z}$ direction, resulting in a more pronounced tail for the $x$-direction distributions.}
    \label{fig:EWWs}
\end{figure}

We can also look at the widths and energy-weighted widths of the energy deposits due to the particles, which are shown in Figs.~\ref{fig:widths} and ~\ref{fig:EWWs}.
The width is defined as
\begin{equation}
    W = \left( \frac{1}{N} \sum_i (x_i - \Bar{x})^2\right)^{\frac{1}{2}},~\textrm{where}~ \Bar{x} = \frac{1}{N} \sum_i x_i,
\end{equation}
and the energy-weighted width is defined as
\begin{equation}
    W_{EW} = \left( \frac{1}{E_{tot}} \sum_i E_i (x_i - \Bar{x})^2\right)^{\frac{1}{2}},~\textrm{where}~ \Bar{x} = \frac{1}{E_{tot}} \sum_i E_i x_i.
\end{equation}
In both cases, the $i$ index runs over the individual energy deposits in the calorimeter, with $x_i$ being a deposit's position in the $x$-direction, and $E_i$ being its energy.
Widths can also be extracted in the $y$-direction by substituting $y$ in the place of $x$ in the above equations.

The width distributions are similar for $e^-$ and $\gamma$, but the energy-weighted width for $e^-$ has a longer tail in in the $x$ distribution, which is due to the $\hat{z}$ orientation of the KMag magnetic field.
The $\pi^+$ and $K^0_L$ distributions are both considerably wider than the $e^-$/$\gamma$ distributions, with proportionally larger spikes near 0cm when the particle traverses the EMCal without decaying.
The $\mu^-$ sample has the least width, one again because $\mu^-$ particles essentially MIP through the EMCal.

Lastly, we look at the total energy deposited in the EMCal as a function of the primary particle energy.
A calorimeter response plot for electrons was shown in Fig.~\ref{fig:emcluster}, which was used to derive the sampling fraction, as discussed in Sec.~\ref{sec:showers}.
The calorimeter response to $e^-$ and $\gamma$ particles is as expected, with linear energy scaling, as the EMCal is designed to fully capture the energy of these particles.

\begin{figure}[htp]
    \centering
    \includegraphics[width=0.4\textwidth]{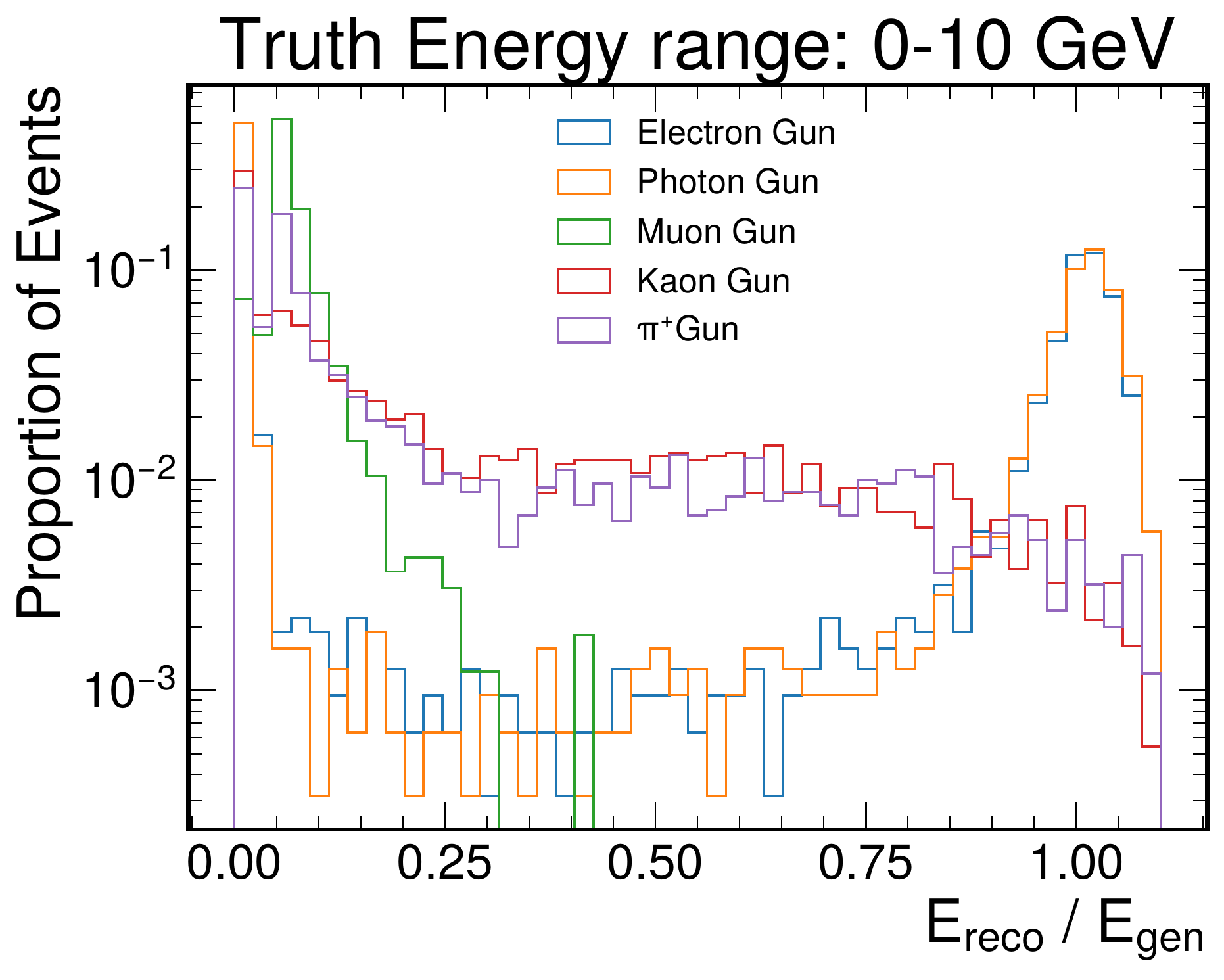}
    \includegraphics[width=0.4\textwidth]{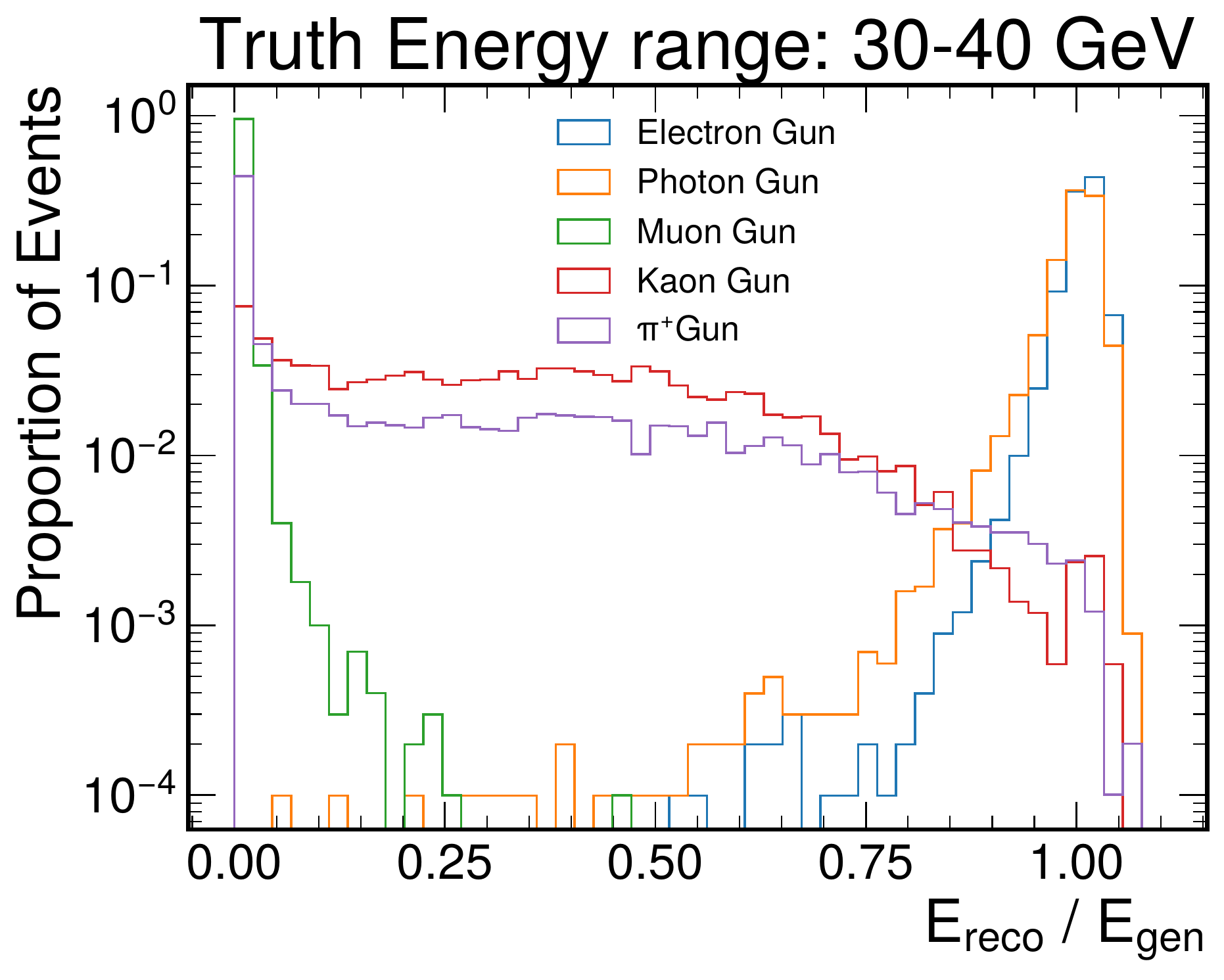}
    \caption{(Left) Distributions of the ratio of total energy collected in the EMCal after scaling ($\textrm{E}_{\textrm{reco}}$) to the energy of the generated particle ($\textrm{E}_{\textrm{gen}}$) for particles with $\textrm{E}_{\textrm{gen}}$ from 0 to 10 GeV.  (Right) The same distribution for particles with $\textrm{E}_{\textrm{gen}}$ from 30 to 40 GeV.}
    \label{fig:EcorrEgen}
\end{figure}

Distributions of the ratio of total energy collected in the EMCal after scaling ($\textrm{E}_{\textrm{reco}}$) to the energy of the generated particle ($\textrm{E}_{\textrm{gen}}$) are shown in Fig.~\ref{fig:EcorrEgen}.
The Figure shows two slices in $\textrm{E}_{\textrm{gen}}$: one from 0 to 10 GeV and one from 30 to 40 GeV.
These slices were chosen as relevant based on the distributions in Figs.~\ref{fig:sigkinematics}~and~\ref{fig:EEPZ}.  The other energy ranges generally have distributions similar to the 30-40 GeV distribution.
In both displayed $\textrm{E}_{\textrm{gen}}$ slices, the $e^-$ and $\gamma$ $\textrm{E}_{\textrm{reco}}$/$\textrm{E}_{\textrm{gen}}$ ratio peaks at 1, as expected.
The $\pi^0$'s distribution also peaks at 1 due to the $\pi^0 \rightarrow \gamma \gamma$, where all of the $\pi^0$'s energy is converted to photons, which are well-reconstructed by the EMCal.
However, there is significant downward smearing for the $\pi^0$ distribution, as decay photons may leave acceptance, and occasionally decay may not occur.
In general $\mu^-$'s leave the least energy in the detector, as they behave as a MIP through the calorimeter and rarely decay or radiate.
When $\pi^+$ or $K^0_L$ particles don't decay before the EMCal, they tend to leave little energy in the EMCal, as it is less than one nuclear interaction length in depth.
However, these particles \textit{can} decay into particles that leave a more significant fraction of energy in the calorimeter, leading to smearing of the distributions.
In the 30-40 GeV slice of $\textrm{E}_{\textrm{gen}}$, the range of $\textrm{E}_{\textrm{reco}}$/$\textrm{E}_{\textrm{gen}}$ between 0.05 and  0.85 is populated almost exclusively by $\pi^+$ and $K^0_L$ events.

For particle identification purposes, the following tentative scheme has been developed:
\begin{itemize}
    \item \textit{Electrons}: Electrons are tagged when a track can be extended into an energetic cluster in the EMCal.  It is also possible to create linear trajectories using the station 2 and 3 drift chambers alone to extend the electron tagging range.  While this does not allow for a momentum measurement using the drift chambers, the calorimeter can be used to extract energy.
    \item \textit{Photons}: Photons can be tagged by finding an energetic EMCal cluster with no track that can be extended into the cluster.
    \item \textit{Muons}: Muons can be tagged by finding a drift chamber track that is extended through a MIP cluster in the EMCal.  It should also be extended through hits in the hodoscopes and proportional tubes behind the absorber behind the EMCal.
    \item \textit{Hadrons}: Hadrons can be tagged through a variety of means.  When a neutral hadron does not decay before the EMCal, it is mostly undetectable.  When a charged hadron does not decay before the EMCal, it looks similar to a muon, with a track extended into a MIP-style cluster in the calorimeter, but there should be little activity behind the absorber.  Hadrons that decay in flight often result in large numbers of clusters, which can be separated from electron, photon, and muon by looking at the number of and spatial distribution of clusters.
\end{itemize}

\begin{figure}[htpb!]
    \centering
    \includegraphics[width=0.5\textwidth]{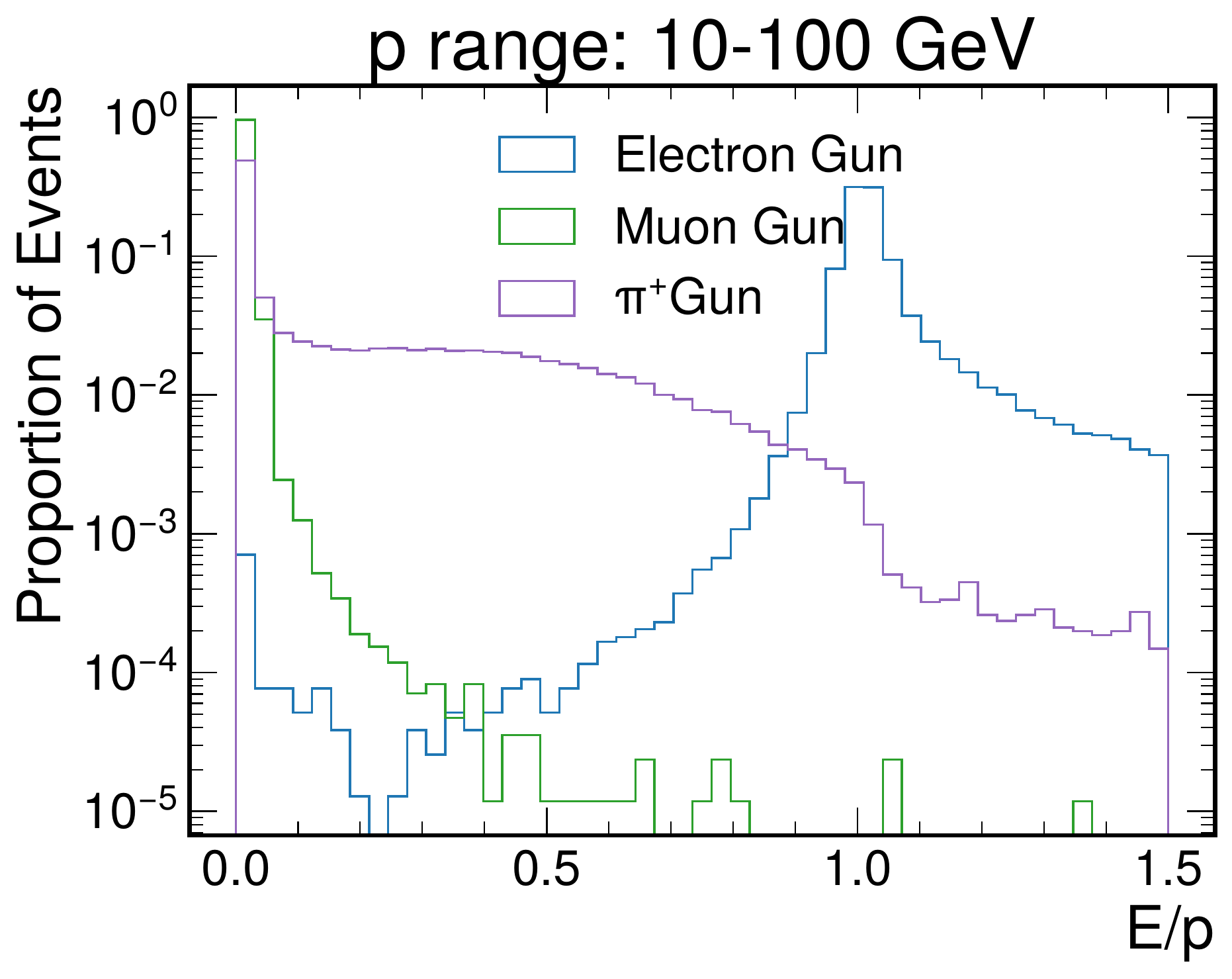}
    \caption{Distributions of the ratio of total energy collected in the EMCal after scaling (E) to the momentum of the particle (p) as measured by drift-chamber tracking for $e^{-}$, $\mu^{-}$, and $\pi^{+}$ particles.  The particles included in this plot were required to have a reconstructed track.  They also had to have a truth-momentum of at least 10 GeV, to ensure high tracking efficiency.}
    \label{fig:EPall}
\end{figure}

The determination of MIP-like cluster can be performed by taking the ratio of the energy deposit in the calorimeter cluster to the momentum of the particle measured by the drift-chamber tracking.
The main particles expected to create both a cluster and measureable track are $e^{\pm}$, $\mu^{\pm}$, and $\pi^{\pm}$.
Distributions of the EMCal energy to momentum ratio for these particles are shown in Fig.~\ref{fig:EPall}.
It is clear from the figure that the E/p ratio alone provides strong discrimination power between $e^{\pm}$ and $\pi^{\pm}$ particles.

\subsection{Signal acceptance to dark sector signatures}
\label{sec:sensitivity}

\begin{figure}[htp]
    \centering
    \includegraphics[width=0.45\textwidth]{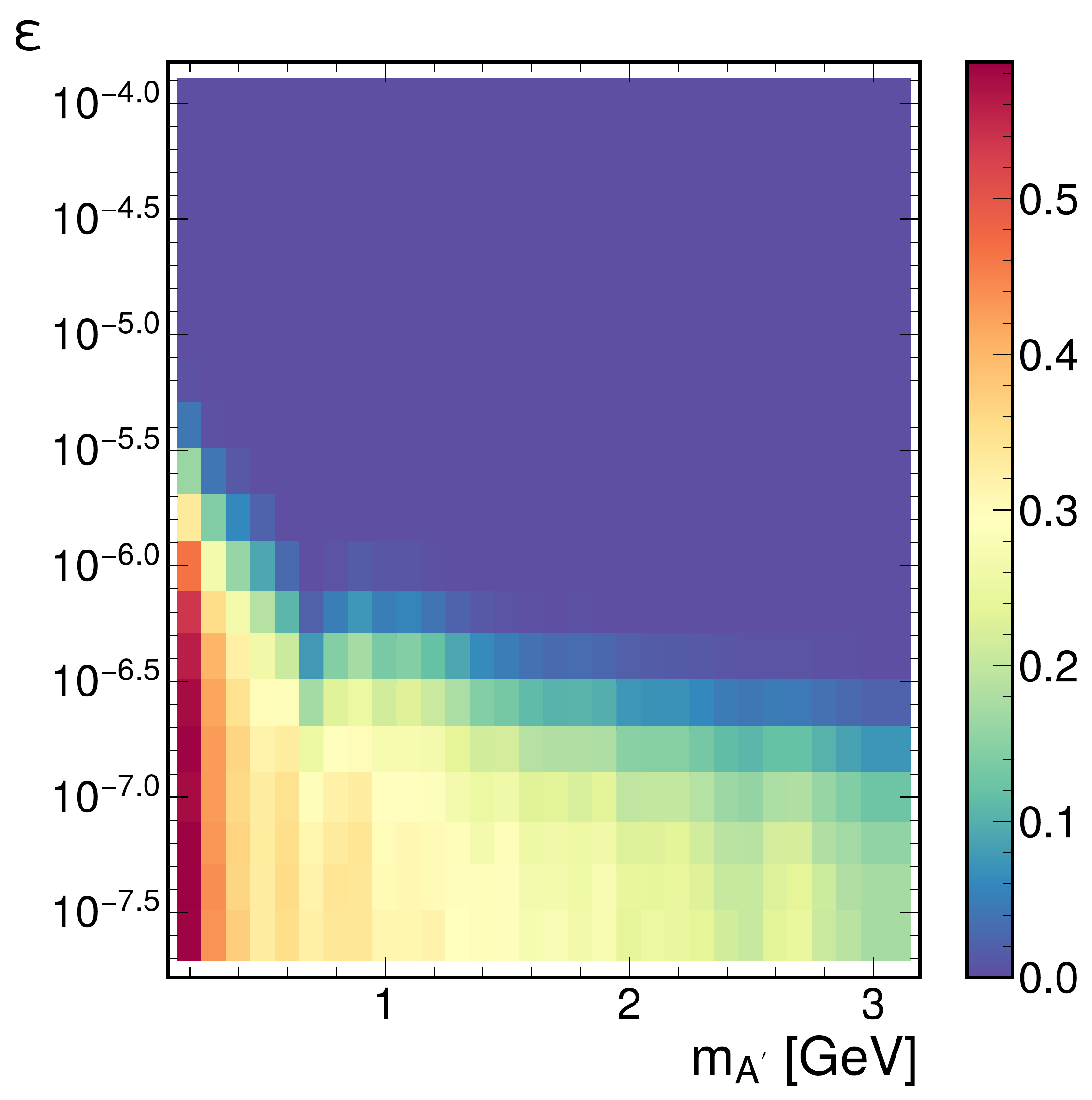}
    \caption{Ratio of events that have 2 tracks in $A^{\prime}\to \mu^+ \mu^-$ decays (between 5-6\,m) in different $A'$ masses and couplings.  
    }
    \label{fig:DP_track_accept}
\end{figure}

To study the dark photon signal sensitivity and acceptance, we generated $A'$ signal samples produced via Proton Bremsstrahlung according to the process outlined in Section \ref{sec:sigsim}.
These samples are scans of different $A'$ in different masses and couplings decaying to muon or electron pairs, with 10k events generated for each specific mass and coupling.
Figure \ref{fig:DP_track_accept} depicts the track-based event acceptance rate for $A^{\prime}\to \mu^+ \mu^-$ decays within 5-6\,m for muons.
Track-based acceptance is determined by whether or not the event has exactly 2 tracks - these being the two leptons from the $A^{\prime}$ decay.
The track acceptance rate ranges from $20-60\%$ in the relevant signal region (Figure \ref{fig:sesitivity_comparison}) for $\mu^+ \mu^-$ decays.
The displaced tracking efficiency is generally quite high for particles in detector acceptance, as shown in Sec.~\ref{sec:trackEff}, and it does not differ strongly when comparing electrons and muons so we expect similar performance.
We leave it to future work to study this in more detail for different final state particles and dark sector signatures. 

\newpage
\section{LongQuest Conceptual Design}
\label{sec:longquest}
\begin{figure}[tbh]
    \centering
    \includegraphics[width=0.9\textwidth]{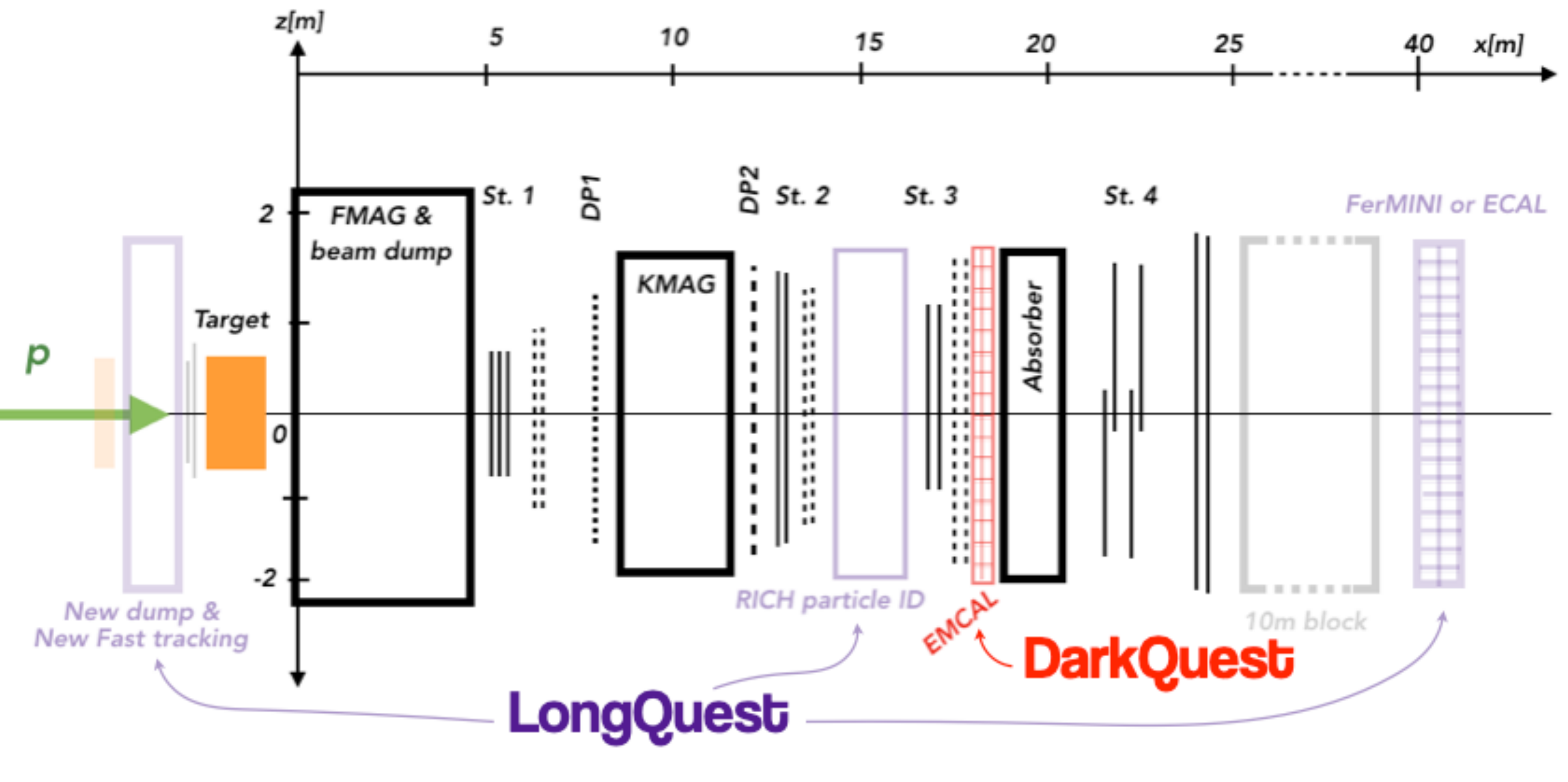}
    \caption{The proposed DarkQuest detector upgrade (in red) to the SpinQuest experiment.}
    \label{fig:longquest}
\end{figure}

This section considers future experimental configurations that can probe new physics models beyond those available to DarkQuest.
We propose LongQuest \cite{Tsai:2019buq}, a new, dedicated set of installations, including adding far detector(s) to search for long-lived and millicharged particles \cite{Kelly:2018brz}; and adding an additional beam-dump as well as tracking detectors.
The improvement for the particle and background identification is especially essential for the future $10^{20}$ POT run for the main injector.

\subsection{Long-Baseline Detectors for Long-lived and Particles}

Here, we propose the placement of detectors further downstream from the present location of the primary SpinQuest detector, utilizing the empty spaces behind the 10-meter iron block downstream from SpinQuest Station 4. This is an environment with much less background, and the installation of new detectors will have minimal interference with SpinQuest scientific projects. In particular, one can install another ECAL from the PHENIX experiment. This would provide an essentially zero-background environment (applying appropriate cuts for beam-related and cosmic-ray backgrounds) in which to search for displaced decays. In addition, one can also study millicharged particles by utilizing a FerMINI setup \cite{Kelly:2018brz}, placing a scintillation detector \cite{Haas:2014dda}  at this location to search for millicharged particles produced at the target.

\subsection{Improved particle ID and background rejection}

There are several additions to the DarkQuest detector that would enhance its capabilities beyond the proposed installation of an ECal.
First, one can install a Ring-Imaging Cherenkov (RICH) detector or a Hadron Blind Detector (HBD) \cite{Akiba:2000up,HBD} to separate pions and electrons. The main advantage of these detectors is to remove or reduce the background from neutral kaon decays, which could lead to $\sim 10^3$ decay events during a $10^{20}$ POT run. More specifically, one can install the PHENIX RICH detector \cite{Liu:2017ryd} between stations 2 and 3 to differentiate between pions and electrons.

Another reason to install RICH or HBD detectors is to allow for the possibility of turning off or tuning the magnetic field of the KMag to improve the sensitivity to new physics models with low-energy charged final states \cite{Berlin:2018pwi}.
The strong magnetic fields of the KMag would remove new physics events that involve 3-body decays and lower-energy visible decay products from new physics models, including inelastic dark matter.
Turning off or tuning the magnetic field of the KMag would allow these final states to be seen by the detector, and the effects of different energy thresholds are discussed in \cite{Tsai:2019buq}.
Since the magnetic field filters out the soft SM background, one would need to compensate for the loss of this background-reduction with additional detector installations (the aforementioned RICH detector or HBD additions).

Finally, one can consider extending the fiducial decay volume, as a large decay volume enhances the power of searches for dark sector particles with visible decay products. For example, one can extend the DarkQuest displaced decay volume from the previously discussed 5\,m – 12\,m to 5\,m – 18\,m, by installing an ECAL between stations 3 and 4 or placing it behind station 4. 
One could even extend the maximum length for the decay volume to beyond the current 18-meter benchmark by improving or replacing the current station 3 detectors with better instrumentation. Increasing the size of the decay volume would roughly increase the event rate proportionally when the decay length of the particle is much longer than the length of the detector. Of course, the background in an extended decay volume could also be larger, but such a configuration is still worth considering as it is low-cost and can be installed in the near future.
The improved DarkQuest detector with reduced background is discussed in \cite{Tsai:2019buq} with specific simulations with several scenarios on minimal dark photon and inelastic dark matter models.

\subsection{New Front-Dump and Fast-Tracking}

One of the strengths of the DarkQuest experiment is its ability to search for dark-sector particles with sizable couplings that would decay before reaching detectors at experiments with longer baselines.
Here, we consider an experimental configuration that can potentially study prompt decays of dark-sector particles beyond the DarkQuest capability. One idea under consideration is the addition of another beam dump in front of the current target and the FMag, with a fast-tracking detector in between the new beam dump and the FMag to help signature analysis and background reduction. The difficulty of this setup is the large SM background, as a large number of dimuons would be produced promptly through SM processes. However, since the measurement of the SM dimuon signature is one of the primary goals for SpinQuest \cite{SeaQuest_slides}, the dedicated simulation of the SM events is an ongoing effort.


\newpage
\section{Outlook}
\label{sec:outlook}
In this paper, we have laid out the physics case for the DarkQuest experiment to search for dark sector long-lived particles at the Fermilab 120\,GeV main injector. Since the experiment builds off existing investment in the SeaQuest and SpinQuest experiments, DarkQuest can be realized for modest cost and on a short timescale.  

We lay out a number of dark sector scenarios that can be searched for with the DarkQuest experiment.  The benchmark dark sector signatures are well-aligned with the accelerator-based fixed-target benchmark scenarios for invisible, visible and rich dark sector signatures.  DarkQuest, because of its compact detector concept on the 120\,GeV proton beam line, is uniquely sensitive to large uncovered parts of the coupling-mass phase space of many dark sector models. DarkQuest is also complementary to other experimental concepts to test dark sector models, as e.g. the LDMX experiment. 

Furthermore, we describe in some detail the key aspects of the detector as well as the modest proposed upgrades.  The EMCal detector relies on mature detector technologies.  The EMCal itself will be a reused Shashlik-style calorimeter from the PHENIX experiment and readout electronics will rely on existing designs from the EMPHATIC experiment.  We also propose an additional tracking layer which will extend the acceptance of DarkQuest for displaced signatures and will provide further background rejection power to improve the detector performance, particularly in high intensity beam bunches.  

A number of novel GEANT-based full simulation studies have been performed to better understand the capabilities of the DarkQuest experiment.  Extensive signal and background simulation sample  generation has been performed, including challenging high statistics background event generation.  Studies of both hodoscope and calorimeter triggers show that we can capture dark sector signatures with reasonable efficiency. New displaced tracking and vertexing algorithms have been introduced with good momentum resolution and efficiency. This demonstrates that we can reconstruct displaced dark sector signatures in busy proton beam dump detector signatures.  We also have demonstrated calorimeter clustering and particle identification which discriminates between electron, pion, and muon detector signatures. 

Finally we also discuss a broader dark sector program building on the initially proposed DarkQuest experiment.  The LongQuest idea extends the DarkQuest concept to bring even further sensitivity of this experiment to dark sector scenarios and could be part of a much more extensive dark sectors program on the Fermilab 120\,GeV beamline.  

\clearpage
\bibliography{references}

\begin{thebibliography}{10}
\urlstyle{rm}
\expandafter\ifx\csname url\endcsname\relax
  \def\url#1{\texttt{#1}}\fi
\expandafter\ifx\csname urlprefix\endcsname\relax\def\urlprefix{URL }\fi
\expandafter\ifx\csname doiprefix\endcsname\relax\def\doiprefix{DOI: }\fi
\providecommand{\bibinfo}[2]{#2}
\providecommand{\eprint}[2][]{\url{#2}}

\bibitem{BRNreport}
\bibinfo{author}{Kolb, R.} \emph{et~al.}
\newblock \bibinfo{title}{{Basic Research Needs for Dark-Matter Small Projects
  New Initiatives}}.
\newblock
  \bibinfo{howpublished}{\url{https://science.energy.gov/~/media/hep/pdf/Reports/Dark_Matter_New_Initiatives_rpt.pdf}}
  (\bibinfo{year}{2019}).

\bibitem{battaglieriUSCosmicVisions2017}
\bibinfo{author}{Battaglieri, M.} \emph{et~al.}
\newblock \bibinfo{journal}{\bibinfo{title}{{{US Cosmic Visions}}: {{New
  Ideas}} in {{Dark Matter}} 2017: {{Community Report}}}}.
\newblock {\emph{\JournalTitle{arXiv:1707.04591 [astro-ph, physics:hep-ex,
  physics:hep-ph]}}}  (\bibinfo{year}{2017}).
\newblock \eprint{1707.04591}.

\bibitem{alexanderDarkSectors20162016c}
\bibinfo{author}{Alexander, J.} \emph{et~al.}
\newblock \bibinfo{journal}{\bibinfo{title}{Dark {{Sectors}} 2016 {{Workshop}}:
  {{Community Report}}}}.
\newblock {\emph{\JournalTitle{arXiv:1608.08632 [astro-ph, physics:hep-ex,
  physics:hep-ph, physics:nucl-ex]}}}  (\bibinfo{year}{2016}).
\newblock \eprint{1608.08632}.

\bibitem{seaquestcollaborationSeaQuestSpectrometerFermilab2017a}
\bibinfo{author}{Collaboration, S.} \emph{et~al.}
\newblock \bibinfo{journal}{\bibinfo{title}{The {{SeaQuest Spectrometer}} at
  {{Fermilab}}}}.
\newblock {\emph{\JournalTitle{arXiv:1706.09990 [nucl-ex, physics:physics]}}}
  (\bibinfo{year}{2017}).
\newblock \eprint{1706.09990}.

\bibitem{Berlin:2018pwi}
\bibinfo{author}{Berlin, A.}, \bibinfo{author}{Gori, S.},
  \bibinfo{author}{Schuster, P.} \& \bibinfo{author}{Toro, N.}
\newblock \bibinfo{journal}{\bibinfo{title}{{Dark Sectors at the Fermilab
  SeaQuest Experiment}}}.
\newblock {\emph{\JournalTitle{Phys. Rev. D}}} \textbf{\bibinfo{volume}{98}},
  \bibinfo{pages}{035011}, \doiprefix\url{10.1103/PhysRevD.98.035011}
  (\bibinfo{year}{2018}).
\newblock \eprint{1804.00661}.

\bibitem{Batell:2020vqn}
\bibinfo{author}{Batell, B.}, \bibinfo{author}{Evans, J.~A.},
  \bibinfo{author}{Gori, S.} \& \bibinfo{author}{Rai, M.}
\newblock \bibinfo{journal}{\bibinfo{title}{{Dark Scalars and Heavy Neutral
  Leptons at DarkQuest}}}.
\newblock {\emph{\JournalTitle{JHEP}}} \textbf{\bibinfo{volume}{05}},
  \bibinfo{pages}{049}, \doiprefix\url{10.1007/JHEP05(2021)049}
  (\bibinfo{year}{2021}).
\newblock \eprint{2008.08108}.

\bibitem{Blinov:2021say}
\bibinfo{author}{Blinov, N.}, \bibinfo{author}{Kowalczyk, E.} \&
  \bibinfo{author}{Wynne, M.}
\newblock \bibinfo{journal}{\bibinfo{title}{{Axion-like particle searches at
  DarkQuest}}}.
\newblock {\emph{\JournalTitle{JHEP}}} \textbf{\bibinfo{volume}{02}},
  \bibinfo{pages}{036}, \doiprefix\url{10.1007/JHEP02(2022)036}
  (\bibinfo{year}{2022}).
\newblock \eprint{2112.09814}.

\bibitem{Berlin:2018tvf}
\bibinfo{author}{Berlin, A.}, \bibinfo{author}{Blinov, N.},
  \bibinfo{author}{Gori, S.}, \bibinfo{author}{Schuster, P.} \&
  \bibinfo{author}{Toro, N.}
\newblock \bibinfo{journal}{\bibinfo{title}{{Cosmology and Accelerator Tests of
  Strongly Interacting Dark Matter}}}.
\newblock {\emph{\JournalTitle{Phys. Rev. D}}} \textbf{\bibinfo{volume}{97}},
  \bibinfo{pages}{055033}, \doiprefix\url{10.1103/PhysRevD.97.055033}
  (\bibinfo{year}{2018}).
\newblock \eprint{1801.05805}.

\bibitem{Gardner:2015wea}
\bibinfo{author}{Gardner, S.}, \bibinfo{author}{Holt, R.~J.} \&
  \bibinfo{author}{Tadepalli, A.~S.}
\newblock \bibinfo{journal}{\bibinfo{title}{{New Prospects in Fixed Target
  Searches for Dark Forces with the SeaQuest Experiment at Fermilab}}}.
\newblock {\emph{\JournalTitle{Phys. Rev.}}} \textbf{\bibinfo{volume}{D93}},
  \bibinfo{pages}{115015}, \doiprefix\url{10.1103/PhysRevD.93.115015}
  (\bibinfo{year}{2016}).
\newblock \eprint{1509.00050}.

\bibitem{Pospelov:2007mp}
\bibinfo{author}{Pospelov, M.}, \bibinfo{author}{Ritz, A.} \&
  \bibinfo{author}{Voloshin, M.~B.}
\newblock \bibinfo{journal}{\bibinfo{title}{{Secluded WIMP Dark Matter}}}.
\newblock {\emph{\JournalTitle{Phys. Lett.}}} \textbf{\bibinfo{volume}{B662}},
  \bibinfo{pages}{53--61}, \doiprefix\url{10.1016/j.physletb.2008.02.052}
  (\bibinfo{year}{2008}).
\newblock \eprint{0711.4866}.

\bibitem{Dobrich:2019dxc}
\bibinfo{author}{D\"obrich, B.}, \bibinfo{author}{Jaeckel, J.} \&
  \bibinfo{author}{Spadaro, T.}
\newblock \bibinfo{journal}{\bibinfo{title}{{Light in the beam dump - ALP
  production from decay photons in proton beam-dumps}}}.
\newblock {\emph{\JournalTitle{JHEP}}} \textbf{\bibinfo{volume}{05}},
  \bibinfo{pages}{213}, \doiprefix\url{10.1007/JHEP05(2019)213}
  (\bibinfo{year}{2019}).
\newblock \bibinfo{note}{[Erratum: JHEP 10, 046 (2020)]}, \eprint{1904.02091}.

\bibitem{TuckerSmith:2001hy}
\bibinfo{author}{Tucker-Smith, D.} \& \bibinfo{author}{Weiner, N.}
\newblock \bibinfo{journal}{\bibinfo{title}{{Inelastic dark matter}}}.
\newblock {\emph{\JournalTitle{Phys. Rev.}}} \textbf{\bibinfo{volume}{D64}},
  \bibinfo{pages}{043502}, \doiprefix\url{10.1103/PhysRevD.64.043502}
  (\bibinfo{year}{2001}).
\newblock \eprint{hep-ph/0101138}.

\bibitem{Hochberg:2014dra}
\bibinfo{author}{Hochberg, Y.}, \bibinfo{author}{Kuflik, E.},
  \bibinfo{author}{Volansky, T.} \& \bibinfo{author}{Wacker, J.~G.}
\newblock \bibinfo{journal}{\bibinfo{title}{{Mechanism for Thermal Relic Dark
  Matter of Strongly Interacting Massive Particles}}}.
\newblock {\emph{\JournalTitle{Phys. Rev. Lett.}}}
  \textbf{\bibinfo{volume}{113}}, \bibinfo{pages}{171301},
  \doiprefix\url{10.1103/PhysRevLett.113.171301} (\bibinfo{year}{2014}).
\newblock \eprint{1402.5143}.

\bibitem{Hochberg:2014kqa}
\bibinfo{author}{Hochberg, Y.}, \bibinfo{author}{Kuflik, E.},
  \bibinfo{author}{Murayama, H.}, \bibinfo{author}{Volansky, T.} \&
  \bibinfo{author}{Wacker, J.~G.}
\newblock \bibinfo{journal}{\bibinfo{title}{{Model for Thermal Relic Dark
  Matter of Strongly Interacting Massive Particles}}}.
\newblock {\emph{\JournalTitle{Phys. Rev. Lett.}}}
  \textbf{\bibinfo{volume}{115}}, \bibinfo{pages}{021301},
  \doiprefix\url{10.1103/PhysRevLett.115.021301} (\bibinfo{year}{2015}).
\newblock \eprint{1411.3727}.

\bibitem{Beacham:2019nyx}
\bibinfo{author}{Beacham, J.} \emph{et~al.}
\newblock \bibinfo{journal}{\bibinfo{title}{{Physics Beyond Colliders at CERN:
  Beyond the Standard Model Working Group Report}}}.
\newblock {\emph{\JournalTitle{J. Phys. G}}} \textbf{\bibinfo{volume}{47}},
  \bibinfo{pages}{010501}, \doiprefix\url{10.1088/1361-6471/ab4cd2}
  (\bibinfo{year}{2020}).
\newblock \eprint{1901.09966}.

\bibitem{Berlin:2018bsc}
\bibinfo{author}{Berlin, A.}, \bibinfo{author}{Blinov, N.},
  \bibinfo{author}{Krnjaic, G.}, \bibinfo{author}{Schuster, P.} \&
  \bibinfo{author}{Toro, N.}
\newblock \bibinfo{journal}{\bibinfo{title}{{Dark Matter, Millicharges, Axion
  and Scalar Particles, Gauge Bosons, and Other New Physics with LDMX}}}.
\newblock {\emph{\JournalTitle{Phys. Rev.}}} \textbf{\bibinfo{volume}{D99}},
  \bibinfo{pages}{075001}, \doiprefix\url{10.1103/PhysRevD.99.075001}
  (\bibinfo{year}{2019}).
\newblock \eprint{1807.01730}.

\bibitem{Kahn:2018cqs}
\bibinfo{author}{Kahn, Y.}, \bibinfo{author}{Krnjaic, G.},
  \bibinfo{author}{Tran, N.} \& \bibinfo{author}{Whitbeck, A.}
\newblock \bibinfo{journal}{\bibinfo{title}{{M$^{3}$: a new muon missing
  momentum experiment to probe (g-2)$_{\mu}$ and dark matter at Fermilab}}}.
\newblock {\emph{\JournalTitle{JHEP}}} \textbf{\bibinfo{volume}{09}},
  \bibinfo{pages}{153}, \doiprefix\url{10.1007/JHEP09(2018)153}
  (\bibinfo{year}{2018}).
\newblock \eprint{1804.03144}.

\bibitem{PHENIX:2003fvo}
\bibinfo{author}{Aphecetche, L.} \emph{et~al.}
\newblock \bibinfo{journal}{\bibinfo{title}{{PHENIX calorimeter}}}.
\newblock {\emph{\JournalTitle{Nucl. Instrum. Meth. A}}}
  \textbf{\bibinfo{volume}{499}}, \bibinfo{pages}{521--536},
  \doiprefix\url{10.1016/S0168-9002(02)01954-X} (\bibinfo{year}{2003}).

\bibitem{berlinDarkSectorsFermilab2018}
\bibinfo{author}{Berlin, A.}, \bibinfo{author}{Gori, S.},
  \bibinfo{author}{Schuster, P.} \& \bibinfo{author}{Toro, N.}
\newblock \bibinfo{journal}{\bibinfo{title}{{Dark Sectors at the Fermilab
  SeaQuest Experiment}}}.
\newblock {\emph{\JournalTitle{Phys. Rev.}}} \textbf{\bibinfo{volume}{D98}},
  \bibinfo{pages}{035011}, \doiprefix\url{10.1103/PhysRevD.98.035011}
  (\bibinfo{year}{2018}).
\newblock \eprint{1804.00661}.

\bibitem{STARFCS}
\bibinfo{author}{{STAR Collaboration}}.
\newblock \bibinfo{title}{{The STAR Forward Calorimeter System and Forward
  Tracking System}}.
\newblock
  \bibinfo{howpublished}{\url{https://drupal.star.bnl.gov/STAR/files/Proposal.ForwardUpgrade.Nov_.2018.Review.pdf}}
  (\bibinfo{year}{2018}).

\bibitem{emphatic}
\bibinfo{author}{Akaishi, T.} \emph{et~al.}
\newblock \bibinfo{title}{{EMPHATIC: A Proposed Experiment to Measure Hadron
  Scattering and Production Cross Sections for Improved Neutrino Flux
  Predictions}} (\bibinfo{year}{2019}).
\newblock \eprint{1912.08841}.

\bibitem{HyperCP:2004kbv}
\bibinfo{author}{Burnstein, R.~A.} \emph{et~al.}
\newblock \bibinfo{journal}{\bibinfo{title}{{HyperCP: A High-rate spectrometer
  for the study of charged hyperon and kaon decays}}}.
\newblock {\emph{\JournalTitle{Nucl. Instrum. Meth. A}}}
  \textbf{\bibinfo{volume}{541}}, \bibinfo{pages}{516--565},
  \doiprefix\url{10.1016/j.nima.2004.12.031} (\bibinfo{year}{2005}).
\newblock \eprint{hep-ex/0405034}.

\bibitem{Blumlein:2013cua}
\bibinfo{author}{Bl\"umlein, J.} \& \bibinfo{author}{Brunner, J.}
\newblock \bibinfo{journal}{\bibinfo{title}{{New Exclusion Limits on Dark Gauge
  Forces from Proton Bremsstrahlung in Beam-Dump Data}}}.
\newblock {\emph{\JournalTitle{Phys. Lett. B}}} \textbf{\bibinfo{volume}{731}},
  \bibinfo{pages}{320--326}, \doiprefix\url{10.1016/j.physletb.2014.02.029}
  (\bibinfo{year}{2014}).
\newblock \eprint{1311.3870}.

\bibitem{Gorbunov:2014wqa}
\bibinfo{author}{Gorbunov, D.}, \bibinfo{author}{Makarov, A.} \&
  \bibinfo{author}{Timiryasov, I.}
\newblock \bibinfo{journal}{\bibinfo{title}{{Decaying light particles in the
  SHiP experiment: Signal rate estimates for hidden photons}}}.
\newblock {\emph{\JournalTitle{Phys. Rev. D}}} \textbf{\bibinfo{volume}{91}},
  \bibinfo{pages}{035027}, \doiprefix\url{10.1103/PhysRevD.91.035027}
  (\bibinfo{year}{2015}).
\newblock \eprint{1411.4007}.

\bibitem{deNiverville:2016rqh}
\bibinfo{author}{deNiverville, P.}, \bibinfo{author}{Chen, C.-Y.},
  \bibinfo{author}{Pospelov, M.} \& \bibinfo{author}{Ritz, A.}
\newblock \bibinfo{journal}{\bibinfo{title}{{Light dark matter in neutrino
  beams: production modelling and scattering signatures at MiniBooNE, T2K and
  SHiP}}}.
\newblock {\emph{\JournalTitle{Phys. Rev. D}}} \textbf{\bibinfo{volume}{95}},
  \bibinfo{pages}{035006}, \doiprefix\url{10.1103/PhysRevD.95.035006}
  (\bibinfo{year}{2017}).
\newblock \eprint{1609.01770}.

\bibitem{Sjostrand:2014zea}
\bibinfo{author}{Sj\"ostrand, T.} \emph{et~al.}
\newblock \bibinfo{journal}{\bibinfo{title}{{An introduction to PYTHIA 8.2}}}.
\newblock {\emph{\JournalTitle{Comput. Phys. Commun.}}}
  \textbf{\bibinfo{volume}{191}}, \bibinfo{pages}{159--177},
  \doiprefix\url{10.1016/j.cpc.2015.01.024} (\bibinfo{year}{2015}).
\newblock \eprint{1410.3012}.

\bibitem{AGOSTINELLI2003250}
\bibinfo{journal}{\bibinfo{title}{Geant4---a simulation toolkit}}.
\newblock {\emph{\JournalTitle{Nuclear Instruments and Methods in Physics
  Research Section A: Accelerators, Spectrometers, Detectors and Associated
  Equipment}}} \textbf{\bibinfo{volume}{506}}, \bibinfo{pages}{250--303},
  \doiprefix\url{https://doi.org/10.1016/S0168-9002(03)01368-8}
  (\bibinfo{year}{2003}).

\bibitem{Sanftl:2014}
\bibinfo{author}{Sanftl, F.}
\newblock \emph{\bibinfo{title}{{Developments of Tracking Methods of Muon Pairs
  with SeaQuest Spectrometer}}}.
\newblock Ph.D. thesis, \bibinfo{school}{Tokyo Institute of Technology}
  (\bibinfo{year}{2014}).

\bibitem{scikit-learn}
\bibinfo{author}{Pedregosa, F.} \emph{et~al.}
\newblock \bibinfo{journal}{\bibinfo{title}{Scikit-learn: Machine learning in
  {P}ython}}.
\newblock {\emph{\JournalTitle{Journal of Machine Learning Research}}}
  \textbf{\bibinfo{volume}{12}}, \bibinfo{pages}{2825--2830}
  (\bibinfo{year}{2011}).

\bibitem{Tsai:2019buq}
\bibinfo{author}{Tsai, Y.-D.}, \bibinfo{author}{deNiverville, P.} \&
  \bibinfo{author}{Liu, M.~X.}
\newblock \bibinfo{journal}{\bibinfo{title}{{Dark Photon and Muon $g-2$
  Inspired Inelastic Dark Matter Models at the High-Energy Intensity
  Frontier}}}.
\newblock {\emph{\JournalTitle{Phys. Rev. Lett.}}}
  \textbf{\bibinfo{volume}{126}}, \bibinfo{pages}{181801},
  \doiprefix\url{10.1103/PhysRevLett.126.181801} (\bibinfo{year}{2021}).
\newblock \eprint{1908.07525}.

\bibitem{Kelly:2018brz}
\bibinfo{author}{Kelly, K.~J.} \& \bibinfo{author}{Tsai, Y.-D.}
\newblock \bibinfo{journal}{\bibinfo{title}{{Proton fixed-target scintillation
  experiment to search for millicharged dark matter}}}.
\newblock {\emph{\JournalTitle{Phys. Rev. D}}} \textbf{\bibinfo{volume}{100}},
  \bibinfo{pages}{015043}, \doiprefix\url{10.1103/PhysRevD.100.015043}
  (\bibinfo{year}{2019}).
\newblock \eprint{1812.03998}.

\bibitem{Haas:2014dda}
\bibinfo{author}{Haas, A.}, \bibinfo{author}{Hill, C.~S.},
  \bibinfo{author}{Izaguirre, E.} \& \bibinfo{author}{Yavin, I.}
\newblock \bibinfo{journal}{\bibinfo{title}{{Looking for milli-charged
  particles with a new experiment at the LHC}}}.
\newblock {\emph{\JournalTitle{Phys. Lett. B}}} \textbf{\bibinfo{volume}{746}},
  \bibinfo{pages}{117--120}, \doiprefix\url{10.1016/j.physletb.2015.04.062}
  (\bibinfo{year}{2015}).
\newblock \eprint{1410.6816}.

\bibitem{Akiba:2000up}
\bibinfo{author}{Akiba, Y.} \emph{et~al.}
\newblock \bibinfo{journal}{\bibinfo{title}{{The PHENIX ring imaging Cherenkov
  detector}}}.
\newblock {\emph{\JournalTitle{Nucl. Instrum. Meth.}}}
  \textbf{\bibinfo{volume}{A453}}, \bibinfo{pages}{279--283},
  \doiprefix\url{10.1016/S0168-9002(00)00645-8} (\bibinfo{year}{2000}).

\bibitem{HBD}
\bibinfo{author}{{Azmoun {\it et al.}}}
\newblock \bibinfo{title}{{Conceptual Design Report on a HBD Upgrade for the
  PHENIX Detector}}.
\newblock
  \bibinfo{howpublished}{{\url{https://www.phenix.bnl.gov/WWW/TPCHBD/HBD_CDR.pdf}}}.

\bibitem{Liu:2017ryd}
\bibinfo{author}{Liu, M.~X.}
\newblock \bibinfo{journal}{\bibinfo{title}{{Prospects of direct search for
  dark photon and dark Higgs in SeaQuest/E1067 experiment at the Fermilab main
  injector}}}.
\newblock {\emph{\JournalTitle{Mod. Phys. Lett.}}}
  \textbf{\bibinfo{volume}{A32}}, \bibinfo{pages}{1730008},
  \doiprefix\url{10.1142/S0217732317300087} (\bibinfo{year}{2017}).

\end{thebibliography}
\end{document}